\documentclass[12pt]{article}
\usepackage[margin=0.7in, paperwidth=8.5in,paperheight=11in]{geometry}
\usepackage{authblk}
\usepackage{amsmath}
\usepackage{amssymb}
\usepackage{breqn}
\usepackage{dsfont}
\usepackage{bm}
\usepackage{color}
\usepackage[makeroom]{cancel}
\usepackage[toc,page]{appendix}
\newcommand{\be}{\begin{equation}}
\newcommand{\ee}{\end{equation}}
\newcommand{\eq}[1]{\begin{align}#1\end{align}}
\newcommand{\bea}{\begin{array}}
\newcommand{\ea}{\end{array}}
\newcommand{\beqa}{\begin{eqnarray}}
\newcommand{\eeqa}{\end{eqnarray}}
\newcommand{\bean}{\begin{eqnarray*}}
\newcommand{\eean}{\end{eqnarray*}}

\newcommand{\del}{\partial}
\newcommand{\Del}{\nabla}

\newcommand{\nn}{\nonumber}
\newcommand{\tr}{\mbox{Tr}}
\newcommand{\xu}{\phantom{x}}

\numberwithin{equation}{section}
\usepackage{graphicx} 
\graphicspath{ {./images/} }
\usepackage{float}
\usepackage{accents}
\usepackage{collectbox}
\usepackage{amsfonts}


\let\chapter\section
\let\section\subsection
\let\subsection\subsubsection

\usepackage{subfiles} % Best loaded last in the preamble

\begin{document}

\title{Step-by-Step Canonical Quantum Gravity:\\
 Part I: Ashtekar's New Variables\\
 \vspace{1cm}}

\author{Lei Lu\footnote{lelu@calpoly.edu},\, Philip A. May\footnote{pamay@calpoly.edu}}
%\vskip 1cm
\affil{Physics Department, California Polytechnic State University,\\ San Luis Obispo, California 93407, USA\\}
\date{}
\maketitle

\abstract{Canonical quantum gravity was first developed by Abhay Ashtekar, Lee Smolin, Carlo Rovelli and their collaborators in the late 1980s. It was a major breakthrough that successfully brought Einstein's theory of General Relativity (GR) into a Yang-Mills-type gauge theory. A new era of quantum gravity research has since started, and with decades of continued efforts from a relatively small community, the area now known as Loop Quantum Gravity (LQG) has flourished, making it a promising theory of quantum gravity. Due to its incredibly high level of complexity, many technical details were left out in introductory texts on LQG. In particular, resources that are appropriate to the undergraduate level are extremely limited. Consequently, there exists a huge gap between the knowledge base of an undergraduate physics major and the necessary readiness to carry out LQG research. 
In an effort to fill this gap, we aim to develop a pedagogical user guide that provides a step-by-step walk-through of canonical quantum gravity, without compromising necessary technical details. We hope that our attempt will bring more exposure to undergraduates on the exciting early developments of canonical quantum gravity, and provide them with the necessary foundation to explore active research fields such as black hole thermodynamics, Wheeler-DeWitt equation, and so on. This work will also serve as a solid base for anyone hoping to pursue further study in LQG at a higher level. 

\clearpage

\tableofcontents

\chapter{Introduction}

A defining moment that kick-started the canonical quantum gravity program may be 
the year 1986, when Abhay Ashtekar published his groundbreaking work on the new canonical variables of General Relativity \cite{Ashtekar:1986, Ashtekar:1987}. Early attempts at a Hamiltonian framework of GR could actually date back to the 1950s, with the pioneering works of Paul A.M. Dirac \cite{Dirac:1950pj, Dirac:1958} and Peter G. Bergmann \cite{Bergman:1949,Bergman:1951}. Significant progress was made soon afterward in the 1960s after Richard Arnowitt, Stanley Deser, and Charles W. Misner introduced their famous ADM variables that carried direct geometric interpretations and greatly simplified the Hamiltonian dynamics of GR \cite{ADM, new ADM}. These attempts led to a well-structured Hamiltonian framework of GR, but upon careful analysis of the constraints, it was faced with a persisting issue of non-polynomial Hamiltonian constraint, rendering its quantization ill-defined. The field of canonical approach remained somewhat stuck for over two decades until the introduction of Ashtekar's new variables, which revived the field and gave new hopes to the canonical approach of quantum gravity. Inspired by the new variables, further developments of quantum gravity occurred in the late 1980s to early 1990s, carried out by Lee Smolin, Ted Jacobson, and Carlo Rovelli, with their foundational work on the Lagrangian formulation of GR in terms of the loop representation \cite{Jacobson,Jacobson:1987,Rovelli:1987df,Samuel:1987td,Jacobson:1988,Rovelli:1989za}, which eventually evolved to a promising quantum theory of gravitation later known as Loop Quantum Gravity\footnote{For a comprehensive overview of the early history of canonical quantum gravity, refer to Thiemann \cite{Thiemann:2001}.}. Today, it continues to be an attractive field of research most notably for its background independence that is at the heart of the Equivalence Principle of GR, and its non-perturbative approach to quantum gravity as compared to other frameworks.

The long and winding history of early LQG development is not accidental. This is because GR, as a theory of space-time geometry, is already highly involved and convoluted as a classical theory. Thus, its quantization is an incredibly difficult undertaking. In the case of LQG, it took generations of brilliant minds over the decades to finally build it into a promising candidate for quantum gravity. Due to the limited resources on this subject in its early days, as well as many explicit details being largely absent (or hard to find) in the literature, learning LQG is an extremely challenging task, especially for those below the graduate level. 
This is in drastic contrast to adjacent frameworks of quantum gravity, such as string theory, where undergraduate research is achievable and relatively commonplace. This is largely due to the fact that string theory is a more studied field and thus, has a more developed infrastructure for introducing those new to the field. 
This being said, there does exist a wide range of graduate-level texts on LQG, such as \cite{Ashtekar, Romano, Rovelli book, covariant, Thiemann:2001, Gambini}, which excellently provide technical details but in many cases pose a formidable barrier to entry. Additionally, there are texts such as \cite{intro,Gambini:2020}, that are specifically targeted at an audience at the undergraduate level that does an exceptional job explaining the underlying concepts in LQG, but in many cases leave the reader with limited information on how to arrive at the important results of the theory. In particular, introductory texts that flesh out the details of the original self-dual formulation of canonical quantum gravity are seldom found.

Besides the lack of sufficient resources at a lower level, the current undergraduate physics curricula do not cover the necessary materials needed to carry out LQG research. To be more specific, the treatment of constrained Hamiltonian systems via the Dirac procedure, which is an essential tool in studying most physical systems, is rarely covered or properly introduced in depth in standard classical mechanics and quantum mechanics texts. The notion of gauge theory is mostly very lightly touched on at the undergraduate level. In the context of GR, advanced topics such as Lie derivative, ADM formalism, tetrad and spin connections, etc, that are essential to understanding LQG, are not part of the core curriculum even at the graduate level! 
%Most of the existing resources in the literature were targeted at the audience of beginning graduate students or experts in the field.

We believe there is room and the demand for LQG texts suitable for the undergraduate level, that is not only pedagogically focused but also meticulously detailed. Thus, we have taken it upon ourselves to create a comprehensible guide to LQG in an attempt to mend the knowledge gap between prospective students and LQG researchers.
Specifically, we follow a lecture-note style similar to \cite{super} and \cite{ads/cft}, that has a heavy focus on providing people with intuitive explanations of physics concepts, as well as step-by-step derivations/calculations of key mathematical results so that the readers do not necessarily need to rediscover every little technical detail from scratch by themselves. We also try to cover all the necessary background material so that the text gives a self-contained delivery of the subject. We are by no means claiming to reinvent the wheel, however, we are making an effort to clearly lay out all of the details, whilst maintaining an accessible presentation of the material.

We want to note that this work is only part 1 of a long-term project, which focuses on the exciting early development of the canonical LQG framework in terms of Ashtekar's new variables. During later developments of LQG, further modifications regarding the originally complex-valued Ashtekar variables have been proposed by Barbero \cite{Barbero,Barbero:1994}, Immirzi \cite{Immirzi}, and later generalized by Holst \cite{Holst}. Real-valued Ashtekar variables are more favored in modern treatments of LQG because of their immediate relevance to real GR and the corresponding Hamiltonian constraint in the quantum theory being better behaved \cite{Thiemann:2001}. However, as the first successful attempt of a canonical formulation of GR, the original self-dual formulation of Ashtekar is more appealing from a pedagogical standpoint. Specifically, because it was built by piecing together evidence bit by bit through trials and errors to eventually form a bigger picture, this type of ``bottom-up'' approach is better motivated by a clear, straightforward flow of logic.  Besides, it is an exciting part of history that's worth more exposure to a much wider audience, at least to the undergraduate level. 

As we have previously alluded to, LQG requires some advanced-level mathematics and physics to understand. Our ultimate goal, and endpoint of this text, is to formulate general relativity into a complex Yang-Mills-like gauge theory. All the necessary tools to understand this process are typically not covered in a conventional undergraduate physics degree which is why we begin with a preliminary section. In this section (section \ref{sec: pre}) we give a brief review of classical Lagrangian and Hamiltonian mechanics, and introduce the Poisson bracket. Next, we introduce the concept of a constrained Hamiltonian system with a few simple examples and introduce what is known as a Dirac constraint analysis. Finally, we lay the foundations for classical field theory, namely the functional derivative, and introduce some tricks useful for handling Poisson brackets in field theory. In section \ref{sec: maxwell}, we introduce gauge theory through the lens of  Maxwell's theory of electromagnetism. Here we derive the Hamiltonian of electromagnetism in terms of the canonical phase space variables, the gauge potential, and the electric field, and from it derive Faraday, and Ampere's law. Next, we introduce gauge transformations and perform a Dirac constraint analysis that unveils a first-class constraint responsible for generating gauge transformations. Using this gauge generator we derive the gauge transformations of the gauge potential and prove that the electric and magnetic fields are invariant under gauge transformations. 
In section \ref{sec yang} we introduce a more technical gauge theory, Yang-Mills theory, where we begin by introducing the basics of Yang-Mills and construct a Hamiltonian in a manner similar to the Maxwell section. Upon doing so, we perform a more in-depth Dirac constraint analysis on the Yang-Mills Hamiltonian and identify a similar generator of gauge transformations. With this generator, we identify the gauge transformations of the gauge potential and show that the electric field and field strength tensor are no longer gauge invariant quantities.
This apparent issue leads us to introduce the concept of holonomy, the trace of which is a  gauge invariant quantity in Yang-Mills theory. 
In section \ref{sec: pal}, we begin by giving an overview of some basic aspects of general relativity such as the metric, the connection, the Riemann curvature tensor, and the Einstein field equation. We then pivot to the Palatini formulation of GR where we introduce its fundamental variables, namely the tetrad and spin connection, and the Palatini action. After showing the equivalence between the newly introduced Palatini action and Einstein-Hilbert action, we introduce the Hamiltonian formulation of GR known as the ADM formalism. In doing so we introduce concepts from differential geometry such as diffeomorphism, Lie derivative, and extrinsic curvature. With these newly equipped concepts, we perform a $3+1$ decomposition of the Palatini action similar to what is done in the ADM formalism. Finally, we briefly explain how a pair of second-class constraints that arise stump any attempt of canonical quantization. 
Lastly in section \ref{sec: ash} we introduce the notion of self-duality and enforce the connection in the Palatini action be self-dual. From this updated form of the Palatini action we again perform a $3+1$ decomposition and ultimately arrive at the self-dual action in terms of the canonical variables, Ashtekar's self-dual connection, and the densitized triad. We then perform a final Dirac constraint analysis, evaluate the constraint algebra, and after imposing reality conditions arrive at a Hamiltonian theory of GR that resembles a Yang-Mills gauge theory.

Before we proceed to the main content, we would like to stress that the focus of our text is achieving Ashtekar's original self-dual formulation of GR and the unique calculations leading up to it. In doing so we come across subjects, such as Yang-Mills theory and GR, that have entire textbooks dedicated to them. Because of this we simply touch on the details that are most relevant to our end goal and leave the reader to find more ``conventional" knowledge on their own. We will point out the key references and make suggestions for further readings at the beginning of each section. %We recommend and encourage those interested to read other sources especially if one finds our work unsatisfactory.
Furthermore, we try to refer to those works to the best of our knowledge and apologize in advance for any unintentional omissions.

\chapter{Preliminary} \label{sec: pre}
We will dedicate the first section of this paper to a preliminary introduction to some important concepts and techniques that will be relevant to the paper. We prefer to introduce them at the very beginning instead of throwing them to the end in terms of appendices or simply referring to other texts so that this work is a self-contained coverage of the subject. In subsection 2.1 we briefly introduce index notation and Einstein summation convention. In subsection 2.2, we give a brief introduction to the constrained Hamiltonian mechanics, where the key concepts such as canonical momenta, Poisson brackets, and constraints are introduced. In subsection 2.3, we introduced the constrained Hamiltonian system in a field theory, where we focused on illustrating the difference brought in by functionals and introduced the functional derivative along with the generalized Poisson brackets in field theories. Readers interested in more details on classical mechanics and classical field theory can refer to \cite{Taylor} and \cite{intro} respectively.

\section{Intro to Index Notation}

In this text we take advantage of what is known as index or tensor notation and Einstein summation convention. We will briefly introduce these conventions here and for those looking for a more in-depth explanation, we direct you to Carroll's book on GR \cite{Carroll}.

We begin with the familiar position vector, which is denoted as $x^i$ where $i=1,2,3$, for example, in Cartesian coordinates
\eq{
x^i = 
\begin{bmatrix}
x^1 \\
x^2 \\ x^3  
\end{bmatrix} = 
\begin{bmatrix}
x \\
y \\ z  
\end{bmatrix}.
}
Notice $x^i$ could also represent position in spherical coordinates i.e. $x^i = (r,\theta,\phi)$, which shows one of the important features index notation has, namely the ability to represent coordinate invariant expressions.
In relativity we commonly use ``four-vectors" with Greek indices to specify position in spacetime denoted as $x^\mu$ where $\mu = 0,1,2,3$, or explicitly as
\eq{
x^\mu = 
\begin{bmatrix}
x^0 \\ x^1 \\
x^2 \\ x^3  
\end{bmatrix} = 
\begin{bmatrix}
ct \\ x \\
y \\ z  
\end{bmatrix}
}
where the factor of $c$, the speed of light, is used to maintain units of distance. However, we will use the popular geometrized units where $c=1$ to simplify calculations. 

Einstein summation notation is simply the implication that any repeated Latin index is summed over from one to three and similarly any repeated Greek index is summed over from zero to three. For example the dot product between two of the same vectors can be written succinctly as
\eq{
x^i x_i := \sum_{i=1}^3 x^i x_i = (x^1)^2 + (x^2)^2 + (x^3)^2.
}
Notice we sum over repeated indices where one is lowered and one is raised. One can lower and raise indices using a tensor quantity known as the metric $g_{\mu\nu}$ and its inverse $g^{\mu\nu}$ as follows
\eq{
x^\mu g_{\mu\nu} &= x_{\nu},
\nn \\
x_\mu g^{\mu\nu} &= x^\nu. \label{raise}
}
The components of this metric and its inverse are unimportant for now, however, those interested can look at the beginning of section \ref{sec: pal}.
What's important is that with our ``most positive" metric convention $(-,+,+,+)$, which we use throughout the text aside from example \ref{rel particle}, we can freely raise and lower spatial indices, and require only a sign flip to raise and lower the temporal index i.e.
\eq{
x^i = x_i \qquad \text{and} \qquad x^0 = - x_0.
}
The pair of expressions above will be used abundantly throughout the text. Coming back to Einstein summation convention, we will occasionally be sloppy and have repeated (summed over) indices that are both high or both low as follows
\eq{x^ix_i = x^ix^i = x_i x_i.}
This however, is not an issue so long as the quantities are spatial since our spatial indices can be freely lowered and raised.

\section{Lagrangian and Hamiltonian Mechanics}
The action is a scalar quantity that is fundamental to the mechanics of a physical system and will almost always be the starting point when analyzing a system.
The action of a classical particle is defined by the integral of the Lagrangian with respect to time
\eq{
S=\int^{t_2}_{t_1} dt\, L(q_i(t),\dot{q}_i(t))
}
where $q_i$ is the configuration variable or generalized coordinates, $\dot{q}_i$ is its associated velocity and the Lagrangian $L$ is the difference between the kinetic and potential energy of a system. Hamilton's principle states that the trajectory actually taken by the particle is the path that extremizes (minimizes or maximizes depending on the system) the action. Equivalently we can say the variation between the extremized action and the actual action is zero or
\eq{
\delta S = 0 \,,
}
which leads us to
\eq{
\frac{\del L}{\del q_i} - \frac{d}{dt} \frac{\del L}{\del \dot{q}_i} = 0.
}
This is known as the Euler-Lagrange equation. One does not necessarily need to understand the derivation, which we will do later, as much as one should realize that the solution to the Euler-Lagrange equation gives the equation of motion of a particle from its respective Lagrangian.

A trivial example can show the power of the Lagrangian formulation. Using the general definition of kinetic and potential energy we can construct a general Lagrangian

\eq{
L = \tfrac{1}{2}m\dot{q}^2_i - V(q_i)
}
where $m$ is the mass of the particle and $V$ is the potential energy which is a function of the generalized coordinates $q_i$. Applying the Euler-Lagrange equation to the following Lagrangian gives us
\eq{
-\frac{\del V}{\del q_i}  &- \frac{d}{dt}m\dot{q}_i = 0
\nn \\ 
-\frac{\del V}{\del q_i}  &- m\ddot{q}_i = 0
\nn \\ 
-\frac{\del V}{\del q_i}  &= m\ddot{q}_i 
}
or in the more familiar form
\eq{
\vec{F} = m\vec{a}
}

As one can see the Lagrangian formulation reproduces Newton's second law which characterizes the dynamics of the system.
 While Lagrangian mechanics is very effective in finding the equations of motion for a particle, Hamiltonian mechanics has proven to be even more versatile to handle more sophisticated situations as we will elaborate on in the next section. 

\subsection{Legendre Transform, Canonical Momenta}
Hamiltonian mechanics makes use of two canonical variables $q_i$ and $p_i$. Where $q_i$ is, once again, the configuration variable and $p_i$ its conjugate momentum defined by 
\eq{
p_i := \frac{\del L}{\del \dot{q}_i} \,.
}
Performing the Legendre transform on the Lagrangian gives us the canonical Hamiltonian
\eq{
H = \sum_{i=1}^N p_i \dot{q}_i - L(q_i,p_i),
}
and Hamilton's equations are
\eq{
\dot{q}_i =  \frac{\del H}{\del p_i} \quad\quad\quad
\dot{p}_i =  -\frac{\del H}{\del q_i}\;.
}
These equations describe a flow or orbit in phase space, a two-dimensional space where one dimension is the configuration variable, or generalized coordinates, $q_i$ and the other dimension is it's conjugate momenta $p_i$. The solutions to these coupled first-order differential equations are equivalent to solutions obtained with the Euler-Lagrange equations. 

While Hamiltonian mechanics is equivalent to Lagrangian mechanics, the advantages of Hamiltonian mechanics begin with the fact that two coupled first-order ODEs are much easier to solve than a second-order ODE especially as problems become increasingly complex. More importantly, Hamiltonian mechanics transfers naturally from classical mechanics to quantum mechanics via the canonical quantization scheme.

\subsection{Poisson bracket}
An important operation in Hamiltonian mechanics is the Poisson Bracket. The Poisson bracket between two functions $f(q_i(t),p_i(t))$ and $g(q_i(t),p_i(t))$ is defined as
\eq{
\{f,g\} = \sum_{i=1}^N \left( 
\frac{\del f}{\del q_i} \frac{\del g}{\del p_i} -
\frac{\del f}{\del p_i} \frac{\del g}{\del q_i}
\right),
}
which gives rise to the fundamental Poisson brackets
\eq{
\{q_i,q_j\} &= 0, \\
\{p_i,p_j\} &= 0, \\
\{q_i,p_j\} &= \delta_{ij}\,.
}
The Poisson brackets between the Hamiltonian and the canonical coordinates reproduce Hamilton's equations of motion
\eq{
\dot{q}_i &= \{q_i\,H\} = \frac{\del H}{\del p_i} \\
\dot{p}_i &= \{p_i\,H\} = -\frac{\del H}{\del q_i}
}
In general for any Quantity $Q$, the Poisson bracket between the Hamiltonian $H$ and $Q$ gives the time evolution of $Q$
\eq{
\{ Q,H \} = \frac{\del Q}{\del t}.
}

\section{Constrained Hamiltonian Systems}
For a comprehensive account of Constrained Hamiltonian Systems, refer to \cite{Dirac,CHS,Henneaux:1994} Some physical systems have redundant variables and require constraints to eliminate non-physical degrees of freedom. 
This is best illustrated by an example, a simple pendulum. A pendulum has one physical degree of freedom, its angular position. This is captured well in polar coordinates by the $\theta$ coordinate variable while the $r$ coordinate variable is fixed. If we wanted to use Cartesian coordinates on the other hand we would have to specify the position of the pendulum with $x$ and $y$, giving us two degrees of freedom. In order to eliminate a degree of freedom we need to introduce a constraint equation namely $x^2+y^2=l^2$, where $l$ is the length of the pendulum. With constraints, we can construct the \textit{total} Hamiltonian as follows
\eq{
H_T = H_{c} + \sum^M_{i=1} \lambda_i \phi_i
}
where $H_{c}$ is the usual canonical Hamiltonian obtained via the Legendre transform, $\lambda_i $ is a Lagrange multiplier, $\phi_i$ is a constraint, and $M$ is the number of constraints. The constraints are written so that $\phi_i(q_i,p_i) = 0$. For example, the constraint for the simple pendulum would be written as
\eq{
\phi(q_i,p_i) = x^2+y^2-l^2 = 0 \,.
}
From the perspective of a constrained Hamiltonian system, proper treatments of the constraints are required. With the constraint of the simple pendulum, even though we set this constraint equal to zero, it is only zero in the ``weak'' sense. What it means is that the constraint is only valid on the reduced phase space it defines, not the entire phase space. Because of this, the constraint may still produce a nonvanishing Poisson bracket. For example, 
\eq{
\{\phi(q_i, p_i), p_x\}&=\{x^2+y^2-l^2, p_x \} \nn \\
&=2x\{x, p_x\}\nn \\
&=2x \neq 0.
}
If we had imposed the constraint equation strongly prior to evaluating the above Poisson bracket, it would become pointless to evaluate it later on because the Poisson bracket of zero with anything is automatically zero. But as we saw it is not the case. It is for this reason that we introduce a new notion of zero for constraints, called weakly zero, indicated by the symbol $\approx$. So now the same constraint for the simple pendulum is to be expressed as the following
\eq{
\phi(q_i,p_i) = x^2+y^2-l^2 \approx 0.
}
From a practical sense, the constraints can only be taken to be strongly zero after all the Poisson brackets have been evaluated.

Let's work out the equations of motion for a simple pendulum in Cartesian coordinates as an example.
We start with the Lagrangian
\eq{
L = \tfrac12 m(\dot{x}^2+\dot{y}^2) - mgy.
}
Next, we define the canonical momenta
\eq{
p_x = \frac{\del L}{\del \dot{x}} = m \dot{x},
\qquad
p_y = \frac{\del L}{\del \dot{y}} = m \dot{y},
}
and construct our total Hamiltonian
\eq{
H_T &= p_x \dot{x} + p_y \dot{y} - L + \lambda \phi
\nn \\ &=
\frac{p_x^2}{m} + \frac{p_y^2}{m} - \left(\tfrac12 m \Big( \frac{p_x^2}{m^2} + \frac{p_y^2}{m^2}  \Big) - mgy \right)
+ \lambda (x^2+y^2-l^2) 
\nn \\ &=
\frac{p_x^2}{2m} + \frac{p_y^2}{2m} + mgy
+ \lambda (x^2+y^2-l^2).
}
Now we can use Hamilton's equations to obtain the equations of motion
\eq{
\dot{x} &= \{x, H_T\}= \frac{\del H_T}{\del p_x} = \frac{p_x}{m}
\qquad \qquad
\dot{p}_x =\{p_x, H_T\}= -\frac{\del H_T}{\del x} = -2\lambda x
\nn \\
\dot{y} &=\{y, H_T\}= \frac{\del H_T}{\del p_y} = \frac{p_y}{m}
\qquad \qquad
\dot{p}_y =\{p_y, H_T\}= -\frac{\del H_T}{\del y} = -mg -2\lambda y.
}
Differentiating the left two equations with respect to time and substituting them into the right two equations gives us the equations of motion for $x$ and $y$
\eq{
\ddot{x} &= -\frac{2\lambda x}{m},
\nn \\
\ddot{y} &= -g - \frac{2\lambda y}{m} \label{xx yy}.
}
These equations in combination with the constraint equation, $l^2 = x^2 + y^2$, give us three equations and three unknowns, $x,y, \lambda$ and completely describe the dynamics of the system. To verify that these solutions are indeed correct we will convert them to the usual polar coordinates.

Starting with the fact that 
\eq{
x&=l\sin \theta,
\nn \\
y &= -l \cos \theta.
}
and differentiating twice with respect to time we get
\eq{
\ddot{x} &= l(-\sin\theta \dot{\theta} + \cos\theta \ddot{\theta}),
\nn \\
\ddot{y} &= l(\cos\theta \dot{\theta} + \sin\theta \ddot{\theta}).
}
Setting the previous equal to Eq. \ref{xx yy} we get
\eq{
\ddot{x}&= 
-\frac{2\lambda x}{m} = 
-\frac{2\lambda l \sin\theta}{m} = 
l(-\sin\theta \dot{\theta} + \cos\theta \ddot{\theta}), 
\nn \\
\ddot{y}&= 
- g -\frac{2\lambda y}{m} = 
-g +\frac{2\lambda l \cos\theta}{m} = 
 l(\cos\theta \dot{\theta} + \sin\theta \ddot{\theta}) \label{ddot}.
}
Solving the first of the previous equations for $\lambda$ yields
\eq{
 \lambda = \tfrac m2 \left( \dot{\theta} - \ddot{\theta} \,\frac{\cos\theta}{\sin\theta} \right),
}
which can be substituted into the second equation of Eq. (\ref{ddot}) to give us
\eq{
-g + l  \left( \dot{\theta} - \ddot{\theta} \,\frac{\cos\theta}{\sin\theta} \right) \cos\theta &= 
 l(\cos\theta \dot{\theta} + \sin\theta \ddot{\theta}),
 \nn \\
 -\frac{g}{l} +  \cos\theta \dot{\theta} -  \frac{\cos^2\theta}{\sin\theta} \ddot{\theta}
 &= \cos\theta \dot{\theta} + \sin\theta \ddot{\theta},
 \nn \\
 -\frac{g}{l} -  \frac{\cos^2\theta}{\sin\theta} \ddot{\theta}
 &=  \sin\theta \ddot{\theta},
 \nn \\
 -\frac{g}{l} \sin\theta  -  \cos^2\theta \ddot{\theta}
 &=  \sin^2\theta \ddot{\theta},
  \nn \\
 -\frac{g}{l} \sin\theta &=
( \cos^2\theta 
 +  \sin^2\theta )\ddot{\theta},
  \nn \\
 -\frac{g}{l} \sin\theta &=
\ddot{\theta}.
}
The last equation is in fact the equation of motion for a pendulum in polar coordinates!

The example of the simple pendulum showed that depending on the choice of variables, a physical system may carry redundant degrees of freedom. Because of that, not all physical variables are associated with physical degrees of freedom. Some are simply arbitrary parameters whose values are irrelevant to the underlying physics. Oftentimes, when building a model, variables are chosen without a very clear understanding of the system, therefore the most efficient variables may not be immediately obvious. In field theories, it is also conventional to choose variables in a way that produces a manifestly Lorentz invariant/covariant action, which, even though handy, does not utilize all of the underlying symmetries. Regardless, in most cases in physics, we are dealing with a system with redundancies indicated by a number of constraints. In order to obtain the true dynamics of the system, special care needs to be taken. The general procedure of handling a constrained system was first developed by Paul Dirac, who laid the foundation of Constrained Hamiltonian Systems \cite{Dirac}. 

According to the Dirac procedure, one first obtains constraints from the definition of the canonical momenta, these constraints are referred to as the primary constraints. Then, the primary constraints are to be preserved under time evolution. The specific conditions are called the consistency conditions of the constraints. If new constraints are found from such conditions, they are called secondary constraints. This procedure of checking the consistency of the constraints needs to be repeated until no more secondary constraints can be found. The next step in the Dirac procedure is to compute the Poisson brackets between each pair of constraints. If a set of constraints only produces zero Poisson brackets among themselves, they are called the first-class constraints. On the other hand, those with nonzero Poisson brackets are called second-class constraints, which always come in pairs.

It's important to note that in many systems we don't already know the constraint equations going into the problem and have to discover them along the way. We will examine two simple examples in the following subsections to showcase the general procedure.

\subsection{non-relativistic free particle}
As another example of a constrained Hamiltonian system, let us take a look at a non-relativistic free particle. For simplicity, we will only consider the one-dimensional case here but higher dimensional case can be generalized straightforwardly. First, we recall the familiar Lagrangian from classical mechanics
\eq{
L=\frac {m}{2} \Big(\frac {dx}{dt}\Big)^2.
}
A proper canonical analysis requires some special treatment with the form of the Lagrangian. Specifically, we will introduce a free parameter $\lambda$ that labels the worldline of our free particle. In terms of this parameter, we define the time derivative of the coordinates
\eq{
\dot t =\frac {d t}{d \lambda}, \\
\quad  \dot x = \frac {d x}{d \lambda}.
}
With this, the ordinary coordinate time derivative in the original Lagrangian can be rewritten as
\eq{
\frac {dx}{dt} = \frac {d\lambda}{dt}\frac {dx}{d\lambda} =\frac {\dot x} {\dot t},
}
and the non-relativistic free-particle Lagrangian becomes
\eq{
L=\frac m 2 \Big(\frac {d\lambda}{dt}\Big)^2\Big(\frac {dx}{d\lambda} \Big)^2.
}
This will subsequently turn the action into 
\eq{
S=\int dt\, L &= \int dt\, \frac m 2 \Big(\frac {d\lambda}{dt}\Big)^2\Big(\frac {dx}{d\lambda} \Big)^2\nn\\
&=\int d\lambda\, \frac m 2 \Big(\frac {d\lambda}{dt}\Big)\Big(\frac {dx}{d\lambda} \Big)^2\nn\\
&=\int d\lambda\, \frac m 2 \frac {\dot x^2}{\dot t}.
}

We can now obtain the canonical momenta, namely
\eq{
p_t&= \frac {\del L}{\del \dot t}= -\frac m 2 \Big(\frac {\dot x}{\dot t}\Big)^2,\\
p_x&= \frac {\del L}{\del \dot x}= m \frac {\dot x}{\dot t}
}
and write the canonical Hamiltonian
\eq{
H_{can} &= p_x \dot{x} + p_t \dot{t} -L 
\nn\\&= 
\Big(m \frac {\dot x}{\dot t}\Big)\dot x -\frac m 2 \Big(\frac {\dot x}{\dot t}\Big)^2 \dot{t} - \tfrac{m}{2} \frac {\dot x^2}{\dot t} = 0
}
The canonical Hamiltonian vanishes, but we know a free particle evolves in time, so where are the dynamics? The dynamics are driven by the constraints which we can find upon inspection of the momenta
\eq{
\phi(x,t,p_x,p_t) = p_t^2 + \frac{p_x^2}{2m} = 0
}
With the constraint, we can construct the total Hamiltonian
\eq{
H = N \left( p_t^2 + \frac{p_x^2}{2m} \right)
}
where $N$ is a Lagrange multiplier known as the lapse. With the total Hamiltonian, we can find our equations of motion
\eq{
\dot x &= \{x,H\} = N\frac{p_x}{m} \quad \quad \dot{p_x}=\{p_x ,H\}=0
\nn \\
\dot t &= \{t,H\} = N \quad \quad \quad \,\, \dot{p_t}=\{p_t, H\}=0.
}
Lagrange multipliers are non-physical and and do not actually change the physics of the system, but rather change how we have to interpret the equations of motion.
Notice if we choose the lapse to be equal to $1$ we will have reproduced the ``usual" equations of motion of a free particle. 
The lapse is actually a special Lagrange multiplier that determines time evolution and will be explored in more detail when we introduce the ADM formalism of general relativity in a later section. 

\subsection{Relativistic Free Particle} \label{rel particle}
Our next example is a relativistic free particle\footnote{Readers unfamiliar with the notion of metric in this context should feel free to skip this example and can come back to it after the concept is introduced in later sections.}. The special relativistic action is
\eq{
S = -m \int d\tau = -m \int \sqrt{\eta_{\mu\nu} \frac{d x^\mu}{d\lambda} \frac{d x^\nu}{d\lambda} } d\lambda ,
}
where we are using geometrized units ($c=1$) and the ``most minus" metric convention (+ - - -). From the action, we can see the Lagrangian takes the following form
\eq{
L_{SR} &= -m \sqrt{\eta_{\mu\nu} \frac{d x^\mu}{d\lambda} \frac{d x^\nu}{d\lambda} },
\nn\\&=
-m \sqrt{\eta_{\mu\nu} \dot{x}^\mu \dot{x}^\nu},
\nn\\&=
-m \sqrt{ (\dot{x}^0)^2 - (\dot{x}^i)^2}.
}
The coordinates conjugate canonical momenta are as follows
\eq{
p_0 &=  \frac{\del L}{\del \dot{x}^0} = 
\tfrac{-m}{2}\frac{2 \dot{x}^0}{ \sqrt{(\dot{x}^0)^2 - (\dot{x}^i)^2}} =
\frac{-m \dot{x}^0}{\sqrt{(\dot{x}^0)^2 - (\dot{x}^i)^2}} =
\frac{-m \dot{x}_0}{\sqrt{(\dot{x}^0)^2 - (\dot{x}^i)^2}},
\\ 
p_i &= \frac{\del L}{\del \dot{x}^i} = 
\tfrac{-m}{2}\frac{-2 \dot{x}^i}{ \sqrt{(\dot{x}^0)^2 - (\dot{x}^i)^2}} =
\frac{m \dot{x}^i}{\sqrt{(\dot{x}^0)^2 - (\dot{x}^i)^2}} =
\frac{-m \dot{x}_i}{\sqrt{(\dot{x}^0)^2 - (\dot{x}^i)^2}}.
}
We now perform the Legendre transform on the Lagrangian to obtain the canonical Hamiltonian
\eq{
H_c &= p_0\dot{x}^0 + p_i\dot{x}^i - L_{R},
\nn\\&=
\frac{-m \dot{x}_0 \dot{x}^0}{\sqrt{(\dot{x}^0)^2 - (\dot{x}^i)^2}} +
\frac{-m \dot{x}_i \dot{x}^i}{\sqrt{(\dot{x}^0)^2 - (\dot{x}^i)^2}} + 
m \sqrt{ (\dot{x}^0)^2 - (\dot{x}^i)^2},
\nn\\&=
\frac{-m}{\sqrt{(\dot{x}^0)^2 - (\dot{x}^i)^2}}
( (\dot{x}^0)^2 - (\dot{x}^i)^2) + 
m \sqrt{ (\dot{x}^0)^2 - (\dot{x}^i)^2},
\nn\\&=
-m \sqrt{ (\dot{x}^0)^2 - (\dot{x}^i)^2} +
m \sqrt{ (\dot{x}^0)^2 - (\dot{x}^i)^2},
\nn\\&=0.
}
The vanishing Hamiltonian indicates a totally constrained system in which the dynamics are generated completely by constraints. From the canonical momenta, we can find
\eq{
p_0^2 &= \frac{m^2 ( \dot{x}_0 )^2}{(\dot{x}^0)^2 - (\dot{x}^i)^2},
\\
p_i^2 &= \frac{m^2 ( \dot{x}_i )^2}{(\dot{x}^0)^2 - (\dot{x}^i)^2},
}
and
\eq{
p_0^2 - p_i^2 = m^2\frac{ (\dot{x}_0)^2 - (\dot{x}_i)^2}{(\dot{x}^0)^2 - (\dot{x}^i)^2} = m^2,
}
which can be rewritten in the form of a constraint
\eq{
\phi = p_0^2 - p_i^2 - m^2 = 0.
}
This leads to our total Hamiltonian being
\eq{
H_T = \lambda (p_0^2 - p_i^2 - m^2 )
}
Rewriting the constraint equation with the components of the four-momentum $p^\mu = (E,\vec{p}\,)$ gives us
\eq{E^2 - \vec{p}\,^2 - m^2=0,}
which after reinserting the $c$'s and a little rearranging produces the all too familiar energy-momentum relation
\eq{
E = \sqrt{\vec{p}\,^2 c\,^2 + m^2 c\,^4}\,.
}
As one can see this famous equation is actually a constraint equation and generates the dynamics of a relativistic free particle. Finally, calculating the Poisson brackets gives us the equations of motion
\eq{
\dot x_0 &= \{x_0, H_T\} = 2 \lambda p_0 \qquad \dot{p_0}=\{p_0, H_T\}=0
\nn \\
\dot x_i &= \{x_i, H_T\} = -2 \lambda p_i \qquad \dot{p_i}=\{p_i, H_T\}=0
}
where $\lambda$ is an arbitrary Lagrange multiplier.

\section{Hamiltonian Formalism of Field Theories}
In field theories, all of the configuration variables are dependent on positions. Where before we had, for example, the momentum of a particle as a function of time $p_i(t)$, we now have something like the electric field as a function of time and position $E^i(t,x,y,z)$. We will use a minimalist notation and simply right our fields as $E^i(x)$ where
\eq{
E^i(x) := E^i(x^0,x^1,x^2,x^3)\,.
}
It is useful to define it this way since many times we will have to compare the same field at different locations. For example, we may have to compare the electric fields $E^i(x)$
and $E^i(y)$ where $y$ simply represents a different location in space-time than $x$. It should be noted in the later sections we will even drop the $(x)$ from each field since there are so many!

Fields being functions of position means that the Lagrangian is dependent on the region of space we are interested in. In order to find the Lagrangian we have to integrate the Lagrangian density $\mathcal{L}$ over a volume
\eq{
L = \int_V d^3x \, \mathcal{L} 
}
where $d^3x$ is an infinitesimal volume element for any choice of coordinates (e.g., $dxdydz$ for Cartesian coordinates). Thus the action is given by an integral over time as well as space
\eq{
S = \int dt L = \int d^4x \, \mathcal{L} \,.
}
Similar to the Lagrangian the Hamiltonian is the Hamiltonian density $\mathcal{H}$ integrated over a volume
\eq{
H = \int_V d^3x \, \mathcal{H} = \int_V d^3x \, \left(\Pi^i(x) \dot{\Phi}_i(x) - \mathcal{L}\right),
}
where $\Phi_i$ is our field, or configuration variable, and $\Pi^i(x)$ is the canonically conjugate momentum \textit{density}, hence the Greek letter for $p$ (pi), which is given by
\eq{
\Pi^i(x) = \frac{\delta L}{\delta \dot{\Phi}_i(x)}.
}
Notice that here the delta symbols indicate we are taking a \textit{functional} derivative which is defined in the following subsection.

\subsection{Functional Variation and Derivative}

In a field theory, physical fields are usually functionals of the configuration variables and their derivatives, or equivalently, functionals of the configuration variables and their conjugate momenta. For example, the Lagrangian density $\mathcal{L}(\Phi(x^\mu), \del_\mu \Phi(x^\mu))$ is generally a functional of some configuration variable $\Phi(x^\mu)$ and its derivative $\del_\mu \Phi(x^\mu)$, whereas the Hamiltonian density $\mathcal{H}(\Phi(x^\mu), \Pi(x^\mu))$ is generally a functional of the configuration variables $\Phi(x^\mu)$ and their canonically conjugate momenta $\Pi(x^\mu)$. When comparing two fields, we have to bear in mind that they are defined locally at different points in space. This brings the need for a generalized notion of taking differentiations, which is called functional derivative, also known as variational derivative.

Suppose that we have an arbitrary field $\Phi(t, \bm{x})$ evaluated locally at a point $\bm{x}$ at an instant $t$. The same field can be evaluated at a different point $\bm{y}$ at the same instant $t$, namely, $\Phi(t,\bm{y})$. When comparing the same field $\Phi(x^\mu)$ evaluated at different points $\bm{x}$ and $\bm{y}$ at the same instant of time $t$, a functional differentiation can be constructed as
\eq{
\frac{\delta \Phi(t,\bm{x})}{\delta \Phi(t,\bm{y})} =  \delta^3(\bm{x}-\bm{y}).
}
Here, $\delta^3(\bm{x}-\bm{y})$ is a 3-dimensional Dirac delta function between the points $\bm{x}$ and $\bm{y}$. The action of $\delta$ generates an infinitesimal variation on the field $\Phi(x^\mu)$, therefore we will call it the variational operator, which acts just like the symbol $d$ in a coordinate differential $dx$. For example,
\eq{
\delta (x^2 + y^2) = 2x \delta x+2y \delta y.
}
Because of this common behavior, the functional derivative also follows a chain rule similar to an ordinary derivative, for example
\eq{
\frac{\delta }{\delta \Phi(y)} (\Phi(x))^2 = 2 \Phi(x) \frac{\delta \Phi(x)}{\delta \Phi(y)} = 2 \Phi(x) \delta^3(\bm{x}-\bm{y}).
}
Notice that we have adopted a somewhat sloppy notation here, namely $\Phi(t,\bm{x}):=\Phi(x)$, where the $x$ inside the parenthesis is not to be confused with the coordinate component $x$ of the vector $\bm{x}=(x,y,z)$, but rather it is a shorthand notation for $x^\mu=(t, x, y, z)$. This treatment will make the technical analysis less cluttered therefore we will adopt the same notation from this point on. While we finish out this section with bold text for the position vector in the Dirac deltas, in the following sections we omit the use of bold text as it becomes obvious it is always the spatial position vector in a 3-dimensional Dirac delta.

The functional derivative and variational operator are commonly used in field theories. In particular, they facilitate the idea of the Least-action Principle in obtaining the equations of motion. Recall that the trajectory actually taken by a particle is when the variation of the action is zero, i.e.
\eq{
\delta S =0 .
}
If we take the variation of a general action we get
\eq{
\delta S &= \int^{t_2}_{t_1} dt\, \delta L(q_i,\dot{q}_i),
\nn \\ &=
\int^{t_2}_{t_1} dt\, \left( \frac{\del L}{\del q_i}\delta q_i + \frac{\del L}{\del \dot{q}_i}\delta \dot{q}_i
\right),
\nn \\ &=
\int^{t_2}_{t_1} dt\, \left( \frac{\del L}{\del q_i}\delta q_i - \frac{d}{dt} \frac{\del L}{\del \dot{q}_i}\delta q_i
\right),
\nn \\ &=
\int^{t_2}_{t_1} dt\, \left( \frac{\del L}{\del q_i} - \frac{d}{dt} \frac{\del L}{\del \dot{q}_i}
\right)\delta q_i,
}
where in the second to third line we integrate by parts on the second term, which takes the time derivative on everything to the left of $q$ and flips the overall sign. As one can see from above, the variation of the action is equal to zero for all variations of $q_i$ when the following equation is satisfied
\eq{
\frac{\del L}{\del q_i} - \frac{d}{dt} \frac{\del L}{\del \dot{q}_i} = 0
}
which is the Euler-Lagrange equation!

\subsection{Poisson Brackets in Field Theories}
We will wrap up this section with a generalized form of the Poisson brackets in field theories. Recall that in mechanics, the Poisson brackets are defined in terms of ordinary differentiation. Because we are dealing with functionals in a field theory, the Poisson brackets are defined in terms of functional differentiation instead. To make this definition more relevant to what we are about to discuss in the later sections, let us consider two general fields $F$ and $G$, which are both functionals of some phase space variables $\Phi_a(x)$ and $\Pi^a(x)$. The Poisson bracket between $F$ and $G$ is defined as
\eq{
&\{F[\Phi_a(x), \Pi^a(x)],G[\Phi_b(y), \Pi^b(y)]\} \nn\\
= &\int d^3z \left( 
\frac{\delta F[\Phi_a(x), \Pi^a(x)]}{\delta \Phi_c(z)} \frac{\delta G[\Phi_b(y), \Pi^b(y)]}{\delta \Pi^c(z)} - \frac{\delta F[\Phi_a(x), \Pi^a(x)]}{\delta \Pi^c(z)} \frac{\delta G[\Phi_b(y), \Pi^b(y)]}{\delta \Phi_c(z)}
\right)\\
:=&\int d^3z\, \left(\frac {\delta F(x)}{\delta \Phi_c(z)}\frac {\delta G(y)}{\delta \Pi^c(z)} - \frac {\delta F(x)}{\delta \Pi^c(z)}\frac {\delta G(y)}{\delta \Phi_c(z)}\right). \label{pb}
}
Now due to the functional dependence of the fields, evaluating Poisson brackets in field theory can become a tedious process. We would like to introduce a trick to simplify the process to a certain degree when some basic Poisson brackets have already been obtained. 

We will claim the Poisson bracket defined in Eq.  (\ref{pb}) is equivalent to the following
\eq{
&\{F[\Phi_a(x), \Pi^a(x)],G[\Phi_b(y), \Pi^b(y)]\}\nn\\
=&\int d^3u\, d^3v \left(\frac {\delta F(x)}{\delta \Phi_c(u)}\{\Phi_c(u),\Pi^d(v)\}\frac {\delta G(y)}{\delta \Pi^d(v)} + \frac {\delta F(x)}{\delta \Pi^d(u)}\{\Pi^d(u),\Phi_c(v)\}\frac {\delta G(y)}{\delta \Phi_c(v)}\right).
\label{pb2}
}
Again, all the functional dependence with variables $x, y, u, v$ are referring to different points in space. As a simple consistency check, let's consider that the pair $\Phi_a(x)$ and $\Pi^b(y)$ satisfy the canonical Poisson brackets relation, i.e.,
\eq{
\{\Phi_c(u), \Pi^d(v)\}=\delta_c^d\delta^3(\bm{u}-\bm{v}).
}
Eq. (\ref{pb2}) now becomes
\eq{
&\int d^3u\, d^3v \left(\frac {\delta F(x)}{\delta \Phi_c(u)}\Big(\delta_c^d\delta^3(\bm{u}-\bm{v})\Big)\frac {\delta G(y)}{\delta \Pi^d(v)} + \frac {\delta F(x)}{\delta \Pi^d(u)}\Big(-\delta_c^d\delta^3(\bm{v}-\bm{u})\Big)\frac {\delta G(y)}{\delta \Phi_c(v)}\right)\\
=&\int d^3u\, \left(\frac {\delta F(x)}{\delta \Phi_c(u)}\frac {\delta G(y)}{\delta \Pi^c(u)} - \frac {\delta F(x)}{\delta \Pi^c(u)}\frac {\delta G(y)}{\delta \Phi_c(u)}\right),
}
which goes back to the original definition in (\ref{pb}), just with a different integration variable $\bm{u}$ instead of $\bm{z}$. We will see how this seemingly trivial trick plays out in the following sections.

We will not be looking into any specific example of a constrained Hamiltonian system in the case of a field theory within this section. Rather, we will dedicate the next two sections to a detailed presentation of two well-known field theories.

\chapter{Maxwell Theory} \label{sec: maxwell}
We will begin our journey of understanding the workings of gauge theories with Maxwell's theory of electromagnetism. As the first gauge theory that describes a realistic, physical system, Maxwell's theory has a simple and clear structure that not only captures nearly all the key elements of a gauge theory but also provides a deep physical intuition. 
In this section, we will use Maxwell's theory as a prototype example to introduce some of the basic ingredients of a gauge theory. We will also go over some technical details that will be useful for later parts of this paper.

Electromagnetism was originally formulated as a theory of interactions of charged particles. The concept of fields was introduced as a way to describe physical alterations in space in the presence of a source. Specifically, the presence of a static electric charge produces an electric field around it, and a moving electric charge produces a magnetic field in its neighborhood. The two seemingly distinct phenomena of electricity and magnetism were unified as electromagnetic interaction by James Clerk Maxwell in the mid-19th century, which revealed a deeper interconnection between electricity and magnetism. Namely, a time-varying electric field generates a magnetic field, and if the magnetic field also happens to vary, it will generate an electric field, which in turn will generate a magnetic field. This leads to a sustainable generation of electric and magnetic fields that propagate through space, as electromagnetic waves. 

The intimate connection between electric and magnetic fields inspired people to incorporate them into the same quantity, the electromagnetic field strength tensor, or simply the field-strength tensor $F_{\mu\nu}$. This covariant form also allows us to define a manifestly Lorentz invariant action for the Maxwell theory
\eq{
S_{EM} = \int d^4 x \, \mathcal{L}_{EM},
}
where the Lagrangian density takes the following form
\eq{
\mathcal{L}_{EM} = -\frac{1}{4}F_{\mu\nu}F^{\mu\nu}.
}
Here field strength tensor is defined as $F_{\mu\nu} = \del_\mu A_\nu - \del_\nu A_\mu$. It is a functional of the gauge potential $A_\mu = (A^0, \vec A)$, where $A^0 = V$ is also commonly referred to as the scalar potential and $\vec A$ the vector potential. We refer those in need of more foundational knowledge of electromagnetism to Griffiths text \cite{Griffiths} and a similar canonical analysis can be found in \cite{intro,Baez}.

\section{Canonical analysis of Maxwell Theory}

The first step in a canonical analysis is to explicitly separate out the Lagrangian density into terms involving time-derivative namely $F_{0a}$, and terms that only involve spatial derivatives namely $F_{ab}$. This is a necessary step to bring the Lagrangian to a desired form appropriate for obtaining the Hamiltonian via the Legendre transform. We have
\eq{
\mathcal{L}_{EM}&=-\frac 1 4  F_{\mu\nu}F^{\mu\nu}\nn\\
&=-\frac 1 4 (2F_{0a}F^{0a} + F_{ab}F^{ab})\nn\\
&=\frac 1 2 F_{0a} F_{0a} -\frac 1 4 F_{ab}F_{ab}\nn\\
&=\frac 1 2 (\del_0 A_a -\del_a A_0)^2 - \frac 1 4 (\del_a A_b - \del_b A_a)^2.
}
This explicit form allows us to obtain the canonical momenta. Right off the bat, we have a somewhat trivial result
\eq{
\Pi^0=\frac {\delta L}{\delta \dot A_0(x)} \approx 0.
}
This condition on the phase space variables provides us with a primary constraint, which we will discuss in detail later. Further, we have
\eq{
\Pi^a(x) = \frac {\delta L}{\delta \dot A_a(x)},
\label{moment1}
}
where
\eq{
\delta L &= \delta \int d^3 y\, (\frac 1 2 (\del_0 A_a -\del_a A_0)^2 - \frac 1 4 (\del_a A_b - \del_b A_a)^2) \nn\\
& \supset \int d^3 y\, \delta \Big(\frac 1 2 (\del_0 A_a -\del_a A_0)^2\Big)\nn\\
&\supset  \int d^3 y\, (\del_0 A_a -\del_a A_0)\delta \dot A_a =\int d^3 y\, F_{0a}\, \delta \dot A_a .
}
The reason for using superset instead of the equal sign in the above steps is because it highlights the only relevant terms that would eventually contribute to the canonical momentum defined in Eq. (\ref{moment1}). It follows that
\eq{
\Pi^a(x)&= \int d^3 y\, F_{0b}(y)\, \frac {\delta \dot A_b(y)} {\delta \dot A_a(x)}\nn\\
&=\int d^3 y\, F_{0b}(y)\, \delta_{ab}\delta^3(y-x)\nn\\
&=F_{0a}(x)=\dot A_a - \del_a A_0
\label{maxwell pi}
}
Note that we would \textit{conventionally} define $F^{0a}$ as the electric field vector $E^a$, so there is a subtle sign difference between $\Pi^a$ and $E^a$, namely
\eq{
E^a =-\Pi^a.
}

In terms of the canonical momentum, we can rewrite the Lagrangian density as 
\eq{
\mathcal{L}_{EM} = \frac 1 2 \Pi_a \Pi_a - \frac 1 4 F_{ab} F_{ab}.
}
$\Pi^a$ together with $A_a$ now define the basic set of phase space variables of the Hamiltonian system, thereby producing canonical Poisson brackets
\eq{
\{A_a(x), A_b(y)\}&=\{\Pi_a(x), \Pi_b(y)\}=0,\\
\{A_0(x), \Pi^0(y)\}&=\delta^3(x-y),\\
\{A_a(x), \Pi_b(y)\}&=\delta_{ab}\delta^3(x-y).
} 

Next, we perform the Legendre transform on the Lagrangian to obtain the canonical Hamiltonian
\eq{
H_{c} &= \int d^3 x\, (\Pi^{a} \dot A_a - \mathcal{L}_{EM})\nn\\
&= \int d^3 x \Big(\Pi^{a} \dot A_a - (\frac 1 2 (\Pi^{a})^2 -\frac 1 4(F_{ab})^2)\Big)\nn\\
&= \int d^3 x \Big(\Pi^{a} (\Pi^a + \del_a A_0) - \frac 1 2 (\Pi^{a})^2 + \frac 1 4(F_{ab})^2\Big)\nn\\
&= \int d^3 x \Big(\frac 1 2 (\Pi^{a})^2 + \frac 1 4(F_{ab})^2)  - A_0 \del_a \Pi^a \Big).
}
In the above steps, between the second and the third lines, we used Eq. (1.7), performed integration by parts on the term $\Pi^{a} \del_a A_0$, and dropped the surface term between the third and the last lines. In the final result of the Maxwell Hamiltonian, the first two terms correspond to the kinetic energy of the electric and magnetic fields, as expected. 

An important point to make here is the role of $A_0$: Due to the total anti-symmetry of the field tensor, $\dot A_0$ never appears in the Lagrangian and thus does not behave as a dynamical field like the other three components of the gauge potential. It is essentially a redundancy introduced in the theory from writing the Lagrangian in a manifestly Lorentz invariant form. Fields like this are generally treated as Lagrange multipliers in the theory. In the Hamiltonian, what a Lagrange multiplier multiplies would generally be a constraint, which also acts as a generator of gauge transformations. This brings us to the last term in the Maxwell Hamiltonian, in which $A_0$ acts as a Lagrange multiplier and $\del_a \Pi^a$ is interpreted as a generator of gauge transformations on the gauge potentials. We will elaborate on this later.

We shall now focus on obtaining Hamilton's equations by taking the Poisson bracket of a field with the canonical Hamiltonian $H_{c}$. First, we have
\eq{
\dot A_a(x) &= \{A_a(x), H_{c}\}\nn\\
&=\int d^3 z\, \{A_a(x), \Pi_b(z)\}\frac {\delta H(y)} {\delta \Pi_b(z)},
}
where 
\eq{
\delta H_{c} &= \delta \int d^3 y\, \Big(\frac 1 2 (\Pi^{a}(y))^2 + \frac 1 4(F_{ab}(y))^2)  - A_0(y) \del_a \Pi^a(y) \Big)\nn\\
&\supset \int d^3y \,\Big(\Pi_c(y)\delta \Pi^c(y) + \del_c A_0(y) \delta \Pi^c(y)\Big).
}
Thus
\eq{
\frac {\delta H(y)} {\delta \Pi_b(z)} &= \int d^3y \,\Big(\Pi_c(y) + \del_c A_0(y) \Big)\frac {\delta \Pi^c(y)} {\delta \Pi_b(z)}\nn\\
&= \int d^3y \,\Big(\Pi_c(y) + \del_c A_0(y) \Big)\delta_{cb}\, \delta^3(y-z)\nn\\
&= \Pi_b(z) + \del_b A_0(z).
}
Inserting the above expression back, we have
\eq{
\dot A_a(x) &=\int d^3 z\, \{A_a(x), \Pi_b(z)\}(\Pi_b(z) + \del_b A_0(z)),\nn\\
&=\int d^3 z\, \delta_{ab}\, \delta^3(x-z) (\Pi_b(z) + \del_b A_0(z))\nn\\
&=\Pi_a(x) + \del_a A_0(x).
}
The same result has been obtained previously in (\ref{maxwell pi}). This is a nice consistency check for our Hamiltonian.

Next, we work on the time evolution of the canonical momenta (or equivalently the electric field vector)
\eq{
\dot \Pi_a(x) &= \{\Pi_a(x), H_{c}\}\nn\\
&=\int d^3 z\, \{\Pi_a(x), A_b(z)\}\frac {\delta H(y)} {\delta A_b(z)}\nn\\
&=\int d^3 z\, (-\delta_{ab}\delta^3(x-z))\frac {\delta H(y)} {\delta A_b(z)}=-\frac {\delta H(y)} {\delta A_a(x)},
}
where 
\eq{
\delta H_{EM} &= \delta \int d^3 y\, \Big(\frac 1 2 (\Pi^{a}(y))^2 + \frac 1 4(F_{ab}(y))^2)  - A_0(y) \del_a \Pi^a(y) \Big)\nn\\
&\supset \int d^3y \,\Big(\frac 1 2 F_{ab}\, \delta (\del_a A_b - \del_b A_a)\Big)\nn\\
&=\int d^3y \,\frac 1 2 \Big(-\del_a F_{ab}\,\delta A_b + \del_b F_{ab}\delta A_a\Big)\nn\\
&=\int d^3y\, \del_b F_{ab}\delta A_a.
}
Inserting the above relation back, we have
\eq{
\dot \Pi_a(x) &=-\int d^3y\, \del_b F_{cb}(y)\frac {\delta A_c(y)}{\delta A_a(x)}\nn\\
&=-\del_b F_{ab}(x).
\label{max1}
}
We can relate the spatial field tensor $F_{ab}$ with the magnetic field vector via 
\eq{
 B^a=\frac 1 2 \epsilon^{abc}F_{bc},
 }
 which is equivalent to
 \eq{
F_{ab} =\epsilon_{abc}B^c,
 }
This along with the relations that $E^a=-\Pi^a$, allow us to recognize Eq. (\ref{max1}) in the more familiar form
\eq{
{\bf \Del \times B = \frac{\del E}{\del t}}
}
which we recognize as Ampere's law in free space (current density $\vec {\bf J}=0$) from Maxwell's equations.

Similarly, we can obtain the time evolution of $F_{ab}$
\eq{
\dot F_{ab}(x) &= \{F_{ab}(x), H_{EM}\}\nn\\
&=\int d^3z \, \frac {\delta F_{ab}(x)}{\delta A_c(z)}\{A_c(z), H_{EM}\}\nn\\
&=\int d^3z \, \frac {\delta F_{ab}(x)}{\delta A_c(z)}\Big(\Pi_c(z) + \del_c A_0(z)\Big),
}
where
\eq{
\frac {\delta F_{ab}(x)}{\delta A_c(z)} &= \frac {\delta (\del_a A_b(x) - \del_b A_a(x))}{\delta A_c(z)}\nn\\
&=\frac {\del}{\del x^a}\Big(\frac {\delta A_{b}(x)}{\delta A_c(z)}\Big) - \frac {\del}{\del x^b}\Big(\frac {\delta A_{a}(x)}{\delta A_c(z)}\Big)\nn\\
&=\delta_{bc} \frac {\del}{\del x^a}\delta^3(x-z) - \delta_{ac} \frac {\del}{\del x^b}\delta^3(x-z), 
}
and substituted back, we have
\eq{
\dot F_{ab}(x) &=\int d^3z \,\Big(\delta_{bc} \frac {\del}{\del x^a}\delta^3(x-z) - \delta_{ac} \frac {\del}{\del x^b}\delta^3(x-z)\Big) \Big(\Pi_c(z) + \del_c A_0(z)\Big)\nn\\
&=\frac {\del}{\del x^a} \int d^3z \, \delta^3(x-z) \Big(\Pi_b(z) + \del_b A_0(z)\Big) - (a\leftrightarrow b)\nn\\
&=\del_a \Pi_b(x) - \del_a \del_b A_0(x) - (\del_b \Pi_a(x) - \del_b \del_a A_0(x))\nn\\
&=\del_a \Pi_b(x) - \del_b \Pi_a(x).
}
Again, upon applying the relations $B^c=\frac 1 2 \epsilon^{abc}F_{ab}$ and $E^a=-\Pi^a$, we have
\eq{
\dot B^c &= -\frac 1 2 \epsilon^{abc}(\del_a E_b - \del_b E_a)\nn\\
&= - \epsilon^{abc}\del_a E_b, 
}
or equivalently
\eq{
\bf \Del \times E = -\frac {\del B}{\del t},
}
which we recognize as Faraday's law from Maxwell's equations.

\section{Gauge Transformations}

In electromagnetism, we learned that the physical fields are the electric and magnetic fields, whose dynamics are dictated by the Maxwell equations. On the other hand, one can introduce a vector potential $\vec A$ that helps us define the magnetic field vector via
\eq{
\vec B = \Del \times \vec A.
}
This vector potential is only defined up to the gradient of an arbitrary scalar field $\lambda$, because the magnetic field vector is invariant under the transformation
\eq{
\vec A \rightarrow \vec A + \Del \lambda.
}
This can be checked easily below
\eq{
B'_i &= \epsilon_{ijk}\del_j A'_k\nn\\
&= \epsilon_{ijk}\del_j (A_k + \del_k \lambda)\nn\\
&=\epsilon_{ijk}\del_j A_k + \epsilon_{ijk}\del_j \del_k \lambda\nn\\
&= B_i,
}
where the second term in the third line vanishes automatically due to the totally antisymmetric epsilon tensor.

Similarly, we can introduce a scalar potential $V$ that helps define the electric field vector
\eq{
\vec E = -\Del V - \frac {\del \vec A}{\del t}.
}
This scalar field is only defined up to the time derivative of an arbitrary scalar field because the electric field vector is invariant under the transformation
\eq{
V\rightarrow V-\frac {\del \lambda}{\del t},
}
along with the transformation on $\vec A$. Again, this can be shown below 
\eq{
 E_i' &= -\del_i V' - \frac {\del A_i'}{\del t}\nn\\
&=-\del_i (V-\frac {\del \lambda}{\del t}) - \frac {\del}{\del t}(A_i + \del_i \lambda)\nn\\
&= -\del_i V+\del_i \frac {\del \lambda}{\del t} - \frac {\del}{\del t}A_i - \frac {\del}{\del t}\del_i \lambda\nn\\
&= -\del_i V - \frac {\del A_i}{\del t}\nn\\
&= E_i.
}
These properties of the scalar and vector potentials can be put together into a covariant form by introducing the four-vector potential $A^\mu = (V, A^i):= (A^0, A^i)$, which is defined up to a four-derivative of an arbitrary scalar field $\lambda$
\eq{
A^\mu \rightarrow A^\mu + \del^\mu \lambda.
}
For any arbitrary choice of $\lambda$, the underlying physics is equivalently described by the same physical variables as the electric and magnetic fields. This type of transformation that involves adding or subtracting derivatives of an arbitrary scalar function while keeping the physical variables unchanged is called a gauge transformation. The Maxwell theory is therefore called gauge invariant, or that it carries a gauge symmetry.

According to the Dirac conjecture \cite{Dirac}, all first-class constraints in a Hamiltonian system act as generators of gauge transformations, which relate equivalent descriptions of the same physical state. This conjecture proved to be true for many physical systems, including the Maxwell theory discussed in this section and the Yang-Mills theory to be discussed in the next section. From the canonical analysis in the previous subsection, we have already identified the primary constraint as $\Pi^0 \approx 0$, which needs to be added to the canonical Hamiltonian to construct the total Hamiltonian
\eq{
H_T &= H_c + \int d^3x \, \mu(x) \Pi^0(x),\\
&=\int d^3x\, \Big(\frac 1 2 (\Pi^{a})^2 + \frac 1 4(F_{ab})^2)  - A_0 \del_a \Pi^a + \mu(x) \Pi^0(x)\Big),
}
where $\mu$ is an arbitrary scalar function called a ``smearing'' such that the added term is well-behaved in practical calculations.
The total Hamiltonian allows us to evaluate the consistency condition for the primary constraint
\eq{
\{\Pi^0, H_T\}&=\{\Pi^0, \int d^3x \,(- A_0 \del_a \Pi^a)\},\\
&=\del_a \Pi^a \approx 0.
}
This consistency condition on the primary constraint defines a secondary constraint that happens to be the familiar Gauss law. The Dirac procedure requires us to further check the consistency of the secondary constraint to possibly discover more secondary constraints. In this case, the Gauss law constraint does not lead to more secondary constraints so the hunt for constraints is complete. 

The next step is to examine the Poisson brackets between all constraints to separate out the first-class and second-class constraints. Specifically, those constraints that produce weakly vanishing Poisson brackets are categorized as first-class, whereas those with non-vanishing Poisson brackets are second-class constraints. Since there are only two constraints in the Maxwell theory, all we have to check is $\{\Pi^0, \del_a \Pi_a\}$, which is clearly zero. Therefore we can conclude that both the primary constraint $\Pi^0 \approx 0$ and the secondary constraint $\del_a \Pi^a\approx 0$ are of first class. We will focus on the Gauss law constraint below. 

To have a well-defined canonical analysis in a field theory, we introduce a smeared constraint of the Gauss law and claim that it generates gauge transformations in Maxwell's theory 
\eq{
G(\lambda) = \int d^3 x\, \lambda (x)\, \del_a \Pi^a(x),
}
where $\lambda(x)$ is an arbitrary function of space-time that is smooth, and differentiable, and it is usually referred to as the gauge parameter. To show that $G(\lambda)$ does indeed behave as a gauge generator, we will act $G(\lambda)$ on the gauge potential $A_a$, electric field vector $E^a$, and magnetic field vector $B^a$ to show their dependence or invariance under gauge transformations.

\section{Gauge dependence of the Gauge Potential}

For an arbitrary function $F(A_a, \Pi^a)$ of the phase variables $A_a(x)$ and $\Pi^a(x)$, its gauge transformation is defined as
\eq{
\delta_\lambda F(A_a, \Pi^a)&\equiv \{G(\lambda), F(A_a, \Pi^a)\}\\
&=\int d^3z\, d^3w \, \frac {\delta G(\lambda, y)}{\delta \Pi^c(z)}\{\Pi^c(z), A_b(w)\}\frac {\delta F(A_a(x), \Pi^a(x))}{\delta A_b(w)},
}
where
\eq{
\delta G(\lambda, y) &= \delta \int d^3y\, \lambda(y)\, \del_d \Pi^d(y)\nn\\
&=  \int d^3y\, \lambda(y)\, \del_d \delta \Pi^d(y)\nn\\
&=  -\int d^3y\, \del_d\lambda(y)\, \delta \Pi^d(y),
}
and thus
\eq{
\frac {\delta G(\lambda, y)}{\delta \Pi^c(z)} &= -\int d^3y\, \del_d\lambda(y)\, \frac {\delta \Pi^d(y)}{\delta \Pi^c(z)}\nn\\
&=-\int d^3y\, \del_d\lambda(y)\, \delta_{cd}\,\delta^3(y-z)\nn\\
&=-\del_c\lambda(z).
}
Now we apply the above on the gauge potential $A_a(x)$, namely
\eq{
\delta_\lambda A_a(x)&\equiv \{G(\lambda), A_a(x)\}\nn\\
&=\int d^3z\, \frac {\delta G(\lambda, y)}{\delta \Pi^b(z)}\{\Pi^b(z), A_a(x)\}\nn\\
&=\int d^3z\,\Big(-\del_b\lambda(z)\Big) \Big(-\delta_{ab}\delta^3(x-z)\Big)\nn\\
&=\del_a\lambda(x).
}
This is what we expect from the gauge transformation on the vector potential in Maxwell's theory.

\section{Gauge invariance of the Electric Field and Magnetic Field}
The electric field vector and magnetic field vector are the central elements in Maxwell's theory. Being the physical observable in the theory, they are invariant under gauge transformations. Here we apply the gauge transformations on the electric field vector and magnetic field vector separately to verify their invariance. Specifically, they should lead to vanishing Poisson brackets with the gauge generator $G(\lambda)$.

Recall that the electric field vector only differs from the canonical momenta by a minus sign, i.e., $E^a=-\Pi^a$. Therefore
\eq{
\delta_\lambda E^a(x)&\equiv \{G(\lambda), -\Pi^a(x)\}\nn\\
&=\int d^3z\, \frac {\delta G(\lambda, y)}{\delta \Pi^b(z)}\{\Pi^b(z), -\Pi^a(x)\}=0.
}
Therefore we have verified the gauge invariance of the electric field. 

Next, we move on to the case of the magnetic field vector $B^a$, which can be expressed in terms of the gauge potential $B^a=\frac{1}{2}\epsilon^{abc}F_{bc}$. To determine whether the magnetic field is gauge invariant we have to perform the following Poisson bracket
\eq{
\{G(\lambda(y)),B_a(x) \} = 
\int d^3z\,d^3w\, \frac{\delta G(\lambda(y))}{\delta \Pi_b(z)} \{\Pi_b(z),A_c(w) \} \frac{\delta B_a(x)}{\delta A_c(w)} 
}
First, we will address the smeared constraint term.
\eq{ 
\frac{\delta G(\lambda(y))}{\delta \Pi_b(z)} = \frac{\delta}{\delta \Pi_b(z)} \int d^3y\, \lambda(y)\del_d \Pi^d(y) 
}
Integrating by parts allows us to act the derivative on $\lambda$ and then variational derivative on $\Pi^c$.
\eq{
\frac{\delta G(\lambda(y))}{\delta \Pi_b(z)}
&= - \frac{\delta}{\delta \Pi_b(z)} \int d^3y\, \del_d\lambda(y) \Pi^d(y)  \nn\\
&= -\int d^3y\, \del_d\lambda(y) \frac{\delta \Pi^d(y)}{\delta \Pi_b(z)} \nn\\
&= -\int d^3y\, \del_d\lambda(y) \delta^{db}\delta^3(y-z)\nn\\
&= -\del_b\lambda(z)
}
Second, we have the Poisson brackets between the canonical variables which is identically
\eq{
\{\Pi_b(z),A_c(w) \}=-\delta_{bc}\delta^3(z-w)
}
Last we expand the term containing the magnetic field
\eq{
\frac{\delta B_a(x)}{\delta A_c(w)} &= \frac{\delta}{\delta A_c(w)} 
\tfrac{1}{2}\epsilon_{ade}(\del_d A_e(x)-\del_e A_d(x))
\nn\\&=
\tfrac{1}{2}\epsilon_{ade}\left(\del_d \frac{\delta A_e(x)}{\delta A_c(w)} -\del_e \frac{\delta A_d(x)}{\delta A_c(w)}\right)
\nn\\&=
\tfrac{1}{2}\epsilon_{ade}\left(\delta_e^c \del_d \delta^3(x-w)
- \delta_d^c \del_e \delta^3(x-w)\right)
}
substituting the following results into the full Poisson bracket gives us
\eq{
\{G(\lambda(y)),B_i(x) \} 
&= \int d^3z\,d^3w\, \del_b\lambda(z) \delta_{bc}\delta^3(z-w)
\tfrac{1}{2}\epsilon_{ade}\left(\delta_e^c \del_d \delta^3(x-w)
- \delta_d^c \del_e \delta^3(x-w)\right)
\nn\\&=
\tfrac{1}{2}\epsilon_{ade} \int d^3w\, \del_c\lambda(w) 
\left(\delta_e^c \del_d \delta^3(x-w)
- \delta_d^c \del_e \delta^3(x-w)\right)
\nn\\&=
\tfrac{1}{2}\epsilon_{ade} \int d^3w\, \del_c\lambda(w) 
\left(\delta_e^c \del_d \delta^3(x-w)
- \delta_d^c \del_e \delta^3(x-w)\right)
\nn\\&=
\tfrac{1}{2}\epsilon_{ade} \int d^3w\, 
\left(\del_e\lambda(w)  \del_d \delta^3(x-w)
- \del_d\lambda(w)  \del_e \delta^3(x-w)\right)
}
now we perform integration by parts and act the partial derivative, previously acting on the Dirac delta function, on $\lambda$ which allows us to factor out the Dirac delta
\eq{
\{G(\lambda(y)),B_i(x) \} &=
-\tfrac{1}{2}\epsilon_{ade} \int d^3w\, 
\left(
\del_d\del_e\lambda(w)   
- \del_e\del_d\lambda(w) \right) \delta^3(x-w)
\nn\\&= 
-\tfrac{1}{2} \epsilon_{ade} \left( \del_d \del_e \lambda(x)  - \del_e \del_d \lambda(x)   \right)
}
and since derivatives commute we have proven the magnetic field is gauge invariant in Maxwell Theory
\eq{
\{G(\lambda(y)),B_a(x) \} = 0
}

With the proofs of gauge invariance of the electric and magnetic fields from the perspective of a constrained Hamiltonian system, we have concluded this section on the Maxwell theory. The gauge field $A_a(x)$ in the Maxwell theory is also called a differential 1-form. Because of its simple form, we can always switch the order in products freely, for example, $A_1 A_2= A_2 A_1$. From a group theory standpoint, the gauge fields have a generator of the Lie algebra as simply the unity matrix. This type of gauge theory is called an abelian gauge theory with an underlying $U(1)$ gauge symmetry. However, this simple fashion is not always the case, as we will see in the next section with the Yang-Mills theory, which is also referred to as the non-abelian gauge theory with an $SU(N)$ gauge symmetry. For example, the electroweak theory has an $SU(2)$ gauge symmetry, and the generators of the Lie algebra are the three $2\times 2$ Pauli matrices. Similarly, quantum chromodynamics (QCD) has an $SU(3)$ gauge symmetry whose Lie generators are the eight $3\times 3$ Gellmann matrices. In those cases, the gauge fields are Lie-algebra-valued 1-forms, with additional structures. We will explore all the nitty-gritty of the Yang-Mills theory in the next section.

\chapter{Yang-Mills Theory} \label{sec yang}

In the previous section, we saw that Maxwell's theory can be formulated as a gauge theory in terms of abelian, or equivalently, $U(1)$ gauge fields $A_\mu(x)$. The theory is invariant under the gauge transformations defined by such gauge fields. The construction of a gauge theory can be generalized to non-abelian or $SU(N)$ gauge fields where individual components of these fields don't commute with each other. This type of gauge theory is called a non-abelian gauge theory, also commonly referred to as the Yang-Mills theory. The electroweak theory and quantum chromodynamics (QCD) of particle physics are both examples of a Yang-Mills theory.

In this section, we will focus on the following aspects of the Yang-Mills theory: In subsection 4.1, we will introduce some basic elements of the Yang-Mills theory to help define the action. In subsection 4.2 we will perform the canonical analysis of the theory to obtain the total Hamiltonian. In subsection 4.3 we discuss the gauge generator and transformations. For a similar introduction to Yang-Mills theory and more insight on holonomy refer to \cite{Gambini,intro, Baez}.  

\section{Basics of Yang-Mills theory}
The central element of the Yang-Mills theory is a generalized, Lie-algebra-valued gauge potential
\eq{A_\mu=A_\mu^i T^i, \quad i=1,2,\dotsb , N^2-1}
where $T^i$ are the generators of the Lie algebra that satisfy the following commutation relation
\eq{
[T^i, T^j]=if^{ijk}T^k. \label{comutator}
}
The number of independent generators depends on the dimension $N$ of the gauge group. For $SU(N)$, the number of generators equals $N^2-1$. For example, $SU(2)$ gauge group has $2^2-1=3$ independent generators which are directly related to the three Pauli matrices:
\eq{
T^i = \frac 1 2 \sigma^i, \quad i=1,2,3.
}

Due to the non-abelian nature of the gauge potential, we have to generalize the notion of partial derivative to a covariant derivative 
\eq{
D_\mu = I \del_\mu - i g A_\mu,
}
where $I$ is the identity matrix (since $A_\mu$ is a matrix). The covariant derivative transforms in the adjoint representation of the Lie algebra. For example, when acting on a gauge potential, it follows
\eq{
D_\mu A_\nu = \del_\mu A_\nu  - i g [A_\mu, A_\nu].
}
Applying Eq. (\ref{comutator}),  we have
\eq{
D_\mu A_\nu^i = \del_\mu A_\nu^i  + g f^{ijk}A_\mu^j A_\nu^k.
}
The covariant derivative (which has a similar counterpart in general relativity) also helps us introduce the non-abelian field strength tensor through the following equation
\eq{
[D_\mu, D_\nu] = -ig F_{\mu\nu} = -ig F_{\mu\nu}^i T^i,
} 
from which we can obtain the component $F_{\mu\nu}^i$ as 
\eq{
F_{\mu\nu}^i = \del_\mu A_\nu^i - \del_\nu A_\mu^i + gf^{ijk}A_\mu^j A_\nu^k
} 

Now we can define the Yang-Mills action similar to that of the Maxwell theory
\eq{
S_{YM}=-\frac 1 2 \int d^4x\, \tr  (F_{\mu\nu}F^{\mu\nu}),
}
where the trace operation is required to produce a scalar quantity. Upon applying the following trace identity of the Lie generators $T^i$
\eq{
\tr (T^i T^j)=\frac 1 2 \delta^{ij},
}
the action becomes 
\eq{
S_{YM}=-\frac 1 4 \int d^4x\, F_{\mu\nu}^i F^{\mu\nu\, i},
}

 The equations of motion of the Yang-Mills theory (called the Yang-Mills equations) can be obtained similarly following the Least-action Principle, which gives
\eq{
\epsilon^{\mu\nu\rho\sigma}D_\nu F_{\rho\sigma}=0.
}
This is known as the non-abelian Bianchi identity. Contrary to Maxwell's equations, the Yang-Mills equations do not associate directly with the physical observables of the theory. This is due to the fact that the Yang-Mills field strength tensor is not a gauge invariant quantity. We will get into this aspect in subsection 4.3.

\section{Canonical analysis of the Yang-Mills action}
As usual, we now express the Lagrangian density in the $3+1$ form explicitly
\eq{
\mathcal{L}_{YM} &= -\frac 1 4 F_{\mu\nu}^i F^{\mu\nu\, i}\nn\\
&=-\frac 1 4 (F_{0a}^iF^{0a\,i} + F_{a0}^iF^{a0\,i} + F_{ab}^iF^{ab\,i})\nn\\
&=\frac 1 2 (F_{0a}^i)^2 -\frac 1 4(F_{ab}^i)^2.
\label{ym lagrangian}
}
For the purpose of obtaining the canonical momenta, only the first term of the above expression is relevant. Specifically, the variation of the Lagrangian density produces
\eq{
\delta \mathcal{L}_{YM} &\supset F_{0a}^i \,\delta F_{0a}^i \nn\\
& \supset F_{0a}^i \,\delta \dot A_a^i.
}
Thus we have the canonical momenta
\eq{
\Pi^{a i} = \frac {\delta L_{YM}}{\delta \dot A_a^i} = F_{0a}^i := E^{ai}.
\label{ym momenta}
}
A careful look at $E^{a i}$ shows 
\eq{
E^{a i}=\dot A_a^i - \del_a A_0 + gf^{ijk}A_0^j A_a^k.
\label{adot}
}
Similar to the Maxwell case, we have the following set of basic Poisson brackets
\eq{
\{A_\mu^i(x), A_\nu^j(y)\}&=0,\\
\{E^{\mu i}(x), E^{\nu j}(y)\}&=0,\\
\{A_0^i(x), E^{0j}(y)\}&=\delta^{ij}\delta^3(x-y),\\
\{A_a^j(x), E^{bj}(y)\}&=\delta_{a}^b\delta_{ij}\delta^3(x-y).
} 
Now the canonical Hamiltonian of Yang-Mills theory follows immediately
\eq{
H_{c} &= \int d^3 x \Big(E^{ai} \dot A_a^i - (\frac 1 2 (E^{ai})^2 -\frac 1 4(F_{ab}^i)^2)\Big)\nn\\
&= \int d^3 x \Big(E^{ai}(E^{ai}+\del_a A_0^i - gf^{ijk}A_0^j A_a^k)  - \frac 1 2 (E^{ai})^2 +\frac 1 4(F_{ab}^i)^2\Big)\nn\\
&=\int d^3 x \Big(\frac 1 2 (E^{ai})^2 + \frac 1 4(F_{ab}^i)^2 - A_0^i \del_a E^{ai} - gf^{ijk}A_0^j A_a^k E^{ai} \Big)\nn\\
&=\int d^3 x \Big(\frac 1 2 (E^{ai})^2 + \frac 1 4(F_{ab}^i)^2 - A_0^i (\del_a E^{ai} + gf^{jik} A_a^k E^{aj}) \Big)\nn\\
&=\int d^3 x \Big(\frac 1 2 (E^{ai})^2 + \frac 1 4(F_{ab}^i)^2 - A_0^i (\del_a E^{ai} + gf^{kij} A_a^j E^{ak}) \Big)\nn\\
&=\int d^3 x \Big(\frac 1 2 (E^{ai})^2 + \frac 1 4(F_{ab}^i)^2 - A_0^i (D_a E^a)^i) \Big).
}
In the above steps, Eq. (\ref{adot}) is used in the first line to the second line, and integration by parts is performed from the second to the third line. It is worth noting the similarity between the Yang-Mills and Maxwell Hamiltonians, with the difference being the last term with the covariant derivative instead of the partial derivative. The role of $A_0$ is again a Lagrange multiplier, which suggests that what it multiplies should be treated as a constraint. 

Now we will follow the standard Dirac procedure for the constraint analysis. First, we have the primary constraint immediately following the canonical momentum in (\ref{ym momenta}), namely,
\eq{
(\phi^0)^i=(E^0)^i=-(F^{00})^i\approx 0.
}
This allows us to construct the total Hamiltonian
\eq{
H_T &= H_c + \int d^3x \, \mu^i(x) (E^0)^i(x)\\
&=\int d^3 x \Big(\frac 1 2 (E^{ai})^2 + \frac 1 4(F_{ab}^i)^2 - A_0^i (D_a E^a)^i + \mu^i (E^0)^i \Big)
}
Next, we impose the consistency condition on the primary constraint to obtain a potential secondary constraint:
\eq{
\{(E^0)^i,H_T\}&=\int d^3x \{(E^0)^i,- A_0^j (D_a E^a)^j)\}\\
&=(D_a E^a)^i,
}
which we recognize as the non-abelian Gauss constraint
\eq{
(\phi^1)^i=(D_a E^{a})^i\approx 0
}

The next step is to impose the consistency condition on the newly obtained secondary constraint, namely the Gauss constraint, to seek potentially additional secondary constraint, namely
\eq{
\{(\phi^1)^i, H_T\}
&=gf^{ijk}(A_0)^j (D_a E^a)^k \nn \\
&=gf^{ijk}(A_0)^j (\phi^1)^k \approx 0.
\label{pb phi1 H}
}
Interestingly, this Poisson bracket is proportional to the Gauss constraint so this consistency condition is satisfied automatically, therefore it does not lead to additional secondary constraints.

Now that we have hunted down all the constraints. The Dirac procedure requires that we evaluate their Poisson brackets to identify whether they are of first or second class. Let's show the results first:
\eq{
\{(\phi^0)^i(x), (\phi^0)^j(y)\} &= 0,\label{pb phi0 phi0}\\
\{(\phi^0)^i(x), (\phi^1)^j(y)\}&=0,\label{pb phi0 phi1}\\
\{(\phi^1)^i(x), (\phi^1)^j(y)\}&=-gf^{ijk}\delta^3(\bm{x}-\bm{y})(\phi^1)^k\approx 0.
\label{pb phi1 phi1}
}
In the above, the first two Poisson brackets follow directly from the fundamental Poisson brackets. The third one is proportional to the Gauss constraint thus it is weakly zero. These results suggest that all the constraints in Yang-Mills theory are first-class.

A curious reader may have noticed that we have left out the proofs of some fairly nontrivial results, specifically Eqs.(\ref{pb phi1 H}) and (\ref{pb phi1 phi1}). These are not by accident. We left them out intentionally because proving them requires some special treatment of the constraints. Specifically, the presence of a derivative in the Gauss constraint means that Dirac delta functions will come about when it is evaluated in Poisson brackets. Being a distribution instead of an ordinary function with well-defined derivatives, the Dirac delta function does not have a well-defined behavior on its own in Poisson brackets. This makes Eq. (\ref{pb phi1 H}) and (\ref{pb phi1 phi1}) generally hard to compute in a straightforward way. 

A distribution is much better handled when it is accompanied by a test function inside an integral. This treatment is commonly known as ``smearing'', which is a standard technique to tame the behavior of distributions, specifically, Dirac delta functions in Poisson brackets (See, for example, references \cite{intro} and \cite{Gambini} for additional explanations.) For this reason, we again introduce a smeared integral for the Gauss constraint:
\eq{
G(\lambda)=\int d^3x\, \lambda^i (D_a E^{a})^i
}
where $\lambda^i=\lambda^i(x)$ is an algebra-valued arbitrary parameter that plays the role of the test function. In terms of the smeared constraint $G(\lambda)$, the Poisson brackets in (\ref{pb phi1 H}) and (\ref{pb phi1 phi1}) are equivalent to
$\{G(\lambda), H_T\}$ and $\{G(\lambda),G(\mu)\}$ respectively. These Poisson brackets are much more manageable now. Unfortunately, they are still quite technical to prove. We will hold off on their proofs until the end of the next subsection when some other useful relations are obtained.

\section{Gauge Transformation of A and E}

Similar to Maxwell's theory, the smeared Gauss constraint $G(\lambda)$ also serves as the generator of gauge transformations. In order to find the actions of the gauge generator we perform the following Poisson brackets. For simplicity, in the following parts, we will take the $SU(2)$ Lie generator $T^i=\frac 1 2 \sigma^i$, where $\sigma^i$ are the Pauli matrices and the structure constant $f^{ijk}$ becomes the totally antisymmetric tensor $\epsilon^{ijk}$. 

We will start with the gauge transformation on the gauge potential, i.e.,
\eq{
\{G(\lambda), A^i_a (x)\} = \int d^3 y\, \frac{\delta G(\lambda)}{\delta E^j_b (y)} \{ E^j_b (y), A^i_a (x) \}
}
where
\eq{
\frac{\delta G(\lambda)}{\delta E^j_b (y)} &= \frac{\delta}{\delta E^j_b (y)} \int d^3 z\, \lambda^k (z) D_c E^k_c (z)
\nn\\
&= 
\frac{\delta}{\delta E^j_b (y)} \int d^3 z\, \lambda^k (z) (\del_c E^k_c (z) + g\epsilon^{kmn} A^m_c (z) E^n_c (z))
\nn\\
&= 
\frac{\delta}{\delta E^j_b (y)} \int d^3 z\, (- \del_c \lambda^k (z) E^k_c (z) + g \epsilon^{kmn} A^m_c (z) E^n_c (z) \lambda^k (z))
\nn\\
&=
\int d^3 z\, \left(- \del_c \lambda^k (z) \frac{\delta E^k_c (z)}{\delta E^j_b (y)} +g\epsilon^{kmn} A^m_c (z) \lambda^k (z) \frac{\delta E^n_c (z)}{\delta E^j_b (y)}\right)
\nn\\
&=
\int d^3 z\, (- \del_c \lambda^k (z) \delta^k_j \delta^b_c \delta^3 (z-y) + g\epsilon^{kmn} A^m_c (z) \lambda^k (z) \delta^n_j \delta^b_c \delta^3 (z-y))
\nn\\
&=
-\del_b \lambda^j (y) + g\epsilon^{kmj} A^m_b (y) \lambda^k (y) 
\nn\\
&=
-\del_b \lambda^j (y) - g\epsilon^{jmk} A^m_b (y) \lambda^k (y)
\nn\\
&=
-D_b \lambda^j (y)
}
substituting the previous result back into the Poisson bracket 
\eq{
\{G(\lambda), A^i_a (x)\} &= \int d^3 y\, \left( -D_b \lambda^j (y) \right) \{ E^j_b (y), A^i_a (x) \}
\nn\\&=
\int d^3 y\, \left(-D_b \lambda^j (y) \right) \left( -\delta^{ij} \delta_{ab} \delta^3(y-x) \right)
}
and finally we arrive at our result
\eq{\{G(\lambda), A^i_a (x)\} = D_a \lambda^i(x)
}
The gauge transformation for the gauge potential $A$ in a Yang-Mills field theory is the covariant derivative of the gauge parameter $\lambda$ i.e.
\eq{{A^{i}_a}'  \to A^i_a + D_a \lambda^i.
}
Notice this is of a similar fashion to that in Maxwell's theory where the gauge transformation of $A$ is simply the gradient of $\lambda$.

Next, we obtain the gauge Transformation of the canonical momenta,i.e.,
\eq{
\{G(\lambda), E^i_a (x)\} = \int d^3 y\, \frac{\delta G(\lambda)}{\delta A^j_b (y)} \{ A^j_b (y), E^i_a (x) \}
}
where
\eq{
\frac{\delta G(\lambda)}{\delta A^j_b (y)} &= \frac{\delta}{\delta A^j_b (y)} \int d^3 z\, \lambda^k (z) D_c E^k_c (z)
\nn\\
&= 
\frac{\delta}{\delta A^j_b (y)} \int d^3 z\, \lambda^k (z) (\del_c E^k_c (z) + g\epsilon^{kmn} A^m_c (z) E^n_c (z))
}
Since the first term is independent of $A$ it will vanish under the functional derivative and we are left with
\eq{
\frac{\delta G(\lambda)}{\delta A^j_b (y)} &=
\int d^3 z\,  g\epsilon^{kmn}  \lambda^k (z) \frac{\delta A^m_c (z)}{\delta A^j_b (y)}  E^n_c (z)
\nn\\ &=
\int d^3 z\,  g\epsilon^{kmn} \lambda^k (z)  \delta^m_j \delta^b_c \delta^3(z-y) E^n_c (z)
\nn\\ &=
g\epsilon^{kjn} \lambda^k (y) E^n_b (y)
\nn\\&=
-g\epsilon^{jkn} \lambda^k (y) E^n_b (y)
}
substituting the previous result back into the Poisson bracket 
\eq{
\{G(\lambda), E^i_a (x)\} &= \int d^3 y\, \left( -g\epsilon^{jkn} \lambda^k (y) E^n_b (y) \right) \{ A^j_b (y), E^i_a (x) \}
\nn\\&=
\int d^3 y\, \left( -g\epsilon^{jkn} \lambda^k (y) E^n_b (y) \right) \delta_{ab} \delta^{ij} \delta^3(y-x)
\nn\\&=
-g\epsilon^{ikn}\lambda^k(x) E_a^n(x),
}
and finally renaming some summed-over indices we arrive at our final result
\eq{
\{G(\lambda), E^i_a (x)\} = -g\epsilon^{ijk}\lambda^j(x) E_a^k(x).
}
We have shown that the electric field analog $E$ in Yang-Mills theory has the gauge transformation 
\eq{
{E^i_a}' \to E^i_a -g\epsilon^{ijk}\lambda^j E_a^k,
}
which means it is no longer a gauge invariant quantity and therefore not a candidate for a physical observable in Yang-Mills theory. As we shall see in a later section, the field strength tensor $F_{ab}^i$ also fails to pass the same test, as we will show in the next subsection.

\section{Gauge Dependent Field Strength Tensor}
In Yang-Mills theory the spatial field tensor $F^i_{ab}(x)$ is not invariant under gauge transformations. To prove this we will calculate the following Poisson bracket:
\eq{
\{ G(\lambda),F^i_{ab}(x) \} = 
\int d^3y \, \frac{\delta G(\lambda)}{\delta E^c_j(y)} \{E^c_j(y),F^i_{ab}(x)\} \label{{G,F}}
}
We will first calculate the Poisson bracket between the electric field and the field tensor.
\eq{
\{E^c_j(y),F^i_{ab}(x)\} = \int d^3z \, \left(
\frac{\delta E^c_j(y)}{\delta A^l_d(z)} \frac{\delta F^i_{ab}(x)}{\delta E^d_l(z)} - 
\frac{\delta E^c_j(y)}{\delta E^d_l(z)} \frac{\delta F^i_{ab}(x)}{\delta A^l_d(z)}
\right)
}
The first term is identically zero and if we expand $F^i_{ab}(x)$ we are left with 
\eq{
&= 
-\int d^3z \frac{\delta E^c_j(y)}{\delta E^d_l(z)} \frac{\delta}{\delta A^l_d(z)} 
( \del_a A^i_b(x) - \del_b A_a^i(x) + g \epsilon^{ijk} A^j_a(x) A^k_b(x) )
\nn\\&=
-\int d^3z \delta^c_d \delta^l_j \delta^3(y-z)
 \frac{\delta}{\delta A^l_d(z)} 
( \del_a A^i_b(x) - \del_b A_a^i(x) + g \epsilon^{ijk} A^j_a(x) A^k_b(x) )
\nn\\&=
-\frac{\delta}{\delta A^j_c(y)} 
( \del_a A^i_b(x) - \del_b A_a^i(x) + g \epsilon^{ilk} A^l_a(x) A^k_b(x) )
\nn\\&=
\del_b \frac{\delta A_a^i(x)}{\delta A^j_c(y)}-
\del_a \frac{\delta A^i_b(x)}{\delta A^j_c(y)} - g \epsilon^{ilk} \left(
\frac{\delta A^l_a(x)}{\delta A^j_c(y)} A^k_b(x)+
A^l_a(x)\frac{\delta A^k_b(x)}{\delta A^j_c(y)}
\right)
\nn\\&=
\delta^i_j (\delta_a^c\, \del_b\, \delta^3(x-y)- \delta_b^c\, \del_a\, \delta^3(x-y) ) -
g \epsilon^{ilk} \left(
\delta^l_j \delta_a^c A^k_b(x)+
A^l_a(x)\delta^k_j \delta_b^c
\right) \delta^3(x-y)
\nn\\&=
\delta^i_j (\delta_a^c\, \del_b\, \delta^3(x-y)- \delta_b^c\, \del_a\, \delta^3(x-y) ) -
g (\epsilon^{ijk} \delta^c_a A^k_b(x) + 
\epsilon^{ilj} \delta^c_b A^l_a(x)) \delta^3(x-y)
\nn \\
&=
\delta^i_j (\delta_a^c\, \del_b\, \delta^3(x-y)- \delta_b^c\, \del_a\, \delta^3(x-y) ) -
g \epsilon^{ijk} (\delta^c_a A^k_b(x) - 
\delta^c_b A^k_a(x)) \delta^3(x-y)
\nn \\
&=2\delta^i_j \delta_{[a}^c\, \del_{b]}\, \delta^3(x-y) -
2g \epsilon^{ijk} \delta^c_{[a} A^k_{b]}(x) \delta^3(x-y)
}
where in the third to second to last line we have renamed the dummy indices from $l$ to $k$ on the second epsilon tensor term and changed the order of the last two indices to flip the sign which allows us to use a single epsilon tensor.

Putting the previous result on hold we will now calculate the functional derivative of $G(\lambda)$ with respect to $E$
\eq{
\frac{\delta G(\lambda)}{\delta E^c_j(y)} &= \frac{\delta}{\delta E^c_j(y)}  \int d^3w \,\lambda^i(w) \left( \del_a E_a^i(w) + g\epsilon^{ilk} A^l_a(w) E^k_a(w) \right)
\nn\\&=
\frac{\delta}{\delta E^c_j (y)} 
\int d^3 w \, (-E^i_a (w)\del_a \lambda^i (w) + g\epsilon^{ilk} A^l_a (w) E^k_a (w) \lambda^i (w))
\nn\\&= 
\int d^3 w \, \left( -\frac{\delta E^i_a (w) \delta}{\delta E^c_j (y)}\del_a \lambda^i (w) + g\epsilon^{ilk} A^l_a (w) \frac{\delta E^k_a (w)}{\delta E^c_j (y)} \lambda^i (w) \right)
\nn\\ &= 
\int d^3 w \, \delta^{ij} \delta_{ac} \delta^3 (w-y) (-\del_a \lambda^i (w)) + g \epsilon^{ilk} A^l_a \delta^{jk} \delta_{ac} \delta^3 (w-y) \lambda^i (w)
\nn\\&=
-\del_c \lambda^j(y) + g\epsilon^{ilj} A_c^l(y)\lambda^i(y)
\nn\\&=
-\del_c \lambda^j(y) - g\epsilon^{jli} A_c^l(y)\lambda^i(y) 
\nn\\&=
-\del_c \lambda^j(y) - g\epsilon^{jkl} A_c^k(y)\lambda^l(y) 
= -D_c \lambda^j(y)
}
where in the last step the summed over indices have been renamed for simplicity later.

Now we can substitute the following results in back into Eq. (\ref{{G,F}})
\eq{
\{ G(\lambda),F^i_{ab}(x) \} =
\int d^3y \, (-D_c \lambda^j(y)) \left( 2\delta^i_j \delta_{[a}^c\, \del_{b]}\, \delta^3(x-y) -
2g \epsilon^{ijk} \delta^c_{[a} A^k_{b]}(x) \delta^3(x-y) \right)
\label{gf 2 terms}.
}
We will perform this calculation by splitting it into two terms. First, we have
\eq{
 \int d^3y\, (&-D_c\lambda^j(y)) \left( 2\delta^i_j \delta_{[a}^c\, \del_{b]}\, \delta^3(x-y) \right)
\nn \\
=&
-\int d^3y\, \left( \del_c \lambda^j(y) + g\epsilon^{jkl} A_c^k(y)\lambda^l(y) \right)
\delta^i_j 
\left( \delta_a^c\, \del_b\, \delta^3(x-y)- \delta_b^c\, \del_a\, \delta^3(x-y) \right)
\nn\\=&
-\int d^3y\, \del_c \lambda^i(y) \delta_a^c \del_b \delta^3(y-x)
- \del_c \lambda^i(y) \delta_b^c \del_a \delta^3(y-x)
\nn\\&
+ g\epsilon^{ikl} \lambda^l(y)
\left( A^k_c(y) \delta^c_a \del_b \delta^3(y-x)
- A^k_c(y) \delta^c_b \del_a \delta^3(y-x)
\right)
\nn\\=&
-\int d^3y\,
\del_a \lambda^i(y) \del_b \delta^3(y-x)
- \del_b \lambda^i(y) \del_a \delta^3(y-x)
\nn\\&+ g\epsilon^{ikl} \lambda^l(y)
\left( 
A^k_a(y) \del_b \delta^3(y-x)
- A^k_b(y) \del_a \delta^3(y-x)
\right)
}
Now we integrate by parts and act the partial derivative, previously on the Dirac delta function, on everything to the left of the partial derivative in each term. This also allows us to group together terms that have the same derivative acting on them as follows
\eq{
&= \int d^3y\, \left[
\del_a \left( \del_b \lambda^i(y) + g\epsilon^{ikl} \lambda^l(y) A_b^k(y) \right) -
\del_b \left( \del_a \lambda^i(y) + g\epsilon^{ikl} \lambda^l(y) A_a^k(y) \right) 
\right] \delta^3(y-x)
\nn\\&=
g\epsilon^{ikl} \left[ \del_a (\lambda^l(x) A_b^k(x)) -
\del_b (\lambda^l(x) A_a^k(x)) \right]
\nn\\&=
g\epsilon^{ikl} \left[ 
\del_a \lambda^l(x) A_b^k(x) +
\lambda^l(x) \del_a A_b^k(x)-
\del_b \lambda^l(y) A_a^k(x) -
\lambda^l(x) \del_b A_a^k(x)
\right].
\label{gf 1}
}
We will stop here and now work on the second term of Eq. (\ref{gf 2 terms}) as follows
\eq{
& \int d^3y\, (-D_c\lambda^j(y)) \left( -
2g \epsilon^{ijk} \delta^c_{[a} A^k_{b]}(x) \delta^3(x-y) \right)
\nn \\
&=
\int d^3y\, D_c\lambda^j(y)
2g \epsilon^{ijn} \delta^c_{[a} A^n_{b]}(x) \delta^3(x-y) 
\nn\\&=
D_c\lambda^j(x)
2g \epsilon^{ijn} \delta^c_{[a} A^n_{b]}(x) 
\nn \\
&=D_c\lambda^j
2g \epsilon^{ijn} \delta^c_{[a} A^n_{b]} 
\nn \\
&=
\left (\del_c \lambda^j + 
g\epsilon^{jkl} A_c^k\lambda^l \right)
g \epsilon^{ijn} (\delta^c_a A^n_b - \delta^c_b A^n_a )
\nn\\&=
g\epsilon^{ijn} ( \del_c \lambda^j \delta^c_a A_b^n - \del_c \lambda^j \delta^c_b A_a^n ) +
g^2 \epsilon^{ijn}\epsilon^{jkl}
( A^k_c \lambda^l \delta^c_a A^n_b - 
A^k_c \lambda^l \delta^c_b A^n_a)
\nn\\&=
g\epsilon^{ijn} 
( \del_a \lambda^j A_b^n - 
\del_b \lambda^j A_a^n) +
g^2 \epsilon^{ijn}\epsilon^{jkl} \lambda^l
( A^k_a  A^n_b - 
A^k_b A^n_a)
\label{gf 2}
}
where in the second line we have renamed the summed over index $k$ to $n$ to prevent later confusion and in the fourth line we have suppressed the function input $(x)$ since all functions in the remainder of the calculation are functions of $x$. 

Now we have solved both terms in Eq. (\ref{gf 2 terms}) and can add them together i.e. Eq.  (\ref{gf 1}) + Eq. (\ref{gf 2}) to get 
\eq{
\{ G(\lambda),F^i_{ab}(x) \}  &=
g\epsilon^{ikl} \left[ 
\del_a \lambda^l A_b^k +
\lambda^l \del_a A_b^k-
\del_b \lambda^l A_a^k -
\lambda^l \del_b A_a^k
\right] \nn
\\& \quad
+ g\epsilon^{ijn} 
( \del_a \lambda^j A_b^n
- \del_b \lambda^j A_a^n ) +
g^2 \epsilon^{ijn}\epsilon^{jkl} \lambda^l
( A^k_a  A^n_b - 
A^k_b A^n_a )
\nn \\
&=
-g\epsilon^{ilk} \left[ 
\del_a \lambda^l A_b^k +
\lambda^l \del_a A_b^k-
\del_b \lambda^l A_a^k -
\lambda^l \del_b A_a^k
\right] \nn
\\& \quad
+ g\epsilon^{ijn} 
( \del_a \lambda^j A_b^n 
- \del_b \lambda^j A_a^n ) +
g^2 \epsilon^{ijn}\epsilon^{jkl} \lambda^l
( A^k_a  A^n_b - 
A^k_b A^n_a )
\nn \\
&=
-g\epsilon^{ilk} \left[ 
\lambda^l \del_a A_b^k-
\lambda^l \del_b A_a^k
\right]
 + g^2 \epsilon^{ijn}\epsilon^{jkl} \lambda^l
( A^k_a  A^n_b - 
A^k_b A^n_a)
\nn\\&=
g\epsilon^{ikl}  
\lambda^l \left[ 
\del_a A_b^k- \del_b A_a^k
\right]
 + g^2 \epsilon^{ijn}\epsilon^{jkl} \lambda^l
( A^k_a  A^n_b - 
A^k_b A^n_a),
}
where in the second line we flipped the last two indices on $g^{ikl}$, and in the second to the third line we cancel the first and third term of the first part, with the first two partial derivative terms of the second part (since $l,k,j,n$ are all summed over). Next, we contract the two Levi-Civita symbols and simplify them as follows
\eq{
&= 
g\epsilon^{ikl}  
\lambda^l \left[ 
\del_a A_b^k- \del_b A_a^k
\right]
 + g^2 (\delta^{kn}\delta^{li} - \delta^{ki}\delta^{ln}) \lambda^l
( A^k_a  A^n_b - A^k_b A^n_a)
\nn\\ &= 
g\epsilon^{ikl}  
\lambda^l(x) \left[ 
\del_a A_b^k- \del_b A_a^k
\right] + g^2 
 \lambda^l(
 A^k_a  A^n_b\delta^{kn}\delta^{li} - 
 A^k_a  A^n_b\delta^{ki}\delta^{ln} - 
 A^k_b A^n_a\delta^{kn}\delta^{li} + 
 A^k_b A^n_a\delta^{ki}\delta^{ln}) 
\nn \\ &= 
g\epsilon^{ikl}  
\lambda^l \left[ 
\del_a A_b^k- \del_b A_a^k
\right]
 +
 g^2 (
 A^n_a  A^n_b\lambda^i - 
 A^i_a  A^n_b\lambda^n - 
 A^n_b A^n_a\lambda^i + 
 A^i_b A^n_a\lambda^n) 
\nn \\
&= 
g\epsilon^{ikl}  
\lambda^l \left[ 
\del_a A_b^k- \del_b A_a^k
\right] +
 g^2 ( 
 -A^i_a  A^n_b\lambda^n +
 A^i_b A^n_a\lambda^n) 
\nn \\
&= 
g\epsilon^{ikl}  
\lambda^l \left[ \del_a A_b^k- \del_b A_a^k\right] +
 g^2 \lambda^l( 
 A^i_b A^l_a - A^i_a  A^l_b )
\nn \\&= 
g\epsilon^{ikl}  
\lambda^l \left[ \del_a A_b^k- \del_b A_a^k\right] +
 g^2 \lambda^l  
 (\delta^i_m \delta^l_n - \delta^l_m \delta^i_n )
 A^m_a A^n_b
\nn \\&= 
g\epsilon^{ikl}  
\lambda^l \left[ \del_a A_b^k- \del_b A_a^k\right] +
 g^2 \lambda^l \epsilon^{mnk} \epsilon^{ikl}
 A^m_a A^n_b 
\nn \\&= 
g\epsilon^{ikl}  
\lambda^l \left[ \del_a A_b^k- \del_b A_a^k +
 g\epsilon^{kmn} A^m_a A^n_b 
 \right]
\nn \\
&= g\epsilon^{ikl}\lambda^l F_{ab}^k 
\nn\\&=
-g\epsilon^{ilk}\lambda^l F_{ab}^k ,
}
where in the fifth line we rename the dummy index $n$ to $l$ in the $g^2$ term to match the first part of the equation. After renaming our indices we have arrived at the final result
\eq{
\{ G(\lambda),F^i_{ab}(x) \} = -g\epsilon^{ijk}\lambda^j(x) F_{ab}^k(x).
}
This shows that the spatial part of the field tensor has the following gauge transformation
\eq{
{F^i_{ab}}' \to F^i_{ab} -g\epsilon^{ijk}\lambda^j F_{ab}^k.
}
Notice this is of a similar form to the gauge transformation of the electric field as the electric field is the remaining component of the field tensor. This means any magnetic field analog in Yang-Mills field theory is also no longer a gauge invariant quantity and cannot be a physical observable. The pressing 
question to ask then, is what would be the physical observables in a Yang-Mills theory? We will answer this question at the very end of this section but for now, we will take a slight detour and address an earlier issue first.

\section{Closure of the Gauge Generators}

As pointed out at the end of subsection 4.2, we intentionally skipped the proofs of Eqs.(\ref{pb phi1 H}) and (\ref{pb phi1 phi1}). The exact reason is that both proofs essentially involve the Poisson brackets between two Gauss constraints, which involves the covariant derivatives of the canonical momenta. A brute-force tackle would become a laborious process. Now, with the important results obtained in the previous sections, specifically the gauge transformation of $A_a^i$, $E_a^i$, and $F^i_{ab}$, we can finally come back to those much-anticipated proofs.

We will start with the Poisson bracket between two gauge generators: 
\eq{
\{G(\lambda),G(\mu)\}&=\int d^3x \Big(\{G(\lambda),E^{ia}(x)\}\frac {\delta G(\mu)}{\delta E^{ia}(x)} + \{G(\lambda),A^i_a(x)\}\frac {\delta G(\mu)}{\delta A^i_a(x)}\, \Big).
 \nn\\
 &= \int d^3x \Big(\
( -g\epsilon^{ijk} \lambda^j(x) E^{ak}(x) ) (-D_a \mu^i(x) ) +( D_a \lambda^i(x)) (-g \epsilon^{ilm} \mu^l(x) E^{am}(x) )
}
where we have taken advantage of the fact that we have already obtained all of the expressions in the first line and substituted them in. Since everything is a function of $x$ we will suppress the function input $(x)$ and simplify
\eq{
\{G(\lambda),G(\mu)\} &= \int d^3x \Big(\
( -g\epsilon^{ijk} \lambda^j E^{ak} ) (-D_a \mu^i ) +( D_a \lambda^i) (-g \epsilon^{ilm} \mu^l E^{am}) \Big)
\nn \\ &= \int d^3x \Big(\
g\epsilon^{ijk} \lambda^j E^{ak} 
(\del_a \mu^i + g\epsilon^{ilm}A^l_a \mu^m ) 
+( \del_a \lambda^i + g\epsilon^{ijk}A^j_a \lambda^k) g \epsilon^{iml} \mu^l E^{am} \Big)
\nn \\ &= 
\int d^3x \Big(\
g\epsilon^{ijk}\lambda^j \del_a \mu^i E^{ak} + 
g\epsilon^{iml} \mu^l \del_a \lambda^i E^{am}
+ g^2 \epsilon^{ijk} \epsilon^{ilm} \lambda^j \mu^m E^{ak} A^l_a 
\nn \\
&\qquad+ g^2 \epsilon^{ijk} \epsilon^{iml} \lambda^k \mu^l E^{am} A^j_a \Big)
\nn \\ &= 
\int d^3x \,g\Big(\
\epsilon^{ijk}\lambda^j \del_a \mu^i E^{ak} + 
\epsilon^{iml} \mu^l \del_a \lambda^i E^{am}
+ g(\delta^{jl}\delta^{mk} - \delta^{jk}\delta^{ml} ) \lambda^j \mu^m E^{ak} A^l_a 
\nn \\
&\qquad + g(\delta^{jk}\delta^{ml} - \delta^{jl}\delta^{mk})  \lambda^k \mu^l E^{am} A^j_a \Big)
\nn \\ &= \int d^3x \,g\Big(\
\epsilon^{ijk}\lambda^j \del_a ( \mu^i \lambda^j) E^{ak} 
+ g( \lambda^j \mu^k E^{ak} A^j_a -  \lambda^j \mu^j E^{ak} A^k_a)
\nn \\
&\qquad+ g( \lambda^k \mu^k E^{aj} A^j_a -  \lambda^k \mu^j E^{ak} A^j_a) \Big)
\nn \\ &= \int d^3x \,g\Big(\
\epsilon^{ijk}\lambda^j \del_a ( \mu^i \lambda^j) E^{ak} 
+ g( \lambda^j \mu^k -  \lambda^k \mu^j)E^{ak} A^j_a\Big)
\nn \\ &= \int d^3x \,g\Big(\
-\epsilon^{ijk}\lambda^j \del_a ( \mu^k \lambda^j) E^{ai} 
+ g \lambda^m \mu^n ( \delta^{mj} \delta^{nk} -  \delta^{mk} \delta^{nj} )E^{ak} A^j_a\Big)
\nn \\ &= \int d^3x \,g\Big(\
\epsilon^{ijk}\lambda^j \mu^k  \del_a E^{ai} 
+ g \lambda^m \mu^n \epsilon^{imn} \epsilon^{ijk} E^{ak} A^j_a\Big)
\nn \\ &= \int d^3x \,g\Big(\
(\epsilon^{ijk}\lambda^j \mu^k)  \del_a E^{ai} 
+  (\epsilon^{imn}\lambda^m \mu^n) g \epsilon^{ijk}  A^j_a E^{ak}\Big)
\nn \\ &= \int d^3x \, g
(\epsilon^{ijk}\lambda^j \mu^k)  D_a E^{ai},
}
where in the fourth line we expand the contracted epsilon tensors as Kronecker deltas, in the fifth line we have ``undone" the chain rule to combine the first two terms into one, and in the eighth line we integrate by parts on the first term to act the derivative on $E^{ia}$ and in the second term have turned the Kronecker deltas into Levi-Civita symbol.
Finally, in the last line, if we recognize that the Levi-Civita symbol contracted with two other quantities is equivalent to the commutator between the two quantities with respect to the free index $i$, we can write the previous line as
\eq{
\{G(\lambda),G(\mu)\} &= g\int d^3x \, 
[\lambda, \mu]^i  D_a E^{ai} 
\nn \\ &=
G(g[\lambda,\mu]).\label{pb GG}
}
This result is equivalent to the claim in Eq. (\ref{pb phi1 phi1}), for the $SU(2)$ case. It shows that the gauge generators form a closed algebra. The physical implication of this closure is that the effect of two consecutive infinitesimal gauge transformations is equivalent to one gauge transformation with a gauge parameter in the specific form of a commutator of the individual gauge parameters. Eq.(\ref{pb GG}), together with Eqs.(\ref{pb phi0 phi0}) and (\ref{pb phi0 phi1}), prove that all the constraints of Yang-Mills theory are of first class, as claimed in subsection 4.2.

With Eq. (\ref{pb GG}), the proof of the consistency condition in Eq. (\ref{pb phi1 H}) also follows straightforwardly. Let's take a closer look here. Eq. (\ref{pb phi1 H}) is now equivalent to the following Poisson bracket
\eq{&\{G(\lambda), H_T\} \nn\\
=&\int d^3 x\, \{G(\lambda),\xu \frac 1 2 (E^{ai})^2 + \frac 1 4(F_{ab}^i)^2 - A_0^i (D_a E^a)^i + \mu^i (E^0)^i \},\\
= &\int d^3 x\, \Big(\{G(\lambda), \frac 1 2 (E^{ai})^2\} + \{G(\lambda), \frac 1 4(F_{ab}^i)^2\} - \{G(\lambda), A_0^i (D_a E^a)^i\} + \{G(\lambda), \mu^i (E^0)^i \}\Big).\nn 
}
There are four Poisson brackets in the last line above, where the first and the second both vanish due to the gauge transformations on $E$ and $A$ found in subsection 4.3. Then the fourth Poisson bracket vanishes trivially from the fundamental Poisson brackets. So the only surviving Poisson bracket from above is the third term, which we immediately recognize simply as a Poisson bracket between two gauge generators, i.e.,
\eq{\{G(\lambda), H_T\} &= -\int d^3 x\, \{G(\lambda), A_0^i (D_a E^a)^i\}\nn\\
&=-\{G(\lambda), G(A_0)\}\nn\\
&=-G([\lambda,A_0])\approx 0.
}
With this, we have proved that the Gauss constraint is preserved by the total Hamiltonian so that there does not exist any additional secondary constraint in Yang-Mills theory.

\section{Holonomy}
In Yang-Mills theory the vector potential $A^i_a$ and its conjugate momenta $E^i_a$ are both subject to change under gauge transformations. This means these quantities are nonphysical and cannot be considered observables. As it turns out in Yang-Mills field theories there is a gauge invariant quantity namely, the trace of a holonomy. Conceptually holonomy can be described as the resulting difference of parallel transporting the vector potential, $A^i_a$, around a closed loop. More generally, a holonomy is the transportation of a \textit{connection} (a notion that will be introduced in the next section) around a closed loop.  

\begin{figure}[ht]
    \centering
    \includegraphics[scale=0.3]{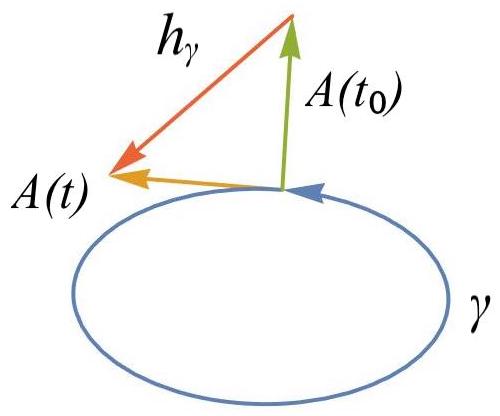}
    \caption{{Holonomy $h_\gamma$ is the difference between the vector potential $A$,  before $A(t_0)$, and after $A(t)$, transporting around the closed loop. $\gamma$}}
    \label{fig:hol}
\end{figure}

Mathematically holonomy is defined as (see, for example, appendix of \cite{Carroll} and chapter 5 of \cite{intro})
\eq{
h_\gamma = \sum_n^\infty
\frac{1}{n!}\oint^t_0
 P\left( 
\dot{\gamma^{a_1}} A_{a_1}(t_1)...\dot{\gamma^{a_n}} A_{a_n}(t_n)
\right) dt_1 ... dt_n,
}
where $P$ is the path-ordered product function and $\gamma$ is a closed path parameterized by $t$. We will now show that the trace of holonomy is gauge invariant by performing the following Poisson bracket
\eq{
\{G(\lambda), \tr\, h_\gamma \} = 
\tr \int d^3y\, \{ G(\lambda), A_b(y) \} \frac{\delta h_\gamma}{\delta A_b (y)} + 
\tr \int d^3y\, \{ G(\lambda), E_b(y) \} \frac{\delta h_\gamma}{\delta E_b (y)},
}
and since $h_\gamma$ is independent of $E$ the second term will vanish and we are left with
\eq{
\{G(\lambda), \tr\,h_\gamma \} = \tr \int d^3y\,
\{ G(\lambda), A_b(y) \} \frac{\delta h_\gamma}{\delta A_b (y)}.
}
When we choose $g=1$ we get
\eq{
\{ G(\lambda), A_b(t) \} = \del_b \lambda + \tfrac{i}{2}\left[ \lambda,A_b \right],
}
and to calculate the functional derivative term we need to address each term in the sum of the holonomy. First, consider the $n=1$ term of holonomy which is simply
\eq{
 h_{\gamma\, n=1} = \oint^t_0
\dot{\gamma^{a}} A_{a}(t,x)\, dt.
}
Using the previous three equations we have
\eq{
\{G(\lambda), \tr\,h_{\gamma} \}_{n=1} &= \tr \int d^3y\, \left(
\del_b \lambda + \tfrac{i}{2}\left[ \lambda,A_b \right] \right)
\frac{\delta}{\delta A_b(t,y)}
\oint^t_0 \dot{\gamma^{a}} A_{a}(t,x)\, dt 
\nn\\&=
\tr \int d^3y\, \left(
\del_b \lambda + \tfrac{i}{2}\left[ \lambda,A_b \right] \right)
\oint^t_0 \dot{\gamma^{a}} \frac{\delta A_{a}(t,x)}{\delta A_b(t,y)} \, dt 
\nn\\&=
\tr \int d^3y\, \left(
\del_b \lambda + \tfrac{i}{2}\left[ \lambda,A_b \right] \right)
\oint^t_0 \dot{\gamma^{a}} \delta^b_a \delta^3(x-y) \, dt 
\nn \\
 &= 
 \tr \int d^3y \,
 \left(\del_b \lambda(y) + \tfrac{i}{2}\left[ \lambda(y) ,A_b(y) \right]\right)
  \delta^3(x-y) \oint^t_0
\dot{\gamma}^{b}  dt 
\nn\\&=
\tr \oint^t_0  \left(
 \del_b \lambda(x) \dot{\gamma}^{b} + \tfrac{i}{2}\left[ \lambda (x) ,A_b(x) \right]
\dot{\gamma}^{b}
\right) dt,
}
where the Dirac delta function is not a function of the parameter $t$ so it can be removed from the integral with respect to $t$.
Since the trace of any commutator between two matrices is zero i.e.
\eq{ \tr \left[ A,B \right]=0 }
the second term vanishes and we are left with (suppressing the function inputs $(x) and(t)$)
\eq{
\{G(\lambda), \tr\,h_{\gamma} \}_{n=1}&=
\tr  \oint^t_0 
 \del_b \lambda \dot{\gamma}^{b}
\, dt
\nn\\&=
 \tr \oint^t_0 
\frac{\del \lambda}{\del \gamma^b} \,
\frac{\del \gamma^b}{\del t}
\, dt
\nn\\&=
\tr  \oint^{\lambda(t)}_{\lambda_0} d \lambda
}
and because we are integrating around a closed loop the previous expression is equal to zero, thus
\eq{
\{ G(\lambda),\tr\,h_\gamma \}_{n=1} =0.
}
Now we will tackle the $n=2$ term in holonomy which is
\eq{
h_{\gamma\, n=2} = \tfrac{1}{2!}\oint^t_0
 P \left( 
\dot{\gamma^{a_1}} A_{a_1}(t_1)\dot{\gamma^{a_2}} A_{a_2}(t_2)
 \right)  dt_1dt_2
}
and using similar substitutions to the $n=1$ term we can say
\eq{
\{ &G(\lambda),\tr\,h_\gamma \}_{n=2} 
\nn \\
&=\tr \int d^3y\, \left(
\del_b \lambda + \tfrac{i}{2}\left[ \lambda ,A_b \right] \right)
\frac{\delta}{\delta A_b(t,y)}
\tfrac{1}{2!} \oint^t_0 P \left( 
\dot{\gamma}^{a_1} A_{a_1}(t_1,x)\dot{\gamma}^{a_2} A_{a_2}(t_2,x)
 \right)  dt_1dt_2
\nn \\
 &=
 \tr \int d^3y\, \left(
\del_b \lambda + \tfrac{i}{2}\left[ \lambda,A_b \right] \right)
\tfrac{1}{2!} \oint_0^t P  \Big(
\dot{\gamma}^{a_1} 
\frac{\delta A_{a_1}(t_1,x)}{\delta A_b(t_1,y)} 
\dot{\gamma}^{a_2} A_{a_2}(t_2,x) 
\nn \\ & \qquad  +
\dot{\gamma}^{a_1} A_{a_1}(t_1,x)\dot{\gamma}^{a_2} 
\frac{\delta A_{a_2}(t_2,x)}{\delta A_b(t_2,y)} \Big) dt_1dt_2
\nn\\&=
 \tr \int d^3y\, \left(
\del_b \lambda + \tfrac{i}{2}\left[ \lambda,A_b \right] \right)
\tfrac{1}{2!} \oint_0^t P  \Big(
\dot{\gamma}^{a_1} \delta^b_{a_1} \delta^3(x-y)
\dot{\gamma}^{a_2} A_{a_2}(t_2,x) 
\nn \\ & \qquad +
\dot{\gamma}^{a_1} A_{a_1}(t_1,x)\dot{\gamma}^{a_2} 
\delta^b_{a_2} \delta^3(x-y) \Big) dt_1dt_2
\nn\\&=
 \tr \int d^3y\, \left(
\del_b \lambda(y) + \tfrac{i}{2}\left[ \lambda(y),A_b(y) \right] \right) \delta^3(x-y)
\tfrac{1}{2!} \oint_0^t P  \Big(
\dot{\gamma}^{b}
\dot{\gamma}^{a_2} A_{a_2}(t_2,x) 
 +
\dot{\gamma}^{a_1} A_{a_1}(t_1,x)\dot{\gamma}^{b}   \Big) dt_1dt_2
\nn\\&=
\tfrac{1}{2!} \, \tr  \oint_0^t \left(
\del_b \lambda(x) + \tfrac{i}{2}\left[ \lambda(x),A_b(x) \right] \right)
 P  \left(
\dot{\gamma}^{b}
\dot{\gamma}^{a_2} A_{a_2}(t_2,x) +
\dot{\gamma}^{a_1} A_{a_1}(t_1,x)\dot{\gamma}^{b}   \right) dt_1dt_2
}
where we have applied the product rule with respect to the functional derivative. Similar to the $n=1$ term, the commutator vanishes under the trace and we are left with (suppressing $(x)$ but not $(t_n)$)
\eq{
\{ &G(\lambda),\tr\,h_\gamma \}_{n=2} 
\nn \\
&=
\tfrac{1}{2!} \, \tr \oint_0^t
\del_b \lambda 
 P  \left(
\dot{\gamma}^{b}(t_1)
\dot{\gamma}^{a_2}(t_2) A_{a_2}(t_2) +
\dot{\gamma}^{a_1}(t_1) A_{a_1}(t_1)
\dot{\gamma}^{b}(t_2)   \right) dt_1dt_2
\nn\\&=
\tfrac{1}{2!} \, \tr \oint_0^t
 P  \left(
\del_b \lambda \dot{\gamma}^{b}(t_1)
\dot{\gamma}^{a_2}(t_2) A_{a_2}(t_2) +
\del_b \lambda \dot{\gamma}^{a_1}(t_1) A_{a_1}(t_1)\dot{\gamma}^{b}(t_2)   \right) dt_1dt_2
\nn\\&=
\tfrac{1}{2!} \, \tr \oint^{t1}_0 
\frac{\del \lambda}{\del \gamma^b} \,
\frac{\del \gamma^b}{\del t_1} dt_1
\oint^{t_2}_0 
\dot{\gamma}^{a_2}(t_2) A_{a_2}(t_2) dt_2 +
\tfrac{1}{2!} \, \tr \oint^{t_2}_0 
\frac{\del \lambda}{\del \gamma^b} \,
\frac{\del \gamma^b}{\del t_2} dt_2
\oint^{t_1}_0 
\dot{\gamma}^{a_2}(t_1) A_{a_1}(t_1) dt_1
\nn\\&=
\tfrac{1}{2!} \, \tr \oint^{\lambda(t)}_{\lambda_0} d \lambda
\oint^{t_2}_0 
\dot{\gamma}^{a_2}(t_2) A_{a_2}(t_2) dt_2 +
\tfrac{1}{2!} \, \tr \oint^{\lambda(t)}_{\lambda_0} d \lambda
\oint^{t_1}_0 
\dot{\gamma}^{a_2}(t_1) A_{a_1}(t_1) dt_1.
}
Once again because we are integrating around a closed loop the integrals containing $d\lambda$ are equal to zero thus
\eq{
\{ G(\lambda),\tr\, h_\gamma \}_{n=2} = 0.
}
As one may already see this pattern continues for every value of $n$. Therefore we can generalize the previous result and say that the Poisson bracket between the trace of a holonomy and the generator of gauge transformations is zero (remember holonomy is an infinite sum of terms, the first two of which we have just calculated) i.e.
\eq{
\{ G(\lambda),\tr\, h_\gamma \} = 0.
}
The trace of a holonomy being a gauge invariant quantity makes it a proper candidate for physical observables in Yang-Mills theory. The closed loops that are integrated around are actually closely related to the loops in loop quantum gravity and the basis for the loop representation.

We will conclude this section with a remark on the similar characteristics of the Yang-Mills theory and Einstein's theory of General Relativity. There are similar notions of a curvature tensor and a covariant derivative, which naturally motivated the attempts of a unified framework where gravity can also be formulated as a gauge theory in the Yang-Mills fashion. In this sense, the theory of general relativity can be thought of as a generalization of a Yang-Mills theory where the ``gauge transformations'' are the coordinate transformations. Efforts inspired by this idea can be traced back to as early as the 1950s by Dirac \cite{Dirac:1958} and even Einstein himself. But we were only close to achieving that goal several decades later. 

In the next two sections, we will be going through the attempts at reformulating GR as a Yang-Mills-type gauge theory in its early years, especially the Palatini formulation and Ashtekar's self-dual formulation of GR.

\chapter{Palatini Formulation of General Relativity} \label{sec: pal}

The Palatini formulation of General Relativity laid the foundation for the Hamiltonian formalism of GR, which is a crucial step in obtaining a quantum theory of gravity via canonical quantization. In this formulation, the fundamental field of the theory is a frame field called the tetrad in place of the familiar metric tensor from the original GR. This change of variables would lead to a drastic change in the phase space structure in the subsequent Hamiltonian theory. 

In the metric case, its conjugate momentum turns out to be related to the extrinsic curvature \cite{ADM}. From the ADM formalism we will introduce later on, it can be shown that the phase space variables are the spatial metric intrinsic to the hypersurface and the extrinsic curvature describing how the 3D spatial hypersurface is embedded in the 4D manifold. In this sense, both phase space variables directly carry geometric interpretations about the space-time manifold. 

In the tetrad case, we may be inclined to consider the tetrad as the configuration variable since it is directly related to the space-time metric, but as it turned out, it is more appropriate to interpret a connection as the configuration variable, which is called the spin-connection. The tetrad field, on the other hand, is interpreted as the canonically conjugate momentum. In this sense, the connection plays a similar role to the vector potentials in Maxwell and Yang-Mills theories, and the tetrad plays the role of the ``electric field" counterpart from those theories.

So now we have a dual interpretation of the theory of general relativity in terms of the phase space variables. The canonically conjugate pair $(q_{ab}, K_{ab})$ gives the theory a complete geometric interpretation, which was referred to as geometrodynamics \cite{Romano}. On the other hand, the canonically conjugate pair $(\vec A, \vec E)$ describes the theory entirely as a gauge theory in the sense of Yang-Mills theories, which was referred to as connectiondynamics. The dual interpretations make the theory of general relativity both attractive and perplexing. They are like two sides of the same coin. Classically, they are essentially equivalent. The geometrodynamics may be more familiar and comfortable to use from a practical standpoint, but the connectiondynamics is just as powerful in producing the key results of GR. The difference only arises when we attempt to develop a quantum theory of gravity.

The Einstein-Hilbert action expressed in terms of the tetrad and spin connection would only be of the first-order derivative of these fields. For this reason, the Palatini formulation is also referred to as the first-order formulation, as opposed to the second-order formulation of the action in terms of the metric.

Since we are introducing some new players in the arena, we will begin this section by introducing some important aspects of conventional GR and exploring some new concepts mentioned above. Lastly, we should mention subsection (\ref{pal 3+1}) closely follows the work of \cite{Ashtekar, Romano}, the latter of which contains additional details.

\section{Aspects of General Relativity}
This subsection will contain a brief discussion on the key elements of general relativity that are essential to understanding the formulations of GR in the following sections. The briefness of this subsection is due in part to the large amount of quality literature that is already available on general relativity such as Carrol \cite{Carroll} and Wald \cite{Wald}. Additionally, although GR is elegantly formalized in the language of differential geometry, we are showing how it can be formalized less conventionally as a Yang-Mills-type gauge theory and do not wish to fully dive into its original treatment.

\subsection{The Metric}
The central element of general relativity (GR) is the metric tensor $g_{\mu\nu}$, and is defined by the dot product of basis vectors
\eq{
g_{\mu\nu} = \mathbf{e}_\mu \cdot \mathbf{e}_\nu.
}
The metric is essentially a map between a geometry and our coordinates where each of its diagonal components contains a ratio between proper distance and coordinate distance and its off-diagonal components contain information about the orthogonality between different coordinate dimensions (we will focus on understanding the diagonal elements since most metrics in GR have vanishing off-diagonal components).
This idea is most intuitively shown via a line element which takes the general form 
\eq{
ds^2 = g_{\mu\nu} dx^\mu dx^\nu ,
\label{line element}
}
where $ds$ is an infinitesimal distance and $dx^\mu$ is a coordinate differential. This equation may appear unfamiliar but anyone who has completed a geometry course has already used this equation in its simplest form known as the Pythagorean theorem. Recall the distance between two points on a two-dimensional plane $\Delta s$ (or the length of the hypotenuse of a right triangle) in Cartesian coordinates is given by
\eq{
 \Delta s^2 = \Delta x^2 + \Delta y^2
}
and if we take the infinitesimal limit 
\eq{
 ds^2 = dx^2 + dy^2,
}
we have a line element that describes an infinitesimal distance in 2 dimensions. From this line element we can use Eq. (\ref{line element}) to identify the components of the metric, namely $g_{xx} = g_{yy} = 1$ and $g_{xy} = g_{yx}=0$, or as represented as a matrix,
\eq{
g_{ab} = \begin{bmatrix}
1 & 0 \\
0 & 1 
\end{bmatrix}.
}
This is the metric of Euclidean space or flat geometry in 2 dimensions (note we are using Latin indices as a general index since Greek indices are reserved for 4-dimensional space-time coordinates). With constant components of value $1$ the metric is telling us that our $\mathbf{e}_x$ and $\mathbf{e}_y$ basis vectors have a magnitude of one at all locations on our coordinate grid.

Another trivial example is the metric of flat geometry in polar coordinates whose line element takes the following form
\eq{
ds^2 = dr^2 + r^2 d\theta^2.
}
If we pick $r$ to be a constant value i.e. $dr=0$ we get
\eq{
ds^2 &= r^2 d\theta^2
\nn \\
ds &= r d\theta
\nn \\
s &= r \int d\theta
}
which is the familiar formula for arc length. Generalizing this concept we can say that any distance $s$ along one dimension $x$ can be calculated with the following equation
\eq{
s = \int \sqrt{g_{xx}}\, dx,
}
where $x$ is a coordinate and is not summed over in $g_{xx}$ despite the repeated indices. As one can intuitively see the square root of the metric is directly proportional to the distance determined by a set of coordinates.

Looking back at the polar line element we can see the metric takes the form 
\eq{
g_{ab} = \begin{bmatrix}
1 & 0 \\
0 & r^2 
\end{bmatrix}.
}
As one can see, as opposed to the prior metric in Cartesian coordinates, this metric is a function of coordinate position i.e. $g_{ab}= g_{ab}(x^c)$. This is equivalent to saying that the basis vectors change with respect to coordinate positions. In this case the angular basis vectors $\mathbf{e}_{\theta}$ have a magnitude proportional to $r$. Another important fact is that despite the differences between the Cartesian and Polar metrics they both describe the exact same geometry, i.e., the flat space, and if we were to perform coordinate transformation we would find that the two metrics are equivalent.

\subsection{Space-time Metric}
The spatial metric is easy to understand, space-time metric on the other hand is far less intuitive as it is four-dimensional and exhibits a hyperbolic geometry. The line element for 4-dimensional space-time is given as follows
\eq{
ds^2 = -dt^2 + dx^2 + dy^2 + dz^2,
}
where the factor of $-1$ on the first term generates the hyperbolic properties of spacetime (note we have adopted the natural units so that the speed of light $c=1$). As one can see from the line element the metric takes the form 
\eq{
\eta_{\mu\nu} := \begin{bmatrix}
-1 & 0 & 0 & 0\\
0 & 1 & 0 & 0\\
0 & 0 & 1 & 0\\
0 & 0 & 0 & 1
\end{bmatrix} ,
}
and is known as the Minkowski metric. This metric is responsible for creating all the physics of special relativity and can alternatively be represented in spherical coordinates as 
\eq{
g_{\mu\nu} = \begin{bmatrix}
-1 & 0 & 0 & 0\\
0 & 1 & 0 & 0\\
0 & 0 & r^2 & 0\\
0 & 0 & 0 & r^2 \sin\theta^2
\end{bmatrix}.
}
Once again both of these metrics describe the same geometry, flat space-time (despite having hyperbolic properties it is conventionally called flat), but differ simply by a coordinate transformation. 

Now we will look at an example of a \textit{curved} geometry known as Schwarzschild geometry, which is characterized by the following line element
\eq{
ds^2 = -\Big(1-\frac{2GM}{r}\Big) dt^2 + \Big(1-\frac{2GM}{r}\Big)^{-1} dr^2 + r^2 d\theta^2 + r^2 \sin^2\theta d\phi^2,
}
where $G$ is Newton's gravitational constant and, $M$ is mass. This line element, or more precisely, the Schwarzschild metric describes the geometry of curved spacetime around a spherically symmetric black hole of mass $M$. A noteworthy feature is that at $r=2GM$ the $g_{rr}$ term diverges, which means that an asymptotic observer would have to see something travel an infinite distance before it reaches $r=2GM$. This turns out to be just a coordinate singularity and if we choose a different set of coordinates that describe the same geometry such as Eddington-Finkelstein coordinates, there is no longer a divergence, and an object can simply pass through $r=2GM$ without anything unusual happening. On the other hand, at $r=0$ there is a real space-time singularity as $g_{tt}$ diverges. This singularity cannot be removed through a change of coordinates and turns out to be an intrinsic feature of the Schwarzschild geometry.

\subsection{Connection}
The notion of a derivative becomes more complicated when dealing with curved geometries. At each point on a curved manifold, there is a locally defined tangent space where we can perform the usual vector analysis. However, when we are comparing vectors at different points, such as when we take a derivative, we have to account for the basis vectors that are changing with respect to location according to the metric. This requires us to introduce the covariant derivative which when acting on a vector is given by
\eq{
\Del_\mu V^\nu = \del_\mu V^\nu + \Gamma^\nu_{\mu\lambda}V^\lambda,
}
where $\Gamma^\nu_{\mu\lambda}$ are known as the Christoffel symbols and are the connection coefficients of the Levi-Civita connection. These connection coefficients allow us to connect neighboring points on our manifold and properly calculate derivatives. On a more conceptual level, while the metric handles the magnitude of the basis vectors, the connection can be thought of as the quantity that determines the orientation of our basis vectors. For example, if we imagine going from a Cartesian to a polar basis, the connection would be responsible for orienting the basis vectors in the familiar circular shape that a polar coordinate grid takes. 
The Christoffel symbols can be calculated by the following equation 
\eq{
\Gamma^\sigma_{\mu\nu} &= - \tfrac{1}{2} g^{\sigma \alpha} ( 
\del_\mu g_{\nu\alpha} + \del_\nu g_{\mu \alpha} - \del_\alpha g_{\mu\nu}),
}
 the derivation of which will be discussed in a later subsection as we analyze two different types of connections. 
 
 Connections allow us to perform an operation known as parallel transportation which is when we transport a vector so that its orientation remains as parallel to itself as possible. For example, if we use the connection of flat space from the line element 
\eq{ds^2=dr^2+r^2d\theta^2}
and parallel transport a vector around a circle in this flat space, the vector will always be pointing in the same direction. On the other hand, if we use the connection from the following line element 
\eq{ds^2=dr^2+d\theta^2}
and parallel transport a vector around a circle, the vector will always point tangent to the circle. An important point here is that in both cases, the transported vector has the exact same orientation before and after parallel transportation. This turns out to not always be the case, in fact, the degree to which the orientation of a vector changes helps us define the notion of \textit{curvature}.

\subsection{Curvature}
Curvature is present in a geometry when the process of parallel transportation fails to preserve the original orientation of a vector. A little experiment that exemplifies this can be done by putting your arm straight out in front of you with your thumb up. If you move your arm 90 degrees so that your arm is now straight out to the side, then move it another 90 degrees so your arm is straight up above you, and finally bring your arm straight down in front of you where you started, you will notice your thumb is now pointing to the side. 
The change of orientation of your thumb indicates that your hand has been transported along a curved surface namely, the surface of a sphere with a radius of one arm's length. 
This experiment can be brought to a larger scale where instead we are traversing the surface of the earth, which locally appears flat, and are transporting a pole that is kept tangent to the surface. The fact that the surface of the Earth is locally flat gets at the notion of \textit{intrinsic} curvature. Intrinsic curvature is curvature we cannot see, as beings bound to the surface of the earth we are stuck moving along the $\theta$ and $\phi$ dimensions while the curvature of the surface is embedded in the $r$ dimension.
This notion can be quantified by the Riemann curvature tensor $R^\lambda_{\mu\nu\rho}$, which is defined as follows
\eq{
R^\lambda_{\rho\mu\nu}V^\rho = ( \Del_\mu \Del_\nu - \Del_\nu \Del_\mu ) V^\lambda .
}
As one can see the magnitude of curvature depends on the degree to which the covariant derivatives fail to commute, and can be seen as representing an infinitesimal parallel transportation around a closed path. Since the Riemann curvature tensor has an upper index, we can contract it with the third index to get what is known as the Ricci curvature tensor, defined as
\eq{
R_{\mu\nu} = R^\lambda_{\mu\lambda\nu}.
}
Additionally, contracting the Ricci tensor with the inverse metric gives what is known as the Ricci curvature scalar,
\eq{
R = R_{\mu\nu} g^{\mu\nu}.
}
The following two quantities appear on the left-hand side of the Einstein field equation of GR,
\eq
{R_{\mu\nu} - \tfrac12 R g_{\mu\nu} = 8 \pi G T_{\mu\nu},
}
where $T_{\mu\nu}$ is the energy-momentum tensor and quantifies the matter-energy content of the respective system. The Einstein equation encapsulates the overarching idea of GR that matter and energy, represented by the right side of the equation, create space-time curvature, represented by the left side.

\section{Tetrad}

The theory of General Relativity was originally formulated in terms of the metric tensor from the perspective of ordinary coordinate basis. For example, in 4-dimensional space-time, these coordinates can be $(t, x, y, z)$, which form a set of orthonormal basis vectors $e_\mu=\del_\mu$. A vector can be then expressed in terms of its components and basis vectors
\eq{
V=V^\mu \del_\mu.
}
Similarly, a covector (or differential 1-form) can be expressed in terms of its components and a dual basis $e^\mu=dx^\mu$:
\eq{
W = W_\mu dx^\mu.
}
The metric tensor via a line element $ds^2 = g_{\mu\nu}dx^\mu dx^\nu$ is an example of differential 2-forms.

Alternatively, we can model space-time from the perspective of a {\bf local} traveler by introducing a set of four orthonormal basis vectors called tetrad $\hat e_I\equiv \{\hat e_0, \hat e_1, \hat e_2, \hat e_3\}$ at each point of the traveler's world-line
\eq{
\hat e_I=e_I^\mu \,\del_\mu,
}
where the uppercase Latin index labels the locally flat tetrad axes. We will refer to these local Lorentz indices simply as internal indices. On the other hand, the Greek letters are again reserved for the space-time coordinates, which we will refer to as external indices. 

Similarly, we have the cotetrad basis
\eq{
\hat e^I=e_\mu^I dx^\mu.
}
Contrary to the general curvilinear coordinate basis mentioned above, the tetrad basis is always locally flat with a Lorentzian signature, namely, it will adopt a Minkowski metric when defining a line element locally
\eq{
ds^2 = \eta_{IJ}\hat e^I \hat e^J.
}

Geometrically, the time-like basis vector $\hat e_0$ is tangent to the world-lines of the traveler at each point by construction, whereas the other 3-dimensional space-like orthonormal basis vectors can be thought of as a local laboratory frame the travelers carry with them at each point, as shown in Fig.\ref{fig:frames}.

\begin{figure}
    \centering
    \includegraphics[scale=0.05]{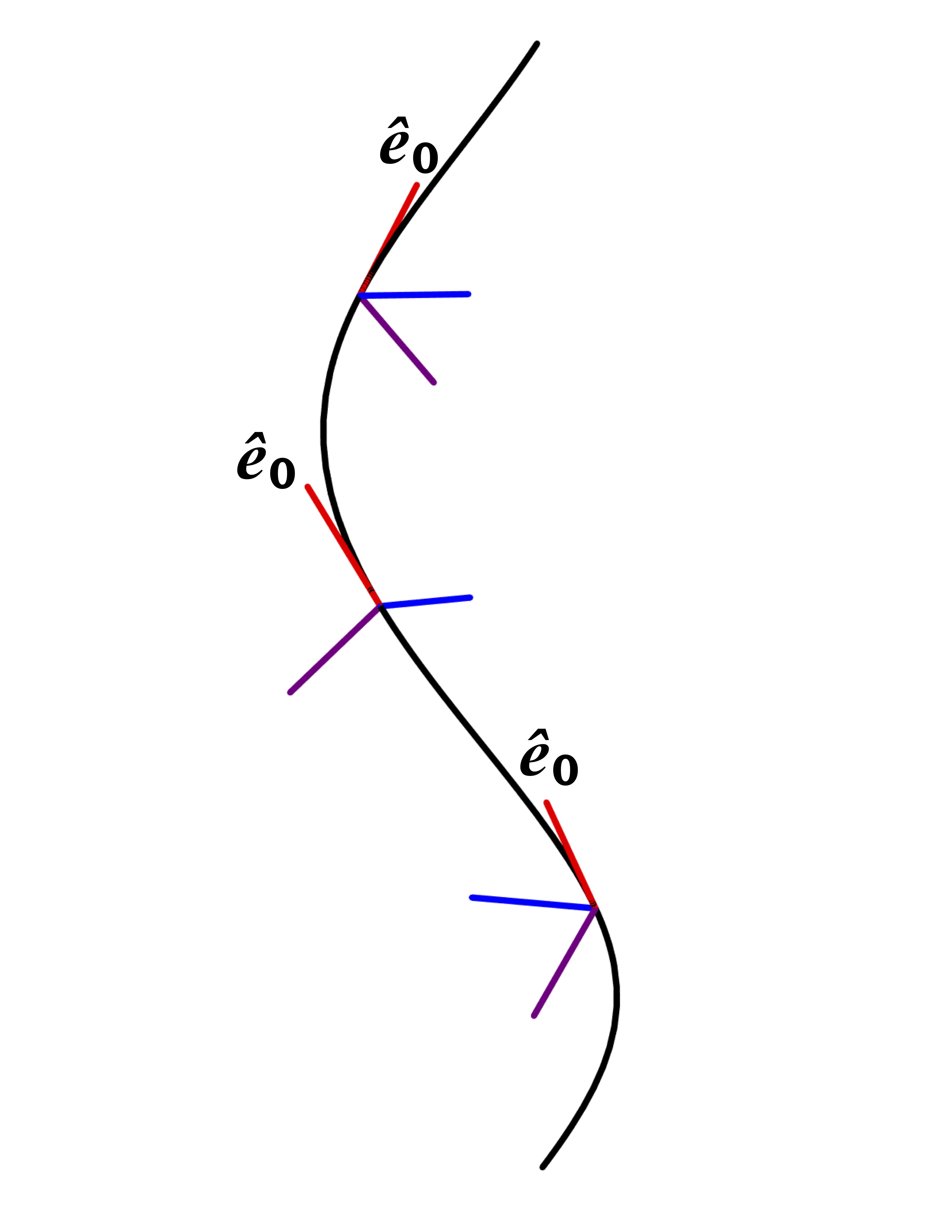}
    \caption{Local orthonormal frames or tetrad frames along a world line. Note the time like $\hat{e}_0$  basis vectors  are always tangent to the world line and the spatial basis vectors are always orthogonal the $\hat{e}_0$  basis vector as well as to each other.}
    \label{fig:frames}
\end{figure}

Here we want to emphasize that the tetrad frame is only valid in the local region around each point along the geodesics of the traveler. It is not a global frame valid everywhere like the frame from the coordinate basis. This is a key difference between the tetrad frame and the coordinate frame! Further elaboration on the tetrad can be found in \cite{Carroll,Baez}.

By setting the space-time intervals in the coordinate basis and orthonormal tetrad basis equal to each other, we get
\eq{
g_{\mu\nu}dx^\mu dx^\nu &= \eta_{IJ}e^I e^J\\
&=\eta_{IJ} (e^I_\mu \,dx^\mu) (e^J_\nu \,dx^\nu)\\
&= (\eta_{IJ}e^I_\mu e^J_\nu) dx^\mu dx^\nu.
}
This allows us to transform between the coordinate basis and a local orthonormal basis 
\eq{
g_{\mu\nu} = e_\mu^I e_\nu^J \eta_{IJ},
}
and vice versa
\eq{
g_{\mu\nu} e^\mu_I e^\nu_J = \eta_{IJ}.
}
The components of a tetrad $e_I^\mu$ and cotetrad $e^I_\mu$ can contract with each other by a pair of internal indices or external indices to produce a Kronecker delta between the internal or external indices
\eq{
e_I^\mu e^I_\nu=\delta_\nu^\mu, \quad e_I^\mu e^J_\mu=\delta_I^J.
}

Tetrads also allow us to write the coordinate vectors in terms of the orthonormal vectors, and orthonormal one-forms in terms of coordinate one forms.
\eq{
V^\mu &=  e^\mu_I V^I \\
W_I &= e^\mu_I W_\mu
}
and the inverse using cotetrads,
\eq{
W_\mu &=  e_\mu^I W_I \\
V^I &= e_\mu^I V^\mu.
}
In a practical sense, we can treat the tetrad field $e_I^\mu$ and cotetrad field $e^I_\mu$ as a sort of ``mixed Kronecker delta'' between internal and external indices. Again, these operations are only valid locally. 

Let's check out an example to get a better sense of working with the tetrad. We will choose the familiar Minkowski line element. In terms of spherical space-time coordinates $x^\mu=\{ct, r, \theta, \phi\}$, we have
\eq{
ds^2 = -c^2 dt^2 + dr^2 + r^2 d\theta^2 + r^2 \sin^2{\theta}d\phi^2.
}
In terms of the cotetrads, we also have
\eq{
ds^2 = \eta_{IJ}\hat e^I \hat e^J = -(\hat e^0)^2 + (\hat e^1)^2 + (\hat e^2)^2 + (\hat e^3)^2.
}
We can then easily identify that
\eq{
\hat e^0 = c dt,\quad \hat e^1 = dr,\quad \hat e^2 = r d\theta,\quad \hat e^3 = r \sin{\theta}d\phi,
}
which leads to the following nonzero cotetrad components
\eq{
e^0_t = 1,\quad e^1_r = 1,\quad e^2_\theta = r,\quad e^3_\phi = r \sin{\theta}.
}
We can express the cotetrad components as a $4\times 4$ matrix
\eq{
e_\mu^I=
\begin{bmatrix}
1 & 0 & 0 & 0\\
0 & 1 & 0 & 0\\
0 & 0 & r & 0\\
0 & 0 & 0 & r \sin{\theta}
\end{bmatrix},
}
which allows us to obtain the tetrad components upon inversion
\eq{
e^\mu_I=
\begin{bmatrix}
1 & 0 & 0 & 0\\
0 & 1 & 0 & 0\\
0 & 0 & \frac 1 r & 0\\
0 & 0 & 0 & \frac {1}{r \sin{\theta}}
\end{bmatrix}.
}
If we now compare the (co)tetrad components to the Minkowski metric and its inverse
\eq{
g_{\mu\nu}=
\begin{bmatrix}
-1 & 0 & 0 & 0\\
0 & 1 & 0 & 0\\
0 & 0 & r^2 & 0\\
0 & 0 & 0 & r^2 \sin^2{\theta}
\end{bmatrix}
}
and
\eq{
g^{\mu\nu}=
\begin{bmatrix}
-1 & 0 & 0 & 0\\
0 & 1 & 0 & 0\\
0 & 0 & \frac {1} {r^2} & 0\\
0 & 0 & 0 & \frac {1}{r^2 \sin^2{\theta}}
\end{bmatrix},
}
we can see that the tetrad components are essentially the ``square roots'' of the metric tensor components in a loose sense.

Note that we have essentially created a local inertial frame at each point of a manifold, we thus have the freedom to perform Lorentz transformations at every point in space. These locally based Lorentz transformations are therefore called Local Lorentz transformations (LLT), as compared to the ordinary General Coordinate transformations in the coordinate basis (GCT).

\section{Spin Connection}

Another key ingredient in the tetrad formulation of GR is a Lorentz connection called the spin connection. To explain the motivation behind the spin connection, let's recall the metric formulation of GR. In order to preserve the parallel transportation of a vector, we need to introduce the notion of an affine connection, which subsequently leads to introducing a covariant derivative as the partial derivative plus additional terms involving the connection coefficients for each external index. Recall the covariant derivative on a vector component is 
\eq{
D_\mu V^\nu = \del_\mu V^\nu + \Gamma^\nu_{\mu\lambda}V^\lambda
}
The additional term would counteract the inhomogeneous term from transforming $\del_\mu V^\nu$ such that $D_\mu V^\nu$ as a whole transforms as a tensor covariantly.

Now we come back to the tetrad formulation. Since we have introduced a locally flat internal space at every point of the manifold, quantities in these spaces will generally carry a mixture of internal and external indices. In this sense, we have tensors with mixed indices $T_{\mu\nu...IJ...}^{\rho\sigma...KL...}$ that need to transform properly under general coordinate transformations and local Lorentz transformations. For this reason, we introduce a dedicated internal connection called the spin connection $\omega_\mu^{IJ}$, which has one external index and two internal indices. The spin connection together with the metric connection allows us to handle differentiations of tensors with both internal and external indices covariantly with a generalized derivative operator $D_\mu$. For example, when acting on a rank-(1,1) mixed tensor $X_{\nu I}$, we have
\eq{
D_\mu X_{\nu I} = \del_\mu X_{\nu I} + \Gamma^\alpha_{\mu\nu} X_{\alpha I} + \omega^J_{\mu I} X_{\nu J}.
}
As one can see, this generalized derivative contains a partial derivative and two ``correction" terms. The metric connection accounts for corrections needed externally and the spin connection accounts for corrections needed internally such that the whole tensor transforms covariantly. 

In GR we demand the covariant derivative on the metric tensor to vanish. This is referred to as the metric compatibility condition
\eq{
D_\mu g_{\alpha\beta
}=0.
}
This has an important consequence on the form of the metric connection, specifically, the metric connection components must be symmetric with its two lower indices, i.e.,
\eq{
\Gamma^\alpha_{\beta \gamma} = \Gamma^\alpha_{\gamma \beta}.
}
Because we can generally express the metric connection as a sum of its symmetric and antisymmetric parts
\eq{
\Gamma^\alpha_{\beta \gamma} = (\Gamma^\alpha_{\beta \gamma})^S + (\Gamma^\alpha_{\beta \gamma})^A,
}
where the latter is defined as the torsion tensor. The condition that the metric connection is symmetric automatically means that the space-time is torsion-free.

Similarly, we can demand the metric compatibility be satisfied for the internal metric $\eta_{IJ}$ of the orthonormal basis, namely
\eq{
D_\mu \eta_{IJ} &= \del_\mu \eta_{IJ} + \omega_{\mu I}^{\xu\xu K} \eta_{KJ} + \omega_{\mu J}^{\xu\xu K} \eta_{IK}\nn\\
&=\omega_{\mu IJ} + \omega_{\mu JI}=0,
}
or equivalently 
\eq{
\omega_{\mu IJ} =- \omega_{\mu JI}.
}
This suggests that the metric compatibility condition essentially demands that the spin connection be antisymmetric with its two internal indices.

To sum up, the metric and tetrad compatibility conditions make it so that the metric connection is symmetric in its lower two indices and the spin connection is anti-symmetric in its two internal indices
\eq{
\Gamma^\alpha_{\beta \gamma} = \Gamma^\alpha_{\gamma \beta}\,, \;\;\;\;\;\;\;\;
\omega_{\alpha IJ} = -\omega_{\alpha JI}.
}
The metric connection $\Gamma^\alpha_{\beta \gamma}$ under the torsion-free condition is a unique connection also known as the Christoffel connection or Levi-Civita connection from conventional GR, whose coefficients are called the Christoffel symbol. 

Under these conditions the generalized derivative $D_\mu$ is no longer general, so $\Del_\mu$ will be used to represent the torsion-free generalized derivative so that
\eq{
\Del_\mu X_{\nu I} = \del_\mu X_{\nu_I} + \Gamma^\alpha_{\mu\nu} X_{\alpha I} + \omega^J_{\mu I} X_{\nu J} \; ,
}
where
\eq{
\Gamma^\sigma_{\mu\nu} &= - \tfrac{1}{2} g^{\sigma \alpha} ( 
\del_\mu g_{\nu\alpha} + \del_\nu g_{\mu \alpha} - \del_\alpha g_{\mu\nu}),
\\
\omega^I_{\mu J} &= -e^{\nu}_J( 
\del_\mu e_{\nu I} + \Gamma^\alpha_{\mu \nu} e_{\alpha I}).\label{compact1}
}
The former equation again results from the metric compatibility condition $\Del_\mu g_{\mu\nu}=0$, whereas the latter is equivalent to a vanishing covariant derivative of the tetrad
\eq{
\Del_\mu e_\nu^I= \del_\mu e_\nu^I + \Gamma_{\mu\nu}^\alpha e_\alpha^I + \omega_{\mu \xu J}^{\xu I}e_\nu^J=0,
}
which is referred to as the tetrad compatibility condition.

With the generalized derivative operator previously introduced, we can define the external and internal curvature tensors
\eq{
2 D_{[\mu} D_{\nu]} X_\sigma &:= {F_{\mu\nu\sigma}}^\alpha X_\alpha,
\\
2 D_{[\mu} D_{\nu]} X_I &:= {F_{\mu\nu I}}^J X_J.
}
The corresponding torsion-free curvature tensors are thus defined by
\eq{
2 \Del_{[\mu} \Del_{\nu]} X_\sigma &:= {R_{\mu\nu\sigma}}^\alpha X_\alpha,
\\
2 \Del_{[\mu} \Del_{\nu]} X_I &:= {R_{\mu\nu I}}^J X_J.
}
The curvature tensor $R_{\mu\nu\sigma}^\alpha$ with four Greek indices is exactly our familiar space-time Riemann curvature tensor. On the other hand, $R_{\mu\nu I}^J$ with a pair of Greek and Latin indices is the curvature tensor expanded in the orthonormal basis and is only valid locally. 

It can be shown that the two curvature tensors are related to each other by the following relation
\eq{
R_{\mu\nu I}^{\xu\xu\xu J} = R_{\mu\nu\alpha}^{\xu\xu\xu\beta} e_I^\alpha e_\beta^J.
}

\section{The Palatini Action}
With our newly introduced tools, that being the tetrad and spin connection, we can formulate GR without directly using the metric tensor. To do this we claim the following \textit{Palatini action} \cite{Palatini} is equivalent to the conventional Einstein-Hilbert action of GR:
\eq{
S_p &=  \int_M \sqrt{-g}\, e^\rho_K e^\sigma_L F_{\rho\sigma}^{\xu KL} \,d^4x 
\\ &=
\tfrac12 \int_M \sqrt{-g}\, (e^\rho_K e^\sigma_L - e^\sigma_K e^\rho_L ) F_{\rho\sigma}^{\xu KL} \,d^4x 
\nn \\ &=
\tfrac12 \int_M \sqrt{-g}\, 
(\delta_\alpha^\rho \delta_\beta^\sigma - \delta_\alpha^\sigma \delta_\beta^\rho  ) 
e^\alpha_K e^\beta_L F_{\rho\sigma}^{\xu KL} \,d^4x .
}
Now using the identity 
\eq{
(\delta_\alpha^\rho \delta_\beta^\sigma - \delta_\alpha^\sigma \delta_\beta^\rho  ) = - \tfrac12 \tilde\epsilon^{\mu\nu\rho\sigma} \tilde\epsilon_{\mu\nu\alpha\beta}
}
where the extra minus sign is due to the Minkowski signature of the metric and the density weights of the contravariant and covariant Levi-Civita symbols cancel. Applying this substitution to the action we have
\eq{
S_p&=
-\tfrac14 \int_M \sqrt{-g}\, 
( \tilde\epsilon^{\mu\nu\rho\sigma} \tilde\epsilon_{\mu\nu\alpha\beta} )
e^\alpha_K e^\beta_L F_{\rho\sigma}^{\xu KL} \,d^4x 
\nn \\ &=
-\tfrac14 \int_M \sqrt{-g}\, 
( \tilde\epsilon^{\mu\nu\rho\sigma} \tilde\epsilon_{\mu\nu KL} )
 F_{\rho\sigma}^{\xu KL} \,d^4x 
 \nn \\ &=
-\tfrac14 \int_M \sqrt{-g}\,  (\tilde\epsilon^{\mu\nu\rho\sigma} \tilde\epsilon_{IJKL} e^I_\mu e^J_\nu)
 F_{\rho\sigma}^{\xu KL} \,d^4x .
}
Now we absorb the $\sqrt{-g}$ into the covariant Levi-Civita symbol to turn it into the Levi-Civita tensor
\eq{
\epsilon_{IJKL} =\sqrt{-g}\, \tilde\epsilon_{IJKL},
}
which gives us
\eq{
S_p&=
-\tfrac14 \int_M   \tilde\epsilon^{\mu\nu\rho\sigma} \epsilon_{IJKL} e^K_\mu e^L_\nu
 F_{\rho\sigma}^{\xu KL}  \,d^4x .
}
With this expanded form of the Palatini action an interesting result will follow when we identify the internal curvature tensor $F_{\rho\sigma}^{\xu KL}$ of the internal connection $\omega_a^{\xu IJ}$ as the (torsion free) Riemann curvature tensor $R_{\rho\sigma}^{\xu KL}$ with two of its spacetime indices converted into internal indices i.e.,
\eq{
F_{cdK}^{\xu\xu\xu L}:=\del_c w_{dK}^{\xu\xu L} - \del_d w_{cK}^{\xu\xu L}+[w_c,w_d]_K^{\xu L} \equiv R_{cdK}^{\xu\xu\xu L}.
}
It then follows that
\eq{
S_p&=
-\tfrac14 \int_M   \tilde\epsilon^{\mu\nu\rho\sigma} \epsilon_{IJKL} e^I_\mu e^J_\nu
R_{\rho\sigma}^{\xu KL}  \,d^4x 
\nn \\ &=
-\tfrac14 \int_M   \tilde\epsilon^{\mu\nu\rho\sigma} \epsilon_{IJKL} e^I_\mu e^J_\nu e_\alpha^K e_\beta^L
R_{\rho\sigma}^{\xu \alpha\beta} \,d^4x 
\nn \\ &=
-\tfrac14 \int_M   \tilde\epsilon^{\mu\nu\rho\sigma} \epsilon_{\mu\nu\alpha\beta}
R_{\rho\sigma}^{\xu \alpha\beta} \,d^4x 
\nn \\ &=
-\tfrac14 \int_M \sqrt{-g}\,  \tilde\epsilon^{\mu\nu\rho\sigma} \tilde\epsilon_{\mu\nu\alpha\beta}
R_{\rho\sigma}^{\xu \alpha\beta} \,d^4x 
\nn \\ &=
-\tfrac14 \int_M \sqrt{-g}\,  
(-2)(\delta_\alpha^\rho \delta_\beta^\sigma - \delta_\alpha^\sigma \delta_\beta^\rho  )
R_{\rho\sigma}^{\xu \alpha\beta} \,d^4x 
\nn\\&=
\tfrac12 \int_M \sqrt{-g}\,  
(R_{\alpha\beta}^{\xu \alpha\beta} - R_{\beta\alpha}^{\xu \alpha\beta}) \,d^4x ,
}
 and since the Riemann curvature tensor is anti-symmetric under exchange of its first two indices i.e. $R_{\mu\nu}^{\xu \rho\sigma} = -R_{\nu\mu}^{\xu \rho\sigma}$ we can write
\eq{
S_p&=
 \int_M \sqrt{-g}\,  
R_{\alpha\beta}^{\xu \xu \alpha\beta} \,d^4x 
\nn \\ &=
 \int_M \sqrt{-g}\,R \,d^4x ,
}
where $R$ is the Ricci scalar. One familiar with the Lagrangian formulation of general relativity will recognize this as the Einstein-Hilbert action! This result shows that the Einstein-Hilbert action of general relativity and the newly introduced Palatini action are equivalent under the torsion-free condition. 
\eq{
S_P = S_{EH}
}
It should also be noted we introduced a factor of $\tfrac{1}{2}$ into the Palatini action 
\eq{ 
S_P = \tfrac{1}{2} \int_M \sqrt{-g}\, e^\mu_I e^\nu_J F_{\mu\nu}^{\xu\xu IJ} \,d^4x .
}
Now we have an action in terms of tetrads and spin connection (contained in the curvature tensor) we are ready to perform its $3+1$ decomposition and construct the Hamiltonian. To better understand the $3+1$ decomposition we must first introduce the original $3+1$ decomposition of GR known as the ADM formalism.

\section{The ADM Formalism}
The ADM formalism is a Hamiltonian formulation of general relativity, with a pair of canonically conjugate variables namely, the spatial metric and its conjugate momentum. 
Geometrically, instead of a pre-selected space-time background with a fixed global coordinate axis (i.e., background-dependent geometry), the space-time manifold is foliated into three-dimensional spatial hypersurfaces evolving through time, and dynamical variables are only defined locally respecting the geometry of the hypersurface \cite{ADM, new ADM}.
This $3+1$ decomposition allows for a dynamical view of general relativity and can be conceptualized by Fig. \ref{fig:adm}.

%\textbf{suppressing figure for now because it increases compiling time}
\begin{figure}[h]
    \centering
    \includegraphics[scale=.15]{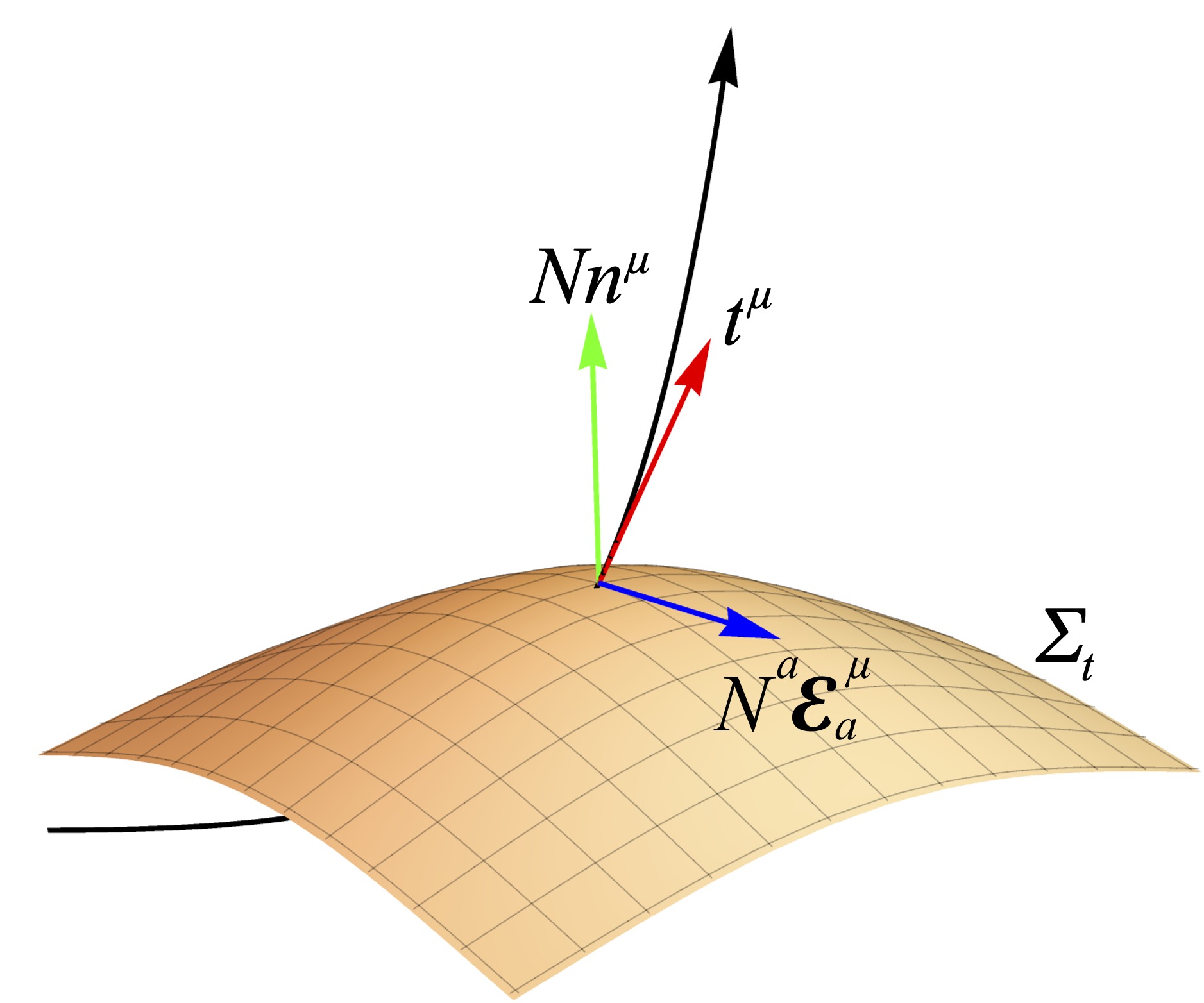}
    \caption{Visual representation of the ADM formalism}
    \label{fig:adm}
\end{figure}

Time evolution is introduced in terms of a dynamical time-like vector field $t^\mu$, which is always tangent to the the particle world-line. It can be decomposed into its normal and tangent components relative to the  spatial hypersurface $\Sigma_t$ as follows
\eq{
t^\mu = Nn^\mu + N^\mu = Nn^\mu + N^a \mathcal{E}^\mu_{a} \label{timelike},
}
where $n^\mu$ is a unit vector field normal  to $\Sigma_t$ and $\mathcal{E}^\mu_a$ is a unit vector field tangent to $\Sigma_t$ which obey the following orthogonality conditions
\eq{
n^\mu n_\mu = -1 \quad\quad\quad n^\mu \mathcal{E}_{\mu a}=0 \,.
}
 $N$ is called the lapse function which measures the rate at which the proper time $\tau$ evolves with respect to the coordinate time $t$ in the direction normal to the spatial hypersurface. On the other hand, $N^a$ is called the shift vector which measures the rate at which local spatial coordinates $y^a$ evolve in the direction tangent to the spatial hypersurface. The timeline vector and tangent vector fields are defined as
\eq{
t^\mu := \left( \frac{\del x^\mu}{\del t} \right)_{y^a}
\quad \quad \quad
\mathcal{E}^\mu_a := \left( \frac{\del x^\mu}{\del y^a} \right)_{t}
}
where $x^\mu$ are 4-dimensional space-time coordinates and $y^a$ are 3-dimensional spatial coordinates on $\Sigma$. Using the previous results we can rewrite the coordinate differential
\eq{
dx^\mu &= 
\frac{\del x^\mu}{\del y^0}dy^0 + \frac{\del x^\mu}{\del y^a}dy^a
\nn\\&= 
\frac{\del x^\mu}{\del t}dt + \frac{\del x^\mu}{\del y^a}dy^a
\nn\\&= 
t^\mu dt + \mathcal{E}^\mu_a dy^a
\nn \\&= 
(Nn^\mu + N^a \mathcal{E}^\mu_{a} ) dt + \mathcal{E}^\mu_a dy^a
\nn \\&= 
(N dt) n^\mu + ( N^a dt + dy^a )\mathcal{E}^\mu_a \,.
}
As one can see we have decomposed the coordinate differential in terms of the basis normal and tangent vector fields. Similarly, we have 
\eq{
dx_\mu = (N dt) n_\mu + ( N^a dt + dy^a )\mathcal{E}_{\mu a} \,
}
where $\mathcal{E}_{\mu a} = \frac{\del x_\mu}{\del y^a}$. This lets us rewrite the line element as follows.
\eq{
ds^2 &= dx^\mu dx_\mu 
\nn\\&=
\Big( (N dt) n^\mu + ( N^a dt + dy^a )\mathcal{E}^\mu_a \Big) \Big((N dt) n_\mu + ( N^b dt + dy^b )\mathcal{E}_{\mu b} \Big)
\nn \\ &=
N^2 dt^2 (n^\mu n_\mu) + N dt ( N^c dt + dy^c )(n^\mu \mathcal{E}_{\mu c} + n_\mu \mathcal{E}^{\mu}_c) + ( N^a dt + dy^a )( N^b dt + dy^b )\mathcal{E}^\mu_a \mathcal{E}_{\mu b}
\nn \\ &= 
-N^2 dt^2 + ( N^a dt + dy^a )( N^b dt + dy^b )\mathcal{E}^\mu_a \mathcal{E}_{\mu b}
}
where in the third to fourth line, the first term gains a minus sign and the second term vanishes under the orthogonality conditions. Additionally, since $\mathcal{E}^\mu_a \mathcal{E}_{\mu b}$ is just the spatial part of the metric $q_{ab}$ as proven below
\eq{
\mathcal{E}^\mu_a \mathcal{E}_{\mu b} &= \frac{\del x^\mu}{\del y^a} \frac{\del x_\mu}{\del y^a}
=
g_{\mu\nu} \frac{\del x^\mu}{\del y^a} \frac{\del x^\nu}{\del y^a}
=
g_{\mu\nu}\delta^\mu_a \delta^\nu_b 
= g_{ab} := q_{ab}
}
the line element becomes
\eq{
ds^2 &= -N^2 dt^2 + q_{ab}( N^a dt + dy^a )( N^b dt + dy^b )
\nn \\ &=
-N^2 dt^2 + q_{ab}dy^a dy^b + q_{ab}N^a N^b dt^2 + 2 q_{ab}N^b dt dy^a
\nn \\ &=
-N^2 dt^2 + q_{ab}dy^a dy^b + N^a N_a dt^2 + 2 N_a dt dy^a
\nn \\ ds^2 &=
(-N^2+N^a N_a) dt^2 + q_{ab}dy^a dy^b + 2 N_a dt dy^a.
}
From the line element, it can be seen that
\eq{
g_{tt} &= -N^2+N^a N_a
\nn \\ 
g_{ta} &= g_{at} = N_a
\nn \\ 
g_{ab} &= q_{ab}.
}
As one can see we have completely decomposed the metric into the lapse function $N$, the shift vector $N_a$, and the spatial part of the metric $q_{ab}$, 
\eq{
g_{\mu\nu}=\begin{bmatrix}
-N^2+N^aN_a & N_1 & N_2 & N_3 \\
N^1 & q_{11} & q_{12} & q_{13} \\
N^2 & q_{21} & q_{22} & q_{23} \\
N^3 & q_{31} & q_{32} & q_{33} \\
\end{bmatrix}.
}
In principle, we can construct a Hamiltonian in terms of these ADM variables and find equations of motion for the spatial metric and its canonical momentum. However, to achieve this we will have to introduce new concepts and mathematical tools which will be done in the following subsections.

Upon introducing the variables in this formalism we can write the metric as the spatial part minus the normal part,
\eq{
g_{\mu\nu} = q_{\mu\nu} - n_\mu n_\nu.
}
Notice our index notation is beginning to fail us as we know the indices on the spatial metric should only run from $1$ to $3$ and should be written as $q_{ab}$. However, in order to keep the indices consistent on our equations whenever there is a purely spatial quantity, such as $q_{ab}$, and it is denoted with Greek indices ($\mu,\nu,$ etc.) we will assume when the indices equal zero the quantity vanishes i.e. $q_{00}=q_{0a}=0.$ This will be the case for all quantities we define as purely spatial such as the shift vector i.e. 
\eq{
N^\mu = \begin{bmatrix}
    0 \\ N^1 \\ N^2 \\ N^3
\end{bmatrix}.
}
Furthermore, we will always explicitly define a purely spatial quantity with its own unique variable. %, and if it has non-zero $\mu=\nu=0$ components in a certain case, it will be indicated by a raised number to the left of the variable. 

Solving the first equation for the spatial metric and raising an index with the metric gives us the \textit{projection operator} which is defined as follows,
\eq{
q^\mu_\nu = \delta^\mu_\nu + n^\mu n_\nu \label{pal proj}.
}
This operator projects a vector field onto the spatial hypersurface, in other words, it takes the inner product between the vector field it acts on, and the tangent vector field. For example, if the vector field is normal to the hypersurface, namely $V^\nu =c n^\nu$, then
\eq{
q^\mu_\nu V^\nu = V^\nu \delta^\mu_\nu + V^\nu n^\mu n_\nu 
= V^\mu - c n^\mu = 0,}
and if the vector is tangent to $\Sigma$, namely $V^\mu n_\mu = 0$, then
\eq{
q^\mu_\nu V^\nu =
V^\nu\delta^\mu_\nu + V^\nu n^\mu n_\nu
=  V^\nu + 0 =  V^\nu\,.
}
This operator will be heavily used throughout the text and will help finish defining the ADM formalism but before we can do that we must introduce the notion of a \textit{diffeomorphism}.

\section{Diffeomorphism and Lie Derivative}

We will dedicate this subsection to a brief introduction to the concepts of diffeomorphism and Lie derivative. The content here follows the Appendix B of reference \cite{Carroll} very closely. Interested readers should go there for more thorough discussions.

In General Relativity, we are used to describing the geometry of a manifold via a specific set of coordinates. For example, if we want to describe a point on a sphere, it is very convenient to choose the spherical coordinates $(r, \theta, \phi)$ that respect the symmetry of the sphere. These coordinates give us a concrete, direct sense of relating different points in the manifold via translations, rotation, and so forth. The choice of spherical coordinates is convenient, but not unique. There is no problem with representing points on a sphere using Cartesian coordinates $(x,y,z)$. This would correspond to a different coordinate map which can be related to the spherical coordinates through the following coordinate transformation:
\eq{
x&=r\sin\theta\cos\phi\nn,\\
y&=r\sin\theta\sin\phi\nn,\\
z&=r\cos\theta\nn.
}
In this example, we keep the manifold fixed and introduce a different coordinate map to describe the same points. For this reason, it is sometimes referred to as ``passive coordinate transformations". 

The passive coordinate transformations are convenient for a manifold with certain global symmetry, like a sphere, a circle, and so on. But when we consider a general manifold without such global symmetry, the notion of specific coordinate maps becomes somewhat less important, and it is more intuitive to think about the points in a manifold directly. If we move a point in the manifold and evaluate the coordinates of that point at a different location, it has the same effect as changing its coordinates. The only difference is that this operation is seen from a more active perspective through the ``eyes'' of the points, as opposed to an external observer. This ``active coordinate transformation" is called diffeomorphism. Formally, it is defined as a map from a manifold to itself, which means moving the points around in a manifold to a new configuration.

Let's make our discussion more concrete by considering a map $\phi$ between two manifolds $M\rightarrow N$, coordinatized by $x^\mu$ and $y^\alpha$ respectively. Notice that the map deals only with the points in the manifolds whereas the coordinatization of a manifold does not get carried over to the other manifold. For example, a vector $V^\mu(p)$ evaluated at a point $p$ in the original manifold $M$ can only be understood in terms of $M's$ own ``language''. If we want to talk about this vector in the target manifold $N$ in any mathematically rigorous manner, some proper ``translation'' is required so that the vector can be understood using $N's$ ``language'', this ``translation'' is formally called a pushforward. Similarly, a function $f(\phi(p))$ evaluated at point $\phi(p)$ in the target manifold $N$ can only be understood in terms of $N's$ own ``language''. If we want to talk about this function in $M$, some proper ``translation'' is required again, and that specific ``translation'' is called a pullback. More generally, the pushforwards and pullbacks are necessary when we want to relate tensor quantities evaluated in different manifolds related to each other by a map. 
Specifically, a pushforward of a map $\phi$, denoted by $\phi_*$, sends a contravariant tensor $T^{\mu_1 \mu_2 \dotsc \mu_N}$ from an existing manifold $M$ to a target manifold $N$. The simplest example of this operation is with a vector field $V^\mu$, for which we have 
\eq{
\phi_* \,V^\mu = \frac {\del y^\mu}{\del x^\alpha} V^\alpha. \label{push}
}
On the other hand, a pullback of a map $\phi$, denoted by $\phi^*$, sends a covariant tensor $T_{\mu_1 \mu_2 \dotsc \mu_N}$, from the tangent space of an existing manifold to the tangent space of a target manifold. An example of this operation is with a differential 1-form $\omega_\mu$, for which we have
\eq{
\phi^*\, \omega_\alpha = \frac {\del y^\mu}{\del x^\alpha} \omega^\mu. \label{pull}
}

It is clear that when $M$ and $N$ happen to be the same manifold, equations (\ref{push}) and (\ref{pull}) are exactly what we would get from ordinary coordinate transformations between two sets of coordinates, $y^\alpha$ and $x^\mu$, describing the same manifold. This points us back to the notion that diffeomorphisms are active coordinate transformations. To see how this machinery is constructed explicitly would require a fairly technical discussion (again, refer to \cite{Carroll} for more details). Here we simply include the above results because they will be relevant to our discussion in what follows next. 

One thing to note is that these maps are generally defined between manifolds of different dimensionalities so they are not necessarily invertible. However, in the case of diffeomorphisms, which are maps from a manifold to itself, they are automatically invertible. We can thus perform pushforwards and pullbacks on any arbitrarily ranked tensors. This would allow us to compare tensors at different points in a manifold. 

Let's consider a vector field $V^\mu(x)$ on a manifold $M$. A vector field can be thought of as the collection of all the tangent vectors to a curve $x^\mu(\lambda)$ in a manifold. Such a curve is called an integral curve, which satisfies
\eq{
\frac {dx^\mu}{d\lambda} = V^\mu.
} 
With this, we can introduce a family of diffeomorphism parameterized by $\lambda$ under which a point can be mapped along the integral curve smoothly and continuously. In this construction, the vector field dictates in which direction the diffeomorphism is to be performed, so it is referred to as the generator of the diffeomorphism. 

The one-parameter family diffeomorphism allows us to introduce a notion of derivative in an active view independent of a specific choice of coordinates, which is called the Lie derivative. In short, the Lie derivative evaluates the change in a given tensor quantity as it moves along the flow (i.e., the integral curve) of some other vector field. For simplicity, let's consider a (2,2) tensor $T_{\xu\xu\rho\sigma}^{\mu\nu}(p)$ evaluated at a point $p$ in a manifold $M$. We can map this point to a different point in the same manifold by a diffeomorphism $\phi(p)$ along an integral curve generated by a vector field $V^\mu$. Naturally, we would like to know how much the tensor has changed compared to the original one. But remember that we can only compare tensors directly at the same point, otherwise, it’s pointless. Therefore, to evaluate the change, one would have to ``pull'' the tensor evaluated at the new point $\phi(p)$ back to point $p$ and compare it with the original tensor evaluated at $p$, namely
\eq{
\Delta_\lambda T_{\xu\xu\rho\sigma}^{\mu\nu}(p) = \phi^*_\lambda[T_{\xu\xu\rho\sigma}^{\mu\nu}(\phi_\lambda(p))]-T_{\xu\xu\rho\sigma}^{\mu\nu}(p).
}
We can now define the Lie derivative of tensor $T_{\xu\xu\rho\sigma}^{\mu\nu}$ along a vector field $V^\mu$ by taking the limit of this change with respect to an infinitesimal change of the parameter $\lambda$, i.e.,
\eq{
\mathcal{L}_V T_{\xu\xu\rho\sigma}^{\mu\nu}= \lim\limits_{\lambda \to 0}\Big(\frac {\Delta_\lambda T_{\xu\xu\rho\sigma}^{\mu\nu}}{\lambda} \Big).
}

We can apply this definition and show that the Lie derivative on a scalar function $f$ would simply reduce it to the action of the ordinary directional derivative
\eq{
\mathcal{L}_V f &=\lim\limits_{\lambda \to 0} \frac {f(x^\mu + V^\mu \lambda) - f(x^\mu)}{\lambda}\\
&=\lim\limits_{\lambda \to 0} \frac {(f(x^\mu) + (V^\mu \lambda) \del_\mu f + \dotsm ) - f(x^\mu)}{\lambda}\\
&=V^\mu\del_\mu f.
}
Similarly, one can obtain the Lie derivative along a vector field $V^\mu$ on another vector field $U^\mu$. To simplify the notation, we will define the image of a point $p$ under the map $\phi$ as $q=\phi(p)$, it follows
\eq{
\mathcal{L}_V U^\mu &=\lim\limits_{\lambda \to 0} \frac {\phi^* [U^\mu(q)] - U^\mu(p)}{\lambda} \\
&=\lim\limits_{\lambda \to 0} \frac {\frac {\del x^\mu}{\del y^\nu}U^\nu(q)-U^\mu(p)}{\lambda}.
}
Let's first work out the pullback matrix $\del x^\mu /\del y^\nu$. The map from point $p$ to its image $q$ along the integral curve generated by $V^\mu$ corresponds to the coordinate transformation
\eq{
y^\mu=x^\mu + V^\mu \lambda,
}
and its infinitesimal form
\eq{
dy^\mu=dx^\mu + V^\mu d\lambda,
}
from which we get
\eq{
\frac {\del x^\mu}{\del y^\nu} &= \frac {\del y^\mu}{\del y^\nu} - \frac {\del V^\mu}{\del y^\nu}d\lambda\\
&= \delta^\mu_\nu -\del_\nu V^\mu d\lambda .
}
Substituting this back into (6.10) gives
\eq{
\mathcal{L}_V U^\mu &=\lim\limits_{\lambda \to 0} \frac {(\delta^\mu_\nu -\del_\nu V^\mu d\lambda)U^\nu(q)-U^\mu(p)}{\lambda}\\
&=\lim\limits_{\lambda \to 0} \frac {U^\mu(q) -U^\nu\del_\nu V^\mu d\lambda - U^\mu(p)}{\lambda}\\
&=\lim\limits_{\lambda \to 0} \frac {U^\mu(x^\mu + V^\mu d\lambda) -U^\nu\del_\nu V^\mu d\lambda - U^\mu(x^\mu)}{\lambda}\\
&=\lim\limits_{\lambda \to 0} \frac {U^\mu(x^\mu) + (V^\nu d\lambda)\del_\nu U^\mu -U^\nu\del_\nu V^\mu d\lambda - U^\mu(x^\mu)}{\lambda}\\
&=V^\nu\del_\nu U^\mu - U^\nu\del_\nu V^\mu.
}
This result can also be expressed as a commutator between the two vector fields, 
\eq{
\mathcal{L}_V U^\mu \equiv [V, U]^\mu.
}
For this reason, the Lie derivative is also commonly referred to as the Lie bracket. We can of course obtain the Lie derivative on a differential 1-form following a similar process, but we will skip the details and simply show the result below
\eq{
\mathcal{L}_V \omega_\mu = V^\nu \del_\nu \omega_\mu + \del_\mu V^\nu\omega_\nu.
}
This expression can be rewritten more abstractly in differential form notation as
\eq{
\mathcal{L}_V \omega = i_V\, d\omega + d\, i_V \omega,\label{cartan1}
}
where the symbol $i_V$ is the inner product operator and the symbol $d$ is the exterior derivative. This formula is referred to as the Cartan's identity. We will come back to it in later sections.

In general, the Lie derivative along a vector field $V^\mu$ on an arbitrary rank tensor can be written as 
\eq{
\mathcal{L}_V T^{\mu_1 \mu_2 ...}_{\xu \xu\xu\xu \nu_1 \nu_2 ...} &=
V^\alpha \Del_\alpha T^{\mu_1 \mu_2 ...}_{\xu \xu\xu\xu \nu_1 \nu_2 ...}
- (\Del_\alpha V^{\mu_1}) T^{\alpha \mu_2 ...}_{\xu \xu\xu\xu \nu_1 \nu_2 ...}
- (\Del_\alpha V^{\mu_2}) T^{\mu_1 \alpha ...}_{\xu \xu\xu\xu \nu_1 \nu_2 ...} - ...
\nn \\
&+ (\Del_{\nu_1} V^\alpha) T^{\mu_1 \mu_2 ...}_{\xu \xu\xu\xu \alpha \nu_2 ...} +
(\Del_{\nu_2} V^{\alpha}) T^{\mu_1 \mu_2 ...}_{\xu \xu\xu\xu \nu_1 \alpha ...} + ...
\label{lie dir}
}
Interestingly, all the partial derivatives are replaced by the (torsion-free) covariant derivatives because the terms involving the connection coefficients will all automatically cancel out. This form of Lie derivative shows its covariant behavior manifestly. 

The Lie derivative helps us define a derivative in a manifold in a coordinate-independent manner, which further provides us with a more natural way of dealing with tensors in a manifold. We will see how this concept applies in the geometric analysis of spatial hypersurfaces in the following sections.

\section{Extrinsic Curvature}

The newly introduced Lie derivative will help us introduce another important (2,0) tensor called the extrinsic curvature. As mentioned previously, while the Ricci curvature tensor and scalar tell us about the curvature intrinsic to the geometry, extrinsic curvature tells us how the 3-dimensional foliations are embedded in 4-dimensional spacetime. Formally, the extrinsic curvature of a hypersurface $\Sigma$ can be defined by the Lie derivative of the spatial metric along the time-like normal vector $n^\mu$ of $\Sigma$:
\eq{
 K_{\mu\nu} :=  \tfrac12 \mathcal{L}_{n} q_{\mu\nu},
}
where $\mathcal{L}_{n}$ is the Lie Derivative with respect to the normal vector. We can thus understand the extrinsic curvature as the rate of change of the spatial metric $q_{\mu\nu}$ as $\Sigma$ evolves along the normal vector field. 

A more geometrically intuitively definition of the extrinsic curvature can be given by the following mathematical form
\eq{
K_{\mu\nu} :=  q^\alpha_\mu q^\beta_\nu \Del_\alpha n_\beta , \label{extrinsic2}
}
which implies that the extrinsic curvature measures how much the time-like normal vector $n^\mu$ changes as it is parallel transported along the hypersurface $\Sigma$, conceptualized by Fig.\ref{fig:extrin}. 
\begin{figure}[h]
    \centering
    \includegraphics[scale=.4]{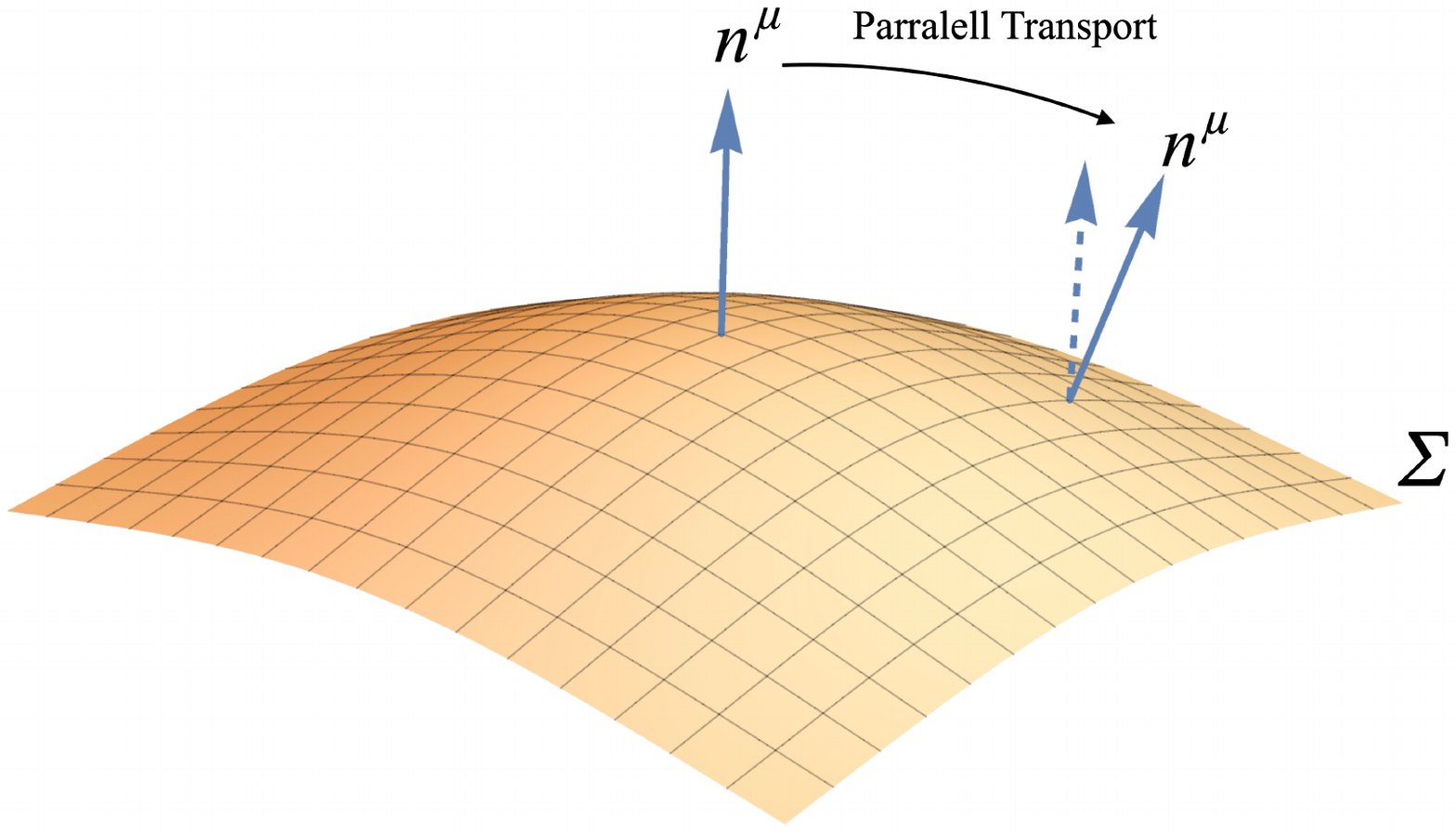}
    \caption{Visual representation of extrinsic curvature}
    \label{fig:extrin}
\end{figure}

These two seemingly different definitions can be shown to be equivalent in the following by applying the explicit form of Lie derivative in  (\ref{lie dir}) for a (2,0) tensor:
\eq{
\mathcal{L}_{n} q_{\mu\nu} &= n^\alpha \Del_\alpha q_{\mu\nu} + q_{\alpha\nu} \Del_\mu n^\alpha + q_{\mu\alpha} \Del_\nu n^\alpha
\nn \\ &=
n^\alpha \Del_\alpha (g_{\mu\nu} +n_\mu n_\nu) +
(g_{\alpha\nu} + n_\alpha n_\nu)\Del_\mu n^\alpha + 
(g_{\mu\alpha} + n_{\mu} n_\alpha) \Del_\nu n^\alpha
\nn \\ &=
n^\alpha \Del_\alpha (n_\mu n_\nu) +
g_{\alpha\nu}\Del_\mu n^\alpha + 
g_{\mu\alpha} \Del_\nu n^\alpha
\nn \\ &=
n_\nu n^\alpha \Del_\alpha n_\mu 
+ n_\mu n^\alpha \Del_\alpha n_\nu +
\Del_\mu n_\nu + \Del_\nu n_\mu
\nn \\ &=
(q^\alpha_\nu - \delta^\alpha_\nu) \Del_\alpha n_\mu + (q^\alpha_\mu - \delta^\alpha_\mu) \Del_\alpha n_\nu + \Del_\mu n_\nu + \Del_\nu n_\mu
\nn \\ &=
q^\alpha_\nu \Del_\alpha n_\mu - \Del_\nu n_\mu + q^\alpha_\mu \Del_\alpha n_\nu - \Del_\mu n_\nu + \Del_\mu n_\nu + \Del_\nu n_\mu
\nn \\ &=
q^\alpha_\nu \Del_\alpha n_\mu + q^\alpha_\mu \Del_\alpha n_\nu ,
\label{extr proof}
}
where in the second line the covariant derivative acting on the metric vanishes due to the metric compatibility condition and we use the identity $n^\alpha \Del_\mu n_\alpha = 0$ proven below
\eq{
n^\alpha \Del_\mu n_\alpha + n_\alpha \Del_\mu n^\alpha &= \Del_\mu(n^\alpha n_\alpha) = \Del_\mu(-1) = 0 
\nn \\ 
\to n_\alpha \Del_\mu n^\alpha &= n^\alpha \Del_\mu n_\alpha = 0.\label{identity x}
}
Here, in the first step we ``undo" the chain rule. With the last line of Eq. (\ref{extr proof}), we have reached a nice intermediate result that looks like Eq. (\ref{extrinsic2}) already, but it is not the exact desired form yet. To prove the equivalence of the two definitions, we just need to repeat the trick of extracting projection operator again following Eq. (\ref{extr proof}):
\eq{
\mathcal{L}_{n} q_{\mu\nu}
&=
q^\alpha_\nu \Del_\alpha (n_\beta \delta^\beta_\mu)
+ q^\alpha_\mu \Del_\alpha (n_\beta \delta^\beta_\nu)
\nn \\ &=
q^\alpha_\nu \delta^\beta_\mu \Del_\alpha n_\beta 
+ q^\alpha_\mu \delta^\beta_\nu \Del_\alpha n_\beta,\nn\\
&=
q^\alpha_\nu (q^\beta_\mu - n^\beta n_\mu) \Del_\alpha n_\beta 
+ q^\alpha_\mu (q^\beta_\nu - n^\beta n_\nu) \Del_\alpha n_\beta
\nn \\ &=
q^\alpha_\nu q^\beta_\mu \Del_\alpha n_\beta - n^\beta n_\mu \Del_\alpha n_\beta 
+ q^\alpha_\mu q^\beta_\nu \Del_\alpha n_\beta - n^\beta n_\nu \Del_\alpha n_\beta
\nn \\ &=
q^\alpha_\nu q^\beta_\mu \Del_\alpha n_\beta
+ q^\alpha_\mu q^\beta_\nu \Del_\alpha n_\beta
\nn \\ &=
2K_{\mu\nu},
}
where the identity in Eq. (\ref{identity x}) is again used in the third to last to second to last line.

The first definition of extrinsic curvature shows that it is a purely spatial quantity as we are using the spatial part of the metric $q_{\mu\nu}$ to define it. If we contract extrinsic curvature with the normal vector itself we can find that
\eq{
n^\mu K_{\mu\nu} &=  n^\mu q^\alpha_\mu q^\beta_\nu \Del_\alpha n_\beta
\nn \\ &=
n^\mu (\delta^\alpha_\mu +  n_\mu n^\alpha) q^\beta_\nu \Del_\alpha n_\beta
\nn \\ &=
 n^\mu q^\beta_\nu ( \Del_\mu n_\beta + n_\mu n^\alpha \Del_\alpha n_\beta)
\nn \\ &=
q^\beta_\mu (n^\mu \Del_\mu n_\beta + n^\mu n_\mu n^\alpha \Del_\alpha n_\beta)
\nn \\ &=
q^\beta_\mu (n^\mu \Del_\mu n_\beta - n^\alpha \Del_\alpha n_\beta)
\nn \\ &= 0,
}
again showing it is a purely spatial quantity so long as we are in a coordinate system adapted to the foliation. In this case, we can say $K_{00} = K_{0a} = 0$, and that is why we typically denote extrinsic curvature with spatial indices as $K_{ab}$. The Lie derivative definition also illuminates that extrinsic curvature is roughly the ``velocity" of the metric and is closely related to the canonically conjugate momentum of the spatial part of the metric. 

In the ADM formalism, the metric components $g_{tt}, g_{ta}$ do not generate their time derivatives in the action and therefore are non-dynamical. The configuration variable is just the spatial metric $q_{ab}$ and its canonical momenta $\Pi_{ab}$ which is related to extrinsic curvature via the following equation, 
\eq{
\Pi^{ab} = \sqrt{g}\left(K q^{ab} - K^{ab}\right).
}
If we proceed to develop the Hamiltonian theory in terms of these variables, the Hamiltonian will take the following form \cite{Poisson}
\eq{
H= \int { d^3 x} \left[ \; 
N \sqrt{q} \Big( 
K^{ab}K_{ab} - K^2 -\, \mathcal{R}
\Big) - 2 N_a \Del_b \left(K q^{ab} - K^{ab}\right)
\right],
}
where $\mathcal{R}$ is the spatial Ricci curvature scalar. From this Hamiltonian, we can derive the equations of motion for the spatial metric and extrinsic curvature which provide a dynamical view of spacetime. Despite the success of this formalism describing classical relativity, all attempts at quantization in this formalism have failed. We will touch on the details of this subject later in this section. For now, let's see another way to unveil the dynamics of GR via the 3+1 decomposition of the Palatini Action.

\section{$3+1$ decomposition of Palatini Action} \label{pal 3+1}
The first step in obtaining a $3 + 1$ decomposition is to introduce the \textit{Triad}, which is defined as a projection operator acting on a tetrad as follows,
\eq{E^\nu_I := q^\nu_\mu e^\mu_I.}
A triad is the same concept as a tetrad except it is purely spatial with respect to its external indices i.e. $E^0_I=0$. With this spatial analog of the tetrad, we can perform the $3+1$ decomposition in terms of the Palatini variables. Starting from the Palatini action
\eq{
S_P = \tfrac{1}{2} \int_M \sqrt{-g}\, e^\mu_I e^\nu_J F_{\mu\nu}^{\xu\xu IJ} \,d^4x ,
}
we can rewrite the interesting part as follows
\eq{
e^\mu_I e^\nu_J F_{\mu\nu}^{\xu\xu IJ} &= e^\mu_I e^\nu_J \delta^\alpha_\mu \delta^\beta_\nu F_{\alpha\beta}^{\xu\xu IJ}
\nn \\
&=
e^\mu_I e^\nu_J (q^\alpha_\mu - n^\alpha n_\mu) (q^\beta_\nu - n^\beta n_\nu) F_{\alpha\beta}^{\xu\xu IJ}
\nn \\ &=
e^\mu_I e^\nu_J (q^\alpha_\mu q^\beta_\nu - q^\alpha_\mu n^\beta n_\nu  -q^\beta_\nu n^\alpha n_\mu  + n^\alpha n_\mu n^\beta n_\nu )F_{\alpha\beta}^{\xu\xu IJ}
\nn \\
&=
E^\alpha_I E^\beta_J F_{\alpha\beta}^{\xu\xu IJ} - 
2 e^\mu_I e^\nu_J q^\alpha_\mu n^\beta n_\nu F_{\alpha\beta}^{\xu\xu IJ}
\nn \\ &=
E^\alpha_I E^\beta_J F_{\alpha\beta}^{\xu\xu IJ} - 
2 E^\alpha_I n^\beta n_J F_{\alpha\beta}^{\xu\xu IJ}
}
where in the second line we use the definition of the projection operator, Eq. (\ref{pal proj}), to expand the Kronecker deltas and in the third to fourth line the first term becomes a pair of triads, the middle two terms combine into one and the last term vanishes due to the antisymmetry in the $\alpha$ and $\beta$ indices when contracted with $F_{\alpha\beta}^{\xu\xu IJ}$.

Recalling the definition of the time-like vector field $t^\mu$ (Eq. (\ref{timelike})) we can rewrite the time like unit vector $n^\beta$ as follows
\eq{
n^\beta = \tfrac{1}{N}( t^\beta - N^\beta )
}
subbing this into the prior expression gives us
\eq{
e^\mu_I e^\nu_J F_{\mu\nu}^{\xu\xu IJ} &=
E^\alpha_I E^\beta_J F_{\alpha\beta}^{\xu\xu IJ} - 
2 E^\alpha_I \tfrac{1}{N}( t^\beta - N^\beta ) n_J F_{\alpha\beta}^{\xu\xu IJ}
\nn \\ &=
E^\alpha_I E^\beta_J F_{\alpha\beta}^{\xu\xu IJ} - 
\tfrac{2}{N} (E^\alpha_I n_J t^\beta F_{\alpha\beta}^{\xu\xu IJ} - E^\alpha_I n_J N^\beta F_{\alpha\beta}^{\xu\xu IJ}).
}
Next, we adopt the following relation: 
\eq{
\mathcal{L}_t\, \omega_\alpha^{\xu IJ} &=
t^\beta F_{\beta\alpha}^{\xu\xu IJ} + D_\alpha (t^\beta \omega_\beta^{\xu IJ}),\label{cartan2}
}
which we recognize as a generalized Cartan's identity to Eq. (\ref{cartan1}). It follows then
\eq{
\mathcal{L}_t\, \omega_\alpha^{\xu IJ} &=
- t^\beta F_{\alpha\beta}^{\xu\xu IJ} + D_\alpha (t^\beta \omega_\beta^{\xu IJ})
\nn \\
t^\beta F_{\alpha\beta}^{\xu\xu IJ} &= -\mathcal{L}_t\, \omega_\alpha^{\xu IJ}  + D_\alpha (t^\beta \omega_\beta^{\xu IJ})
\nn \\ &:=
-\dot{\omega}_\alpha^{\xu IJ}  + D_\alpha (t^\beta \omega_\beta^{\xu IJ}),
}
where $\mathcal{L}_t $ is the Lie derivative with respect to the time-like vector field (which will be denoted with a dot to reduce clutter).
Subbing the previous result into the interesting part of the action gives us
\eq{
e^\mu_I e^\nu_J F_{\mu\nu}^{\xu\xu IJ} & =
E^\alpha_I E^\beta_J F_{\alpha\beta}^{\xu\xu IJ} - 
\tfrac{2}{N} (E^\alpha_I n_J ( -\dot{\omega}_\alpha^{\xu IJ}  + D_\alpha (t^\beta \omega_\beta^{\xu IJ}) ) - E^\alpha_I n_J N^\beta F_{\alpha\beta}^{\xu\xu IJ})
\nn \\ &=
E^\alpha_I E^\beta_J F_{\alpha\beta}^{\xu\xu IJ} +
\tfrac{2}{N} E^\alpha_I n_J \dot{\omega}_\alpha^{\xu IJ}  
- \tfrac{2}{N} E^\alpha_I n_J D_\alpha (t^\beta \omega_\beta^{\xu IJ} ) 
+ \tfrac{2}{N} E^\alpha_I n_J N^\beta F_{\alpha\beta}^{\xu\xu IJ}
\nn \\ &=
E^\alpha_I E^\beta_J F_{\alpha\beta}^{\xu\xu IJ} 
- \tfrac{2}{N} n_I E^\alpha_J \dot{\omega}_\alpha^{\xu IJ}  
+ \tfrac{2}{N} n_I E^\alpha_J  D_\alpha (t^\beta \omega_\beta^{\xu IJ} ) 
- \tfrac{2}{N} n_I E^\alpha_J  N^\beta F_{\alpha\beta}^{\xu\xu IJ}
\nn \\ &=
E^\alpha_I E^\beta_J F_{\alpha\beta}^{\xu\xu IJ} 
- \tfrac{2}{N} n_I E^\alpha_J \dot{\omega}_\alpha^{\xu IJ}  
+ \tfrac{2}{N} n_I E^\alpha_J  D_\alpha (t^\beta \omega_\beta^{\xu IJ} ) 
+ \tfrac{2}{N} n_I N^\alpha E^\beta_J   F_{\alpha\beta}^{\xu\xu IJ}
}
where in the third line we have swapped the $I$ and $J$ indices between the triads and normal vector on the last 3 terms
and in the final line, we have swapped the $\alpha$ and $\beta$ indices between the shift vector and the tetrad of the last term.

Coming back to the full action and substituting in the previous line gives us
\eq{
S_P = \tfrac{1}{2} \int_M d^4x\, 
\sqrt{-g}\, \Big( 
E^\alpha_I E^\beta_J F_{\alpha\beta}^{\xu\xu IJ} 
- \tfrac{2}{N} n_I E^\alpha_J \dot{\omega}_\alpha^{\xu IJ}  
+ \tfrac{2}{N} n_I E^\alpha_J  D_\alpha (t^\beta \omega_\beta^{\xu IJ} ) 
+ \tfrac{2}{N} n_I N^\alpha E^\beta_J   F_{\alpha\beta}^{\xu\xu IJ}
\Big).
}
Now using the relation, 
\eq{
\sqrt{-g} = N\sqrt{q},
}
where $q$ is the determinant of the spatial metric. It allows us to eliminate $\sqrt{-g}$ and most presence of the lapse function as follows (this relation is proven in the next section as some later introduced identities are needed)
\eq{
S_P = \tfrac{1}{2} \int_M d^4x\, 
\sqrt{q}\, \Big( 
N E^\alpha_I E^\beta_J F_{\alpha\beta}^{\xu\xu IJ} 
- 2 n_I E^\alpha_J \dot{\omega}_\alpha^{\xu IJ}  
+ 2 n_I E^\alpha_J  D_\alpha (t^\beta \omega_\beta^{\xu IJ} ) 
+ 2 N^\alpha n_I E^\beta_J   F_{\alpha\beta}^{\xu\xu IJ}
\Big).
}
It is appropriate now to define the densitized triad
\eq{
\tilde{E}^\alpha_I &:= \sqrt{q}E^\alpha_I,
}
and another convenient quantity
\eq{
\tilde{E}^\alpha_{IJ} &:= \tilde{E}^\alpha_{[I} n_{J]} =  \sqrt{q} E^\alpha_{[I} n_{J]}.
}
Utilizing the anti-symmetry of the internal indices along with our newly defined quantities we can rewrite the action as follows
\eq{
S_P &= \tfrac{1}{2} \int_M d^4x\, 
\sqrt{q}\, \Big( 
N E^\alpha_{I} E^\beta_{J} F_{\alpha\beta}^{\xu\xu IJ} 
- 2 n_{[I} E^\alpha_{J]} \dot{\omega}_\alpha^{\xu IJ}  
+ 2 n_{[I} E^\alpha_{J]}  D_\alpha (t^\beta \omega_\beta^{\xu IJ} ) 
+ 2 N^\alpha n_{[I} E^\beta_{J]}   F_{\alpha\beta}^{\xu\xu IJ}
\Big)  
\nn \\ &=
\tfrac{1}{2} \int_M d^4x\,  \Big( 
-N\frac{4}{\sqrt{q}} \tilde{E}^\alpha_{IJ} \tilde{E}^{\beta JK} F_{\alpha\beta K}^{\xu\xu\xu I} 
- 2 \tilde{E}^\alpha_{JI} \dot{\omega}_\alpha^{\xu IJ}  
+ 2 \tilde{E}^\alpha_{JI}  D_\alpha (t^\beta \omega_\beta^{\xu IJ} ) 
+ 2 N^\alpha \tilde{E}^\beta_{JI}   F_{\alpha\beta}^{\xu\xu IJ}
\Big).
}
where the first term can be shown to be equivalent between the previous two lines as follows
\eq{
\tilde{E}^\alpha_{IJ} \tilde{E}^{\beta JK} F_{\alpha\beta K}^{\xu\xu\xu I} &= (\sqrt{q})^2 E^\alpha_{[I}n_{J]}E^{\beta[J}n^{K]} F_{\alpha\beta K}^{\xu\xu\xu I}
\nn \\ &=
q \tfrac{1}{2}(E^\alpha_I n_J - E^\alpha_J n_I ) \tfrac{1}{2}(E^{\beta J} n^K - E^{\beta K} n^J )F_{\alpha\beta K}^{\xu\xu\xu I}
\nn \\ &=
\frac{q}{4}(E^\alpha_I E^{\beta J}n_Jn^K - E^\alpha_I E^{\beta K} n_J n^J - E^\alpha_J E^{\beta J} n_I n^K + E^\alpha_J E^{\beta K} n_I n^J) F_{\alpha\beta K}^{\xu\xu\xu I}
\nn \\ &=
\frac{q}{4}( - E^\alpha_I E^{\beta K} n_J n^J ) F_{\alpha\beta K}^{\xu\xu\xu I}
\nn \\ &=
\frac{q}{4} E^\alpha_I E^{\beta K}  F_{\alpha\beta K}^{\xu\xu\xu I}
\nn \\ &=
\frac{q}{4} E^\alpha_I E^{\beta}_K  F_{\alpha\beta }^{\xu\xu KI}\nn\\
&=
-\frac{q}{4} E^\alpha_I E^{\beta}_J  F_{\alpha\beta }^{\xu\xu IJ}.
}
In the above, between the third to the fourth line, all terms except for the second vanish due to the following boxed identities, and in the fourth line the normal vectors contracted with each other give an extra factor of $-1$.

\begin{center}
 \framebox{ 
\parbox[t][10.0cm]{15.50cm}{
\addvspace{0.2cm} %\centering%
\begin{itemize}
\item
\textbf{Triad Identity Proofs} \label{triad proofs}
\eq{
E^\alpha_J n^J &= q_\beta^\alpha e^\beta_J n^J
\qquad \qquad\qquad  \text{Similarly} \qquad\qquad E^{\beta J}n_J = 0
\nn \\ &=
(\delta^\alpha_\beta+n^\alpha n_\beta) e^\beta_J n^J
\nn \\ &=
(\delta^\alpha_\beta+n^\alpha n_\beta) n^\beta
\nn \\ &=
n^\alpha -n^\alpha
\nn \\ &=0
}
\eq{
E^\alpha_J E^{\beta J} &= q^\alpha_\mu e^\mu_J q^\beta_\nu e^{\nu J}
\nn \\ &=
q^\alpha_\mu  q^\beta_\nu g^{\mu\nu}
\nn \\ &=
q^\alpha_\mu  q^{\beta \mu}
\nn \\ &=
(\delta^\alpha_\mu+n^\alpha n_\mu)(g^{\beta\mu}+n^\beta n^\mu)
\nn \\ &=
\delta^\alpha_\mu g^{\beta \mu} + \delta^\alpha_\mu n^\beta n^\mu + g^{\beta\mu}n^\alpha n_{\mu} + n^\alpha n_\mu n^\beta n^\mu
\nn \\ &=
g^{\beta\alpha}+n^\beta n^\alpha+ n^\alpha n^\beta -
n^\alpha n^\beta
\nn \\ &=
g^{\alpha \beta}+n^\alpha n^\beta
\nn \\ &= q^{\alpha\beta}
}
\end{itemize}
}}
\end{center}

Finally, we can put it altogether and  reach the final form of the Palatini action:
\eq{
S_P &=  \int_M d^4x\,  \Big( 
 \tilde{E}^\alpha_{IJ} \dot{\omega}_\alpha^{\xu IJ} -\frac{2N}{\sqrt{q}} \tilde{E}^\alpha_{IJ} \tilde{E}^{\beta JK} F_{\alpha\beta K}^{\xu\xu\xu I} 
- N^\alpha \tilde{E}^\beta_{IJ}   F_{\alpha\beta}^{\xu\xu IJ}
- \tilde{E}^\alpha_{IJ}  D_\alpha (t^\beta \omega_\beta^{\xu IJ} )
\Big),\nn\\
&= \int_M d^4x\,  \Big( 
 \tilde{E}^\alpha_{IJ} \dot{\omega}_\alpha^{\xu IJ} -2\underset{\sim}{N} \tilde{E}^\alpha_{IJ} \tilde{E}^{\beta JK} F_{\alpha\beta K}^{\xu\xu\xu I} 
- N^\alpha \tilde{E}^\beta_{IJ}   F_{\alpha\beta}^{\xu\xu IJ}
+ (t\cdot \omega)^{IJ} (D_\alpha \tilde{E}^\alpha_{IJ})  \label{sp}
\Big),\\
&= \int_M d^4x\, \tr \Big( 
 -\tilde{E}^\alpha \dot{\omega}_\alpha -2\underset{\sim}{N} \tilde{E}^\alpha \tilde{E}^\beta F_{\alpha\beta} 
+ N^\alpha \tilde{E}^\beta   F_{\alpha\beta}
- (t\cdot \omega) (D_\alpha \tilde{E}^\alpha)  
\Big),
}
where in the second term of (\ref{sp}) we absorbed the factor $\sqrt{q}$ into the lapse function $N$ and denoted it as $\underset{\sim}{N}$, and in the last term we performed integration by parts to expressed it as a scalar quantity $(t\cdot \omega)$ times the covariant derivative of the canonical momentum (integration by parts with a covariant derivative is proven in the next chapter). At this point, we can identify the configuration variables of the standard Palatini theory of GR as $\omega_\alpha^{IJ}, \tilde{E}^\alpha_{IJ}, \underset{\sim}{N}, N^\alpha, (t\cdot \omega)^{IJ}$. Among them, the first two are the canonically conjugate pair of phase space variables, and the latter three are all Lagrange multipliers thus non-dynamical fields. By varying the action with respect to these Lagrange multipliers we immediately obtain the following constraints of the Hamiltonian theory:
\eq{
S &= \tr (\tilde{E}^\alpha \tilde{E}^{\beta} F_{\alpha\beta}) \approx 0, \label{con 1}
\\
V_\alpha &= \tr (\tilde{E}^\beta F_{\alpha\beta}) \approx 0, \label{con 2}
\\
G_{IJ} &= D_\alpha \tilde{E}^\alpha_{IJ}\approx 0. \label{con 3}
}
It is clear that all of these constraints are of polynomial form in the canonically conjugate variables $\omega_\alpha^{\xu IJ}$ and $\tilde{E}^\alpha_{IJ}$, and it can be further shown that they form a closed algebra under Poisson brackets.

If (\ref{con 1} - \ref{con 3}) were indeed the complete set of constraints, the attempt at a Quantum theory of gravity via Palatini formulation would have been successful as a classical theory, and ready to proceed to canonical quantization. However, this was not the case. Upon careful examination, one notices that there exist additional constraints. More specifically, the form of canonical momentum $\tilde{E}^\alpha_{IJ}$ is not completely arbitrary, and it satisfies the following constraint:
\eq{
\phi^{\alpha\beta} = \epsilon^{IJKL}\tilde{E}^\alpha_{IJ}\tilde{E}^\beta_{KL}\approx 0. 
}
To see this, one just needs to apply the specific form of $\tilde{E}^\alpha_{IJ}=\tilde{E}^\alpha_{[I}\, n_{J]}$, then it follows that
\eq{
\phi^{\alpha\beta} 
&=\frac 1 4 \epsilon^{IJKL} (\tilde{E}^\alpha_I n_J - \tilde{E}^\alpha_J n_I)(\tilde{E}^\beta_K n_L - \tilde{E}^\beta_L n_K)\nn\\
&=\frac 1 4 \epsilon^{IJKL} (\tilde{E}^\alpha_I \tilde{E}^\beta_K n_J n_L - \tilde{E}^\alpha_I \tilde{E}^\beta_L n_J n_K - \tilde{E}^\alpha_J \tilde{E}^\beta_K n_I n_L + \tilde{E}^\alpha_J \tilde{E}^\beta_L n_I n_k)=0,
}
where each term in the last line vanishes due to the total antisymmetry of the Levi-Civita tensor. Because this constraint follows directly from the definition of the canonical momenta $\tilde{E}^\alpha_{IJ}$, it is categorized as a primary constraint. 

Furthermore, the consistency condition of this new constraint via its Poisson brackets with the total Hamiltonian would lead to yet another constraint:
\eq{
\chi^{\alpha\beta} = \epsilon^{IJKL} (D_\gamma \tilde{E}^\alpha_{IJ})[\tilde{E}^\beta, \tilde{E}^\gamma]_{KL}+(\alpha \leftrightarrow \beta) \approx 0. 
}
This constraint comes from the consistency condition of a primary constraint, so it is categorized as a secondary constraint. Luckily, this time there are no more constraints resulting from the consistency condition on $\chi^{\alpha\beta}$, we finally have the complete set of constraints ($S,\, V_\alpha,\, G_{IJ},\, \phi^{\alpha\beta},\, \chi^{\alpha\beta}$), and all of them have to be gone through Dirac's constraint analysis together. A thorough breakdown of this analysis would require a highly technical discussion at length, which exceeds the scope of this paper. Therefore we will skip it here. More details can be found in references \cite{Ashtekar} and \cite{Romano}. 

As it turns out, the original three constraints ($S,\, V_\alpha,\, G_{IJ}$) all generate weakly vanishing Poisson brackets among themselves, whereas the new constraints $\phi^{\alpha\beta}$ and $\chi^{\alpha\beta}$ do not produce weakly vanishing Poisson brackets with ($S,\, V_\alpha,\, G_{IJ}$) or between themselves. This means the original three constraints are first-class and the new ones form a second-class pair! It is the presence of this second-class pair that rendered the standard Palatini formulation of GR unsuccessful in constructing a quantum theory of gravity. Specifically, contrary to the first-class constraint, the second-class constraints do not generate gauge transformations on the fields in a gauge theory. They represent redundancies in the canonical variables that need to be eliminated before making the transition to a quantum theory \cite{Dirac}. The standard procedure here is to solve the second-class constraints and impose their solutions on the original phase space variables. In the case we are dealing with right here, solving the second-class constraint is most convenient if we choose a gauge fixing of the internal normal vector $n^I$. This choice comes with a side effect that requires us to solve the boost part of the constraint (\ref{con 3}) as well because an uncontracted $\tilde{E}^\alpha_{IJ}$ appears in that equation, namely, we have to solve the following equation
\eq{
(D_\alpha \tilde{E}^\alpha_{IJ}) n^J\approx 0,
}
in addition to the remaining second-class constraint $\chi^{\alpha\beta}$.

Long story short, after eliminating the second-class constraints by solving these equations, one arrives at a reduced phase space that is characterized by the canonically conjugate pair $(\tilde{E}_i^a, K_a^i)$ with the indices $a=1,2,3$, and $i=1,2,3$. Therefore $\tilde{E}_i^a$ is a densitized triad and $K_i^a$ is a reduced connection variable out of $\omega_\alpha^{IJ}$ which now behaves like a three-dimensional Lie-algebra-valued 1-form. A direct consequence of this change of phase space variables is that the original set of first-class constraints (\ref{con 1} - \ref{con 3}) has turned into the following:
\eq{
S' &= (\tilde{E}_i^b \tilde{E}_j^a - \tilde{E}_i^a \tilde{E}_j^b)K_a^i K_b^j/\sqrt{q} - \sqrt{q}\mathcal{R}\approx 0\\
V'_\alpha &= D_\beta(K_a^i \tilde{E}_i^b - K_c^i \tilde{E}_i^c\delta_a^b) \approx 0\\
G'_i &= \epsilon_{ijk}K_a^j \tilde{E}^{ak},
}
where $\mathcal{R}$ is the scalar spatial curvature. These constraints are referred to as the scalar, vector, and Gauss constraints of the $3+1$ Palatini theory. Surprisingly, they coincide exactly with the same set of constraints in ADM variables from the tetrad gravity\cite{Deser:1976ay, Henneaux} that is equivalent to the standard Einstein-Hilbert theory of general relativity \cite{new ADM}. Essentially, by solving the second-class constraints, the Palatini theory, motivated as a connection-dynamics theory, is brought back to the geometrodynamics theory of GR. So we are facing the same devastating situation: Due to the presence of $\mathcal{R}$ in the scalar constraint, it is no longer a polynomial form in the phase space variables, and the subsequent quantization procedure will become problematic.

In summary, the standard Palatini formulation of General Relativity had its historical significance in bringing GR into a form that resembles a Yang-Mills gauge theory. However, this standard formulation was fatally flawed: Performing a Dirac constraint analysis reveals the presence of second-class constraints. Upon solving these constraints the remaining first-class constraints became non-polynomial and we are essentially forced back to using the geometrodynamical variables i.e. the metric and its conjugate momenta. Therefore in its standard form, the Palatini formulation is equivalent to the metric formulation and did not bring about any substantial breakthrough in quantum gravity.

In the next chapter, we will explore the Self-dual formulation of General Relativity by Abhay Ashtekar, which is a modification of the Palatini formulation, with a new type of variable called the self-dual connection.

\chapter{Ashtekar's New Variables} \label{sec: ash}

As elaborated at the end of the last section, the Palatini formulation provided us with a new perspective on understanding gravity as a connection-dynamics theory, where the fundamental variables are gauge connections, similar to that of a Yang-Mills theory. On the other hand, the metric becomes somewhat of a secondary, derived concept. As illuminating as it was, the standard Palatini formulation did not end up succeeding in building a well-behaved Hamiltonian theory of general relativity. Specifically, after eliminating the second-class constraints, the remaining first-class constraints became non-polynomial. Since the early 1960s, this problem remained a long-standing obstacle to building a quantum theory of gravity in the canonical approach for well over two decades. The breakthrough occurred in the mid-1980s when Abhay Ashtekar introduced a similar formulation of GR  with a new set of variables, called the self-dual formulation. In this section, we will focus on showing how this formulation addressed the issues that existed in the standard Palatini formulation, and led to a much better behaved Hamiltonian theory of GR. Many of the calculations present closely follow \cite{Ashtekar, Romano} which contain
further discussion of the results and additional details.

\section{A Change of Notation}
Before going forward, we would like to address a change of notation in this section. For reasons explained below, all the Greek indices $\mu,\nu,\rho,\sigma,etc$ that we used previously for external spacetime will now be denoted by Latin letters from the beginning of the alphabet $a,b,c,d,...$ i.e. $A_\mu \to A_a$. With this change, we lose the obvious indication of what external indices run from $0-3$ or from $1-3$, but we gain the advantage that our equations have consistent indices despite having a mix of 3-d and 4-d quantities.
To indicate 4-dimensional, or spacetime, external indices we will put a raised ``$(4)$" to the left of the variable. For example
\eq{
A_a = \begin{bmatrix}
    0\\ A_1\\A_2 \\A_3 \\
\end{bmatrix},
\qquad\qquad
^{(4)}A_a = \begin{bmatrix}
    A_0\\ A_1\\A_2 \\A_3 \\
\end{bmatrix}.
}
It's important to note that the raised ``$(4)$" will only be used on quantities that \textbf{can} have indices running from $1-3$ or $0-3$ such as the covariant derivative $^{(4)}D_a$ or curvature tensor $^{(4)}F^a_{bcd}$. For quantities that are explicitly 4-dimensional, such as the tetrad $e^a_I$, normal vector $n^a$, and metric $g_{ab}$ we will omit the use of the raised ``(4)", to reduce clutter. Another reason for this change in notation is the fact that in the end we intend for all quantities to become purely spatial and would therefore end up using Latin indices if we were to continue using Greek alongside Latin. Furthermore, historically this has been the choice for those responsible for pioneering this formalism.

Internal indices on the other hand are much simpler where 4-dimensional internal indices are upper-case Latin letters from the middle of the alphabet $I,J,K,L,...$ as in the previous chapters. When the internal indices are 3-dimensional, purely spatial, lower-case Latin letters from the middle of the alphabet $i,j,k,l,...$ will be used.

In summary, we have
\eq{
&^{(4)}A^{I\, = \,\{0,1,2,3\}}
_{a\, = \,\{0,1,2,3\}}
}
\eq{
&A^{i\, = \,\{1,2,3\}}
_{a\, = \,\{1,2,3\}}
}
where we point out that the raised ``(4)" is only to indicate that the external indices are running from $0-3$.

\section{Intro to Self Duality}

Let $A_a^{IJ}$ be a tensor with anti-symmetric internal indices. Its dual, denoted by the Hodge star operator, is defined as
\eq{
*A_a^{IJ} := \tfrac{1}{2}\epsilon^{IJ}_{\xu\xu KL}A^{KL}_a\,.
}
If the dual of a tensor satisfies the following relation the tensor is referred to as \textit{self-dual}
\eq{
*A_a^{IJ} = i A_a^{IJ} \,.
}
Similarly, the tensor is \textit{anti-self-dual} if the following is true
\eq{
*A_a^{IJ} = -i A_a^{IJ} \,.
}
Any antisymmetric tensor can be separated into its self-dual and anti-self-dual parts denoted by the raised plus and minus sign respectively as shown below
\eq{
A_a^{IJ} = \tfrac12 ({^+}A_a^{IJ} + {^-}A_a^{IJ})\,.
}
Additionally, it can be shown that the self-dual and anti-self-dual parts can be constructed as follows 
\eq{
{^+}A_a^{IJ} &= \tfrac{1}{2}(A_a^{IJ}-i*A_a^{IJ})
\label{sd part} \nn \\
{^-}A_a^{IJ} &= \tfrac{1}{2}(A_a^{IJ}+i*A_a^{IJ}).
}
We will prove the former relation below 
\eq{
*{^+}A_a^{IJ} &= \tfrac{1}{2}\epsilon^{IJ}_{\xu\xu KL}{^+}A_a^{KL}
\\ \nn &=
\tfrac{1}{2}\epsilon^{IJ}_{\xu\xu KL}\tfrac{1}{2}(A_a^{KL}-i*A_a^{KL})
\\ \nn &=
\tfrac{1}{2}\epsilon^{IJ}_{\xu\xu KL}\tfrac{1}{2}(A_a^{KL}-\tfrac{i}{2}\epsilon^{KL}_{\xu\xu MN}A_a^{MN})
\\ \nn &=
\tfrac{1}{2}*A_a^{IJ}-\tfrac{i}{8}\epsilon^{IJ}_{\xu\xu KL}\epsilon^{KL}_{\xu\xu MN}A_a^{MN}
\\ \nn &=
\tfrac{1}{2}*A_a^{IJ}-\tfrac{i}{8}(-2)(\delta^I_M \delta^J_N-\delta^I_N \delta^J_M)A_a^{MN}
\\ \nn &=
\tfrac{1}{2}*A_a^{IJ}+\tfrac{i}{4}(A_a^{IJ} - A_a^{JI})
\\ \nn &=
\tfrac{1}{2}(*A_a^{IJ}+iA_a^{IJ})
\\ &=
i{{^+}A_a^{IJ}}.
}
The latter relation can be proved similarly (which we will skip here), namely 
\eq{
*{^-}A^{IJ}_a = -i {^-}A^{IJ}_a\,.
}

Another important property of self-duality is that when an anti-self-dual tensor is contracted with a self-dual tensor (and vice versa) the result is automatically zero. To see see, consider two anti-symmetric tensors $S_{IJ}$ and $T_{IJ}$, we have  
 \eq{
 T_{IJ}{^+}S^{IJ} &= 
 ({^+}T_{IJ}+{^-}T_{IJ}){^+}S^{IJ}
 \nn \\ &=
 ({^+}T_{IJ}+{^-}T_{IJ})(-i*{^+}S^{IJ})
 \nn \\ &=
 ({^+}T_{IJ}+{^-}T_{IJ})(-\tfrac{i}{2}\epsilon^{IJ}_{\xu\xu MN} {^+}S^{MN})
 \nn \\ &=
 -i(\tfrac{1}{2}\epsilon^{IJ}_{\xu\xu MN}{^+}T_{IJ} + \tfrac{1}{2}\epsilon^{IJ}_{\xu\xu MN}{^-}T_{IJ}){^+}S^{MN}
 \nn \\ &=
 -i(\tfrac{1}{2}\epsilon^{IJ}_{\xu\xu MN}{^+}T_{IJ} + \tfrac{1}{2}\epsilon^{IJ}_{\xu\xu MN}{^-}T_{IJ}){^+}S^{MN}
 \nn \\ &=
 -i(i{^+}T_{MN} - i{^-}T_{MN}){^+}S^{MN}
 \nn \\ &=
 ({^+}T_{MN} - {^-}T_{MN}){^+}S^{MN}.
 }
 When looking back at the first line of the above equations, one can see in order for the final line to be true it must be the case that an anti-self-dual part contracted with a self-dual part is always zero
 \eq{
 {^-}T_{MN}{^+}S^{MN} = 0 \label{-+sd}.
 }
 Effectively, when an arbitrary antisymmetric tensor contracts with a self-dual tensor, only the self-dual part contributes to the contraction, and vice versa. This will be a very useful property later on. We will now claim the mixed index curvature tensor is a self-dual tensor if the connection is self-dual and will prove this as follows.
 \eq{
 *F_{ab}^{\xu\xu IJ} &= \tfrac12 \epsilon^{IJ}_{\xu\xu MN} F_{ab}^{\xu\xu MN}
 \nn \\ &=
 \tfrac12 \epsilon^{IJ}_{\xu\xu MN} (2\del_{[a}A_{b]}^{\xu MN} + A_a^{\xu MK}A_{bK}^{\xu \xu N} - A_b^{\xu MK}A_{aK}^{\xu \xu N})
 \nn \\ &=
 2 i \del_{[a}A_{b]}^{\xu IJ}
 + \tfrac12 \epsilon^{IJ}_{\xu\xu MN}(2 A_{[a}^{\xu MK}A_{b]K}^{\xu \xu N})
 }
where
\eq{
\epsilon^{IJ}_{\xu\xu MN}(2 A_{a}^{\xu MK}A_{bK}^{\xu \xu N}) & = 
\epsilon^{IJ}_{\xu\xu MN}(-i*A_a^{\xu MK}) A_{bK}^{\xu \xu N}
 \nn \\ & = 
\epsilon^{IJ}_{\xu\xu MN}(-i\tfrac{1}{2} \epsilon^{MK}_{\xu\xu RS} A_a^{\xu RS}) A_{bK}^{\xu \xu N}
\nn \\ & = 
-\tfrac{i}{2} \epsilon^{MIJN} \epsilon_{M RSK} A_a^{\xu RS} A_{b\xu N}^{\xu K}
\nn \\ & = 
\tfrac{i}{2}( \delta^I_R \delta^J_S \delta^N_K - \delta^J_S \delta^I_R \delta^N_K 
- \delta^I_R \delta^J_K \delta^N_S
+ \delta^I_K \delta^J_R \delta^N_S
\nn \\
&\quad + \delta^I_S \delta^J_K \delta^N_R
- \delta^I_K \delta^J_S \delta^N_R
)A_a^{\xu RS} A_{b\xu N}^{\xu K}
\nn \\ & = 
\tfrac{i}{2}(
A_{a}^{\xu IJ} A_{b \xu N}^{\xu N}
- A_{a}^{\xu IJ} A_{b \xu N}^{\xu N}
- A_{a}^{\xu IN} A_{b \xu N}^{\xu J}
+ A_{a}^{\xu JN} A_{b \xu N}^{\xu I}
\nn \\
&\quad  + A_{a}^{\xu NI} A_{b \xu N}^{\xu J}
- A_{a}^{\xu NJ} A_{b \xu N}^{\xu I}
)
}
Due to the anti-symmetry in the internal indices, the first two terms vanish. Swapping the $J$ and $N$ indices on the third, fourth and fifth terms as well as swapping the $N$ and $I$ indices on the fifth term allows us to combine like terms as follows
\eq{
\epsilon^{IJ}_{\xu\xu MN}(2 A_{a}^{\xu MK}A_{bK}^{\xu \xu N}) 
&= \tfrac{i}{2}(2A_a^{\xu IN} A_{bN}^{\xu\xu J} - 2A_a^{\xu NJ} A_{b \xu N}^{\xu I})
\nn \\ &=
i[A_a,A_b]^{IJ}.
}
Subbing in the following result leads to 
\eq{
*F_{ab}^{\xu\xu IJ} &= 2i \del_{[a}A_{b]}^{\xu IJ} + i[A_a,A_b]^{IJ}
\nn \\ &=
i \left( 2 \del_{[a}A_{b]}^{\xu IJ} + [A_a,A_b]^{IJ}\right)
\nn \\
 &= i F_{ab}^{\xu\xu IJ}
}
Similarly, it follows that for an anti-self-dual connection, the corresponding curvature tensor will also be anti-self-dual.

\section{Self-dual Formulation}
With the mathematics of self-duality in our possession, we can begin constructing the self-dual action where the starting point is the Palatini action
\eq{
S_p = \int d^4x \sqrt{-g} \, e^a_I e^b_J {^{(4)}}F^{\xu\xu IJ}_{ab} \,.
}
Here we enforce that the connection $A_a^{IJ}$ to be only the self-dual part of the original spin-connection $\omega_a^{IJ}$ from the standard Palatini theory 
\eq{A^{IJ}_a = {^{+}}\omega^{IJ}_a,
\label{w+}
}
such that
\eq{
*A_a^{IJ} = i A_a^{IJ},
}
and thus its corresponding curvature tensor is also self-dual
\eq{
 *F_{ab}^{\xu\xu IJ} = i F_{ab}^{\xu\xu IJ}.
}
Next we reexpress $\sqrt{-g}$ as the determinant of a tetrad, proven below,
\eq{
g = \det(g_{ab}) = \det(e^I_a e^J_b\eta_{IJ}) 
&= \det(e^I_a)\det(e^J_b)\det(\eta_{IJ})
= (e)(e)(-1) =-e^2
\nn \\ &\rightarrow \sqrt{-g} = e
}

\begin{center}
 \framebox{ 
\parbox[t][9.0cm]{15.50cm}{
\addvspace{0.2cm} %\centering%
\begin{itemize}
\item 
\bf{Identities and Definitions From Previous Chapter} 
\eq{
q_{ab} &= g_{ab} + n_a n_b \label{id: qab} \\
q_b^a &= \delta^a_b + n^a n_b  \label{id:proj} \\
t^a &= Nn^a + N^a \label{id:time like} \\
E^a_I &= q^a_b e^b_I  \label{id:triad} \\
\tilde{E}^a_I &= \sqrt{q}E^a_I \label{id:dens} \\
\mathcal{L}_t {^{(4)}}A_{a}^{MN} &= t^b {^{(4)}}F^{\xu\xu MN}_{ba} + {^{(4)}}D_a(t^b \,{^{(4)}}A_b^{MN}) \label{lie}
\\
\underaccent{\tilde}{N} &= \frac{N}{\sqrt{q}} \label{id:lapse}
}
\item 
\bf{New Identities and Definitions} 
\eq{
n_I &:= e^a_I n_a \label{id:internal norm} \\
n_0 &:= N \label{id:n0} \\
\epsilon^{IJK} &:= \epsilon^{IJKL}n_L \label{id:epsi} 
}
\end{itemize}
}}
\end{center}

Notice that in this construction, we have an action in terms of complex variables so we have a complex action to begin with. Next, we will apply the same trick to rewrite the action in the $3+1$ form with the help of the projection operator $q_b^a$
\eq{
S &= \int d^4x \, (e)  e^a_I e^b_J {^{(4)}}F^{\xu\xu IJ}_{ab} 
\nn \\ &=
\int d^4x \, (e)  e^a_I e^b_J \delta^c_a \delta^d_b {^{(4)}}F^{\xu\xu IJ}_{cd}
\nn \\ &=
\int d^4x \, (e)  e^a_I e^b_J (q^c_a - n^cn_a) (q^d_b - n^d n_b) {^{(4)}}F^{\xu\xu IJ}_{cd}
\nn \\ &=
\int d^4x \, (e)  e^a_I e^b_J (q^c_a q^d_b - q^c_a n^d n_b - q^d_b n^cn_a + n^cn_a n^d n_b)
{^{(4)}}F^{\xu\xu IJ}_{cd}
}
Taking a closer look at the third term, one can see it is equivalent to the second term after some index manipulation as shown below
\eq{
- e^a_I e^b_J q^d_b n^c n_a {^{(4)}}F^{\xu\xu IJ}_{cd} &= - e^a_I e^b_J q^c_b n^d n_a (- {^{(4)}}F^{\xu\xu IJ}_{dc} )
\nn \\ &= 
- e^b_I e^a_J q^c_a n^d n_b (- {^{(4)}}F^{\xu\xu IJ}_{dc} )
\nn \\ &= 
- e^b_J e^a_I  q^c_a n^d n_b (+ {^{(4)}}F^{\xu\xu JI}_{dc} )
\nn \\ &= 
- e^a_I e^b_J   q^c_a n^d n_b {^{(4)}}F^{\xu\xu IJ}_{cd}
}
This allows us to combine the middle two terms. Additionally, the last term vanishes due to that the product $n^c n^d$ commutes but the curvature tensor $F_{cd}^{\xu\xu IJ}$ is antisymmetric when switching the $c$ and $d$ indices or more explicitly
\eq{
n^cn_a n^d n_b
{^{(4)}}F^{\xu\xu IJ}_{cd} &= - n^cn_a n^d n_b
{^{(4)}}F^{\xu\xu IJ}_{cd}
\nn \\ \therefore n^cn_a n^d n_b
{^{(4)}}F^{\xu\xu IJ}_{cd} &= 0.
}
We are thus left with
\eq{
S&=
\int d^4x \, (e)  e^a_I e^b_J (q^c_a q^d_b -2 q^c_a n^d n_b)
{^{(4)}}F^{\xu\xu IJ}_{cd}
\nn \\ &=
\int d^4x \, (e) \Big( e^a_I e^b_J q^c_a q^d_b 
- 2 e^a_I e^b_J q^c_a n^d n_b \Big) {^{(4)}}F^{\xu\xu IJ}_{cd} 
\nn \\ &=
\int d^4x \, (e) \Big( E^c_I E^d_J 
- 2 E^c_I n^d n_J \Big) {^{(4)}}F^{\xu\xu IJ}_{cd}, 
}
where in the last step, the tetrads are combined with projection operators to create triads, and the normal vector field $n^d$ combines with a tetrad to produce an internal normal vector $n_J$ (Identities \ref{id:triad} and \ref{id:internal norm} respectively). Next we swap the $c$ and $d$ indices on the second term but also swap $I$ and $J$ indices to keep the sign the same
\eq{
S&=
\int d^4x \, (e) \Big( E^c_I E^d_J 
- 2 E^d_J n_I n^c  \Big) {^{(4)}}F^{\xu\xu IJ}_{cd} 
\nn \\ &=
\int d^4x \, (e) \Big( E^a_I E^b_J 
- 2 E^b_J  n_I n^a \Big) {^{(4)}}F^{\xu\xu IJ}_{ab} ,
}
and we use the identity proven in the following boxed section (\ref{e=qn}) so we can rewrite $e$ as $\sqrt{q}N$ and simplify as follows
\eq{
S &=
\int d^4x \, \sqrt{q}N \Big( E^a_I E^b_J 
- 2 E^b_J  n_I n^a \Big) {^{(4)}}F^{\xu\xu IJ}_{ab} 
\nn \\ &=
\int d^4x \, \frac{N}{\sqrt{q}} \Big(\sqrt{q} E^a_I \sqrt{q} E^b_J 
- 2 \sqrt{q} \sqrt{q} E^b_J  n_I n^a \Big) {^{(4)}}F^{\xu\xu IJ}_{ab} 
\nn \\ &=
\int d^4x \, \frac{N}{\sqrt{q}} \Big(\Tilde{ E}^a_I \Tilde{E}^b_J 
- 2 \sqrt{q} \Tilde{E}^b_J  n_I n^a \Big) {^{(4)}}F^{\xu\xu IJ}_{ab} 
\nn \\
&=
\int d^4x \, \Big( \underaccent{\tilde}{N} \Tilde{ E}^a_I \Tilde{E}^b_J {^{(4)}}F^{\xu\xu IJ}_{ab}
- 2 N \Tilde{E}^b_J  n_I n^a  {^{(4)}}F^{\xu\xu IJ}_{ab} \Big)
\nn \\ &=
\int d^4x \, \Big( \underaccent{\tilde}{N} \Tilde{ E}^a_I \Tilde{E}^b_J {^{(4)}}F^{\xu\xu IJ}_{ab}
- 2 N \Tilde{E}^b_J  n_I n^a  (-i*{^{(4)}}F^{\xu\xu IJ}_{ab}) \Big)
\nn \\ &=
\int d^4x \, \Big( \underaccent{\tilde}{N} \Tilde{ E}^a_I \Tilde{E}^b_J {^{(4)}}F^{\xu\xu IJ}_{ab}
- 2 N \Tilde{E}^b_J  n_I n^a  \left( -\frac{i}{2} \epsilon^{IJ}_{\xu \xu MN} {^{(4)}}F^{\xu\xu MN}_{ab} \right) \Big)
\nn \\ &=
\int d^4x \, \Big( \underaccent{\tilde}{N} \Tilde{ E}^a_I \Tilde{E}^b_J {^{(4)}}F^{\xu\xu IJ}_{ab}
+i N \Tilde{E}^b_J  n^a n_I \epsilon^{IJ}_{\xu \xu MN} {^{(4)}}F^{\xu\xu MN}_{ab} \Big)
\nn \\ &=
\int d^4x \, \Big( \underaccent{\tilde}{N} \Tilde{ E}^a_I \Tilde{E}^b_J {^{(4)}}F^{\xu\xu IJ}_{ab}
+i N \Tilde{E}^b_J  n^a n_I (-\epsilon^{J\xu\xu \xu I}_{\xu MN}) {^{(4)}}F^{\xu\xu MN}_{ab} \Big).
}
Contracting the Levi-Civita tensor with the internal index (identity \ref{id:epsi}) and using identity \ref{id:time like} to rewrite $n^a$, the previous expression becomes
\eq{
S&=
\int d^4x \, \Big( \underaccent{\tilde}{N} \Tilde{ E}^a_I \Tilde{E}^b_J {^{(4)}}F^{\xu\xu IJ}_{ab}
+i N \Tilde{E}^b_J  \frac{(t^a-N^a)}{N}(-\epsilon^{J}_{\xu MN}) {^{(4)}}F^{\xu\xu MN}_{ab} \Big)
\nn \\ &=
\int d^4x \, \Big( \underaccent{\tilde}{N} \Tilde{ E}^a_I \Tilde{E}^b_J {^{(4)}}F^{\xu\xu IJ}_{ab}
+i \Tilde{E}^b_J  N^a \epsilon^{J}_{\xu MN}{^{(4)}}F^{\xu\xu MN}_{ab} - i \Tilde{E}^b_J   \epsilon^{J}_{\xu MN} t^a {^{(4)}}F^{\xu\xu MN}_{ab} \Big)\label{action x}
}

\begin{center}
 \framebox{ 
\parbox[t][21.0cm]{15.50cm}{
\addvspace{0.2cm} %\centering%
\subsection{Proving $e=\sqrt{q}N$} \label{e=qn}
First, we need to show the determinant of $g_{ab}$ and $q_{ab}$ respectively can be written with the Levi-Civita tensors as follows 
\eq{
g &= \epsilon^{abcd}g_{0a}g_{1b}g_{2c}g_{3d} 
\nn \\
q &= \epsilon^{abc}q_{1a}q_{2b}q_{3c} 
}
To show this we will do a trivial example with a general 2 dimensional matrix $M_{ab}$. If we let
\eq{
M_{ab}=
\begin{bmatrix}
M_{11} & M_{12} \\
M_{21} & M_{22}
\end{bmatrix},
\qquad 
\epsilon^{ab}=
\begin{bmatrix}
0 & 1 \\
-1 & 0
\end{bmatrix} ,
}
then
\eq{
\epsilon^{ab}M_{a1}M_{b2}
&=
\epsilon^{11}M_{11}M_{12} + \epsilon^{12}M_{11}M_{22} +
\epsilon^{21}M_{21}M_{12} +
\epsilon^{22}M_{21}M_{22}
\nn \\ &=
(0)M_{11}M_{12} + (1)M_{11}M_{22} +
(-1)M_{21}M_{12} +
(0)M_{21}M_{22}
\nn \\ &=
M_{11}M_{22} -
M_{21}M_{12}
\nn \\ &=
\det(M_{ab}) = M
}
This result generalizes to higher dimensions. To complete this proof we start with the determinant of the metric $g$ expanded with the Levi-Civita tensors, and in the following line we expand each metric with identity \ref{id: qab}
\eq{
g &= \epsilon^{abcd}g_{0a}g_{1b}g_{2c}g_{3d} 
\nn \\ &= \epsilon^{abcd}(q_{0a}-n_0n_a) (q_{1b}-n_1n_b) (q_{2c}-n_2 n_c) (q_{3d}-n_3n_d)
}
The $q_{0a}$ term vanishes since its indices are only spatial. Then upon expanding the parenthesis, all the terms involving the product of two normal vectors such as $n_a n_b, n_a n_c$, etc. will vanish automatically because of the total antisymmetry of $\epsilon^{abcd}$.  Thus we are only left with
\eq{
g&= -\epsilon^{abcd}n_0n_aq_{1b}q_{2c}q_{3d}
\nn \\ &=
-\epsilon^{0bcd}n_0n_0q_{1b}q_{2c}q_{3d}
\nn \\ &=
-(n_0)^2 \epsilon^{bcd}q_{1b}q_{2c}q_{3d}
\nn \\ &=
-N^2 \epsilon^{bcd}q_{1b}q_{2c}q_{3d}
\nn \\ g &=
-N^2 q,
}
where in the second line we set $a=0$ because $b,c,d$ are all spatial as imposed by the spatial metric $q_{1b}q_{2c}q_{3d}$, therefore $a$ can only be zero. Now $\epsilon^{0bcd}$ would behave just like a 3d epsilon tensor so we identify it simply as $\epsilon^{bcd}$.  In the next line we use identity \ref{id:n0}. From this result we can finally deduce that
\eq{
e=\sqrt{-g} = N\sqrt{q}
}
}}
\end{center}

Again, using the generalized Cartan's identity (\ref{lie}), we can rewrite $t^a {^{(4)}}F^{\xu\xu MN}_{ab}$ in the last term of Eq. (\ref{action x}) as follows
\eq{
S&=
\int d^4x \, \Big( \underaccent{\tilde}{N} \Tilde{ E}^a_I \Tilde{E}^b_J {^{(4)}}F^{\xu\xu IJ}_{ab}
+i \Tilde{E}^b_J  N^a \epsilon^{J}_{\xu MN}{^{(4)}}F^{\xu\xu MN}_{ab} - i \Tilde{E}^b_J   \epsilon^{J}_{\xu MN} \left( \mathcal{L}_t\, {^{(4)}}A_b^{MN} {-^{(4)}}D_b ({^{(4)}}A^{MN}_a t^a) \right) \Big)
\nn \\ &=
\int d^4x \, \Big( \underaccent{\tilde}{N} \Tilde{ E}^a_I \Tilde{E}^b_J {^{(4)}}F^{\xu\xu IJ}_{ab}
+i \Tilde{E}^b_J  N^a \epsilon^{J}_{\xu MN}{^{(4)}}F^{\xu\xu MN}_{ab} 
- i \Tilde{E}^b_J   \epsilon^{J}_{\xu MN}  \mathcal{L}_t\, {^{(4)}}A_b^{MN} 
\nn \\
&\quad
+ (- i \Tilde{E}^b_J   \epsilon^{J}_{\xu MN} )  \left({-^{(4)}}D_b ({^{(4)}}A^{MN}_a t^a) \right) \Big)
}
Now we integrate by parts (proven in the following boxed section \ref{byparts}) in the last term and act the covariant derivative on densitized triad and epsilon tensor giving us the following. 

\eq{
S&=
\int d^4x \, \Big( \underaccent{\tilde}{N} \Tilde{ E}^a_I \Tilde{E}^b_J {^{(4)}}F^{\xu\xu IJ}_{ab}
+i \Tilde{E}^b_J  N^a \epsilon^{J}_{\xu MN}{^{(4)}}F^{\xu\xu MN}_{ab} 
- i \Tilde{E}^b_J   \epsilon^{J}_{\xu MN}  \mathcal{L}_t\, {^{(4)}}A_b^{MN} 
\nn \\
&\quad
+ \, {^{(4)}}A^{MN}_a t^a\, {^{(4)}}D_b (-i\Tilde{E}^b_J   \epsilon^{J}_{\xu MN})  \Big)
}

\begin{center}
 \framebox{ 
\parbox[t][15.0cm]{15.50cm}{
\addvspace{0.2cm} %\centering%
\subsection{Proving integration by parts with a covariant derivative} \label{byparts}
Here we will prove the following integration by parts is valid up to surface terms
\eq{
\int d^4x (-i \tilde{E}^b_J \epsilon^J_{\xu MN}) ( {-^{(4)}} D_a( {^{(4)}} A_a^{\xu MN} t^a ) ) = \int d^4x ({^{(4)}} A_a^{\xu MN}t^a ) D_a( -i \tilde{E}^b_J \epsilon^K_{\xu MN} ) \,,
}
in order to do so we will first define,
\eq{
\Pi^a_{\xu IJ} := -i \tilde{E}^b_K \epsilon^K_{\xu IJ} \,,
}
and recall,
\eq{
D_a A_b^{\xu IJ} &= \del_a A_b^{\xu IJ} + A^{\xu I}_{a \xu K} A_b^{\xu KJ} + A_{aK}^{\xu \xu J}A_b^{\xu IK} 
%\nn \\
%D_a \Pi^b_{\xu IJ} &= \del_a \Pi^b_{\xu IJ} + A_{aI}^{\xu\xu K} \Pi^b_{\xu KJ} + A_{a \xu J}^{\xu K} \Pi^b_{\xu IK}\,.
}
Now we can show
\eq{
 - \Pi^a_{\xu IJ} D_a (A_b^{\xu IJ }) &= - \Pi^a_{\xu IJ} \left(\del_a (A_b^{\xu IJ}t^b )
 + A^{\xu I}_{a \xu K} (A_b^{\xu KJ}t^b) + A_{aK}^{\xu \xu J} (A_b^{\xu IK} t^b)
 \right)
 \nn \\ &=
 - \Pi^a_{\xu IJ} \del_a (A_b^{\xu IJ}t^b )
 - \Pi^a_{\xu IJ} A^{\xu I}_{a \xu K} (A_b^{\xu KJ}t^b) 
 - \Pi^a_{\xu IJ} A_{aK}^{\xu \xu J} (A_b^{\xu IK} t^b)
 \nn \\ &=
 - \Pi^a_{\xu IJ} \del_a (A_b^{\xu IJ}t^b )
 + \Pi^a_{\xu JI} A^{\xu I}_{a \xu K} (A_b^{\xu KJ}t^b) 
 - \Pi^a_{\xu IJ} A_{a \xu K}^{\xu  J} (A_b^{\xu KI} t^b),
}
where in the third line we flipped $I$ and $J$ indices on the second term which changes its sign, on the last term we flipped $J$ and $K$ indices on the first $A$, and flipped the $I$ and $K$ indices on the second $A$, keeping the sign the same. Notice the last two terms differ only by a naming of summed over indices, namely $I$ and $J$, which means they cancel out, leaving us with just the partial derivative
\eq{
- \Pi^a_{\xu IJ} D_a (A_b^{\xu IJ }t^b) &= - \Pi^a_{\xu IJ}\del_a (A_b^{\xu IJ}t^b ).
}
Showing the covariant derivative is equivalent to the partial derivative, in this case, shows that we can preform integration by parts on the covariant derivative just like we normally would with a partial derivative.
}}
\end{center}

Next, we rearrange the order of the terms and highlight $- i \Tilde{E}^b_J \epsilon^{J}_{\xu MN}$ wherever it's present. 
\eq{
S&=
\int d^4x \, \Big( (- i \Tilde{E}^b_J   \epsilon^{J}_{\xu MN} )  \mathcal{L}_t\, {^{(4)}}A_b^{MN}
+ \underaccent{\tilde}{N} \Tilde{ E}^a_I \Tilde{E}^b_J {^{(4)}}F^{\xu\xu IJ}_{ab}
\nn \\
&\quad -N^a (-i \Tilde{E}^b_J  \epsilon^{J}_{\xu MN}){^{(4)}} F^{\xu\xu MN}_{ab} 
+ \, {^{(4)}}A^{MN}_a t^a\, {^{(4)}}D_b (-i\Tilde{E}^b_J   \epsilon^{J}_{\xu MN})  \Big).
}
We are very close to the point where we can identify the configuration variable and its canonical momenta, but before that we will further simplify the action by taking the 3-dimensional projections on $\Sigma$ of our 4-dimensional fields.

\section{External Pullback}
Now we are tasked with performing a $3+1$ decomposition which requires us to make all of the external indices spatial i.e. pull them back to the 3-dimensional hypersurface. To begin we first introduce some identities as follows.
\vspace{-0.5cm}
\begin{center}
 \framebox{ 
\parbox[t][9.0cm]{15.50cm}{
\addvspace{0.2cm} %\centering%
\subsection*{Identities for pullback} \label{}
First we have two projection operators, which when contracted with each other act as one
\eq{
q^a_c q^c_b &= (\delta^a_c + n^an_c)(\delta^c_b + n^c n_b)
\nn \\ &= \delta^a_b + n^an_b + n^an_b + n^a n_c n^c n_b
\nn \\ &= \delta^a_b + 2n^an_b - n^a n_b
\nn \\ &= \delta^a_b + n^an_b
\nn \\ q^a_c q^c_b&= q^a_b \label{id:2q}.
}
Next, we have the Lie derivative with respect to time-like vector field $t^a$ to vanish when acting on the projection operator 
\eq{
\mathcal{L}_t q^a_b &= 0.
}
This is due to the combined effects of acting the projection operator $q_b^a$, which forces an object onto a constant-time spatial hypersurface $\Sigma_t$, and then followed by the Lie derivative along $t^a$, which obviously vanishes.}
}
\end{center}

Starting with the first term in the action, with the Lie derivative, we can show (Suppressing the $-i$)
\eq{
 \Tilde{E}^b_J   \epsilon^{J}_{\xu MN}  \mathcal{L}_t\, {^{(4)}}A_b^{MN} &=
q^b_a \Tilde{e}^a_J   \epsilon^{J}_{\xu MN}  \mathcal{L}_t\, {^{(4)}}A_b^{MN}
\nn \\ &=
q^b_c q^c_a \Tilde{e}^a_J   \epsilon^{J}_{\xu MN}  \mathcal{L}_t\, {^{(4)}}A_b^{MN}
\nn \\ &=
 \Tilde{E}^c_J   \epsilon^{J}_{\xu MN}  \mathcal{L}_t ( {q^b_c \, ^{(4)}}A_b^{MN} )
\nn \\ &=
 \Tilde{E}^c_J   \epsilon^{J}_{\xu MN}  \mathcal{L}_t A_c^{MN} ,
}
where in the last line the projection operator acts on the connection and gives us the projection of $A_a^{IJ}$ on the spatial hypersurface $\Sigma$ thus removing the superscript $(4)$. Next, we look at the second term in the action (suppressing the $\underaccent{\tilde}{N}$ )
\eq{
\Tilde{ E}^a_I \Tilde{E}^b_J {^{(4)}}F^{\xu\xu IJ}_{ab} &=
q^a_c q^b_d \Tilde{e}^c_I \Tilde{e}^d_J {^{(4)}}F^{\xu\xu IJ}_{ab}
\nn \\  &=
q^a_e q^e_c q^b_f q^f_d \Tilde{e}^c_I \Tilde{e}^d_J {^{(4)}}F^{\xu\xu IJ}_{ab}
\nn \\  &=
\Tilde{ E}^e_I \Tilde{E}^f_J F^{\xu\xu IJ}_{ef}
,
}
where we again used the projection operator to give us the projection of $F_{ab}^{\xu\xu IJ}$ on the spatial hypersurface. We rewrite the third term in the action as follows (again suppressing the $i$)
\eq{
N^a  \Tilde{E}^b_J  \epsilon^{J}_{\xu MN}{^{(4)}} F^{\xu\xu MN}_{ab}  & = 
q^a_d N^d  q^b_c \Tilde{e}^c_J  \epsilon^{J}_{\xu MN}{^{(4)}} F^{\xu\xu MN}_{ab}  
\nn \\ &=
q^a_d N^d  q^b_e q^e_c \Tilde{e}^c_J  \epsilon^{J}_{\xu MN}{^{(4)}} F^{\xu\xu MN}_{ab}
\nn \\ &=
 N^d   \Tilde{E}^e_J  \epsilon^{J}_{\xu MN} F^{\xu\xu MN}_{de},
}
where similar to the last calculation we use the projection operators on the curvature tensor. Next we look at the covariant derivative in the final term of the action  (suppressing $-i$)
\eq{
 {^{(4)}}D_b (\Tilde{E}^b_J   \epsilon^{J}_{\xu MN}) &= {^{(4)}}D_b (q^b_c \Tilde{e}^c_J   \epsilon^{J}_{\xu MN})
 \nn \\ &=
 {^{(4)}}D_b (q^b_e q^e_c \Tilde{e}^c_J   \epsilon^{J}_{\xu MN})
 \nn \\ &=
 {^{(4)}}D_b (q^b_d q^d_c \Tilde{e}^c_J   \epsilon^{J}_{\xu MN})
 \nn \\ &=
 q^b_d{^{(4)}}D_b (\Tilde{E}^d_J   \epsilon^{J}_{\xu MN})
 \nn \\ &=
 D_d (\Tilde{E}^d_J   \epsilon^{J}_{\xu MN}),
}
where this time around we use the projection operator on the covariant derivative.
Utilizing the 4 previous results, and renaming of summed over indices, we can now write the action as follows
\eq{
S&=
\int d^4x \, \Big( (- i \Tilde{E}^a_J   \epsilon^{J}_{\xu MN} )  \mathcal{L}_t A_a^{MN}
+ \underaccent{\tilde}{N} \Tilde{ E}^a_I \Tilde{E}^b_J F^{\xu\xu IJ}_{ab}
-N^a (-i \Tilde{E}^b_J  \epsilon^{J}_{\xu MN})F^{\xu\xu MN}_{ab}  
\nn \\
&\quad + \, {^{(4)}}A^{MN}_a t^a D_b (-i\Tilde{E}^b_J   \epsilon^{J}_{\xu MN})  \Big).
}
Noticing the common occurrences of the densitized triad contracted with an epsilon tensor in three out of four terms above, we will rewrite the second term to continue this pattern as follows
\eq{
\underaccent{\tilde}{N} (-i\Tilde{ E}^a_M \epsilon^{M \xu K}_{\xu I} ) (-i\Tilde{ E}^b_N \epsilon^{N}_{\xu K J} ) F_{ab}^{\xu \xu IJ}
}
which we prove below
\eq{
\underaccent{\tilde}{N} (-i\Tilde{ E}^a_M \epsilon^{M \xu K}_{\xu I} ) (-i\Tilde{ E}^b_N \epsilon^{N}_{\xu K J} ) F_{ab}^{\xu \xu IJ} 
&=
\underaccent{\tilde}{N} \Tilde{E}^a_M \Tilde{ E}^b_N \epsilon^{M \xu K}_{\xu I} \epsilon^{N}_{\xu JK1`} F_{ab}^{\xu \xu IJ}
\nn \\ &=
\underaccent{\tilde}{N} \Tilde{E}^a_M \Tilde{ E}^b_N ( \delta^{MN} \delta_{IJ} - \delta^M_J \delta^N_I ) F_{ab}^{\xu \xu IJ}
\nn \\ &= 
\underaccent{\tilde}{N} \Tilde{E}^a_M \Tilde{ E}^b_N (\delta^{MN} F_{ab}^{\xu \xu II}  - F_{ab}^{\xu \xu NM})
\nn \\ &= 
\underaccent{\tilde}{N} \Tilde{E}^a_M \Tilde{ E}^b_N ( - F_{ab}^{\xu \xu NM})
\nn \\ &= 
\underaccent{\tilde}{N} \Tilde{E}^a_M \Tilde{ E}^b_N  F_{ab}^{\xu \xu MN}
}
where in the third line the $F_{ab}^{\xu \xu II}$ vanishes due to the anitsymmetry in the internal indices. Subbing the previous result into the action gives us
\eq{
S=
\int d^4x &\, \Big( (- i \Tilde{E}^a_J   \epsilon^{J}_{\xu MN} )  \mathcal{L}_t A_a^{MN}
+ \underaccent{\tilde}{N} (-i\Tilde{ E}^a_M \epsilon^{M \xu K}_{\xu I} ) (-i\Tilde{ E}^b_N \epsilon^{N}_{\xu K J} ) F_{ab}^{\xu \xu IJ}
\nn \\
&\quad - N^a (-i \Tilde{E}^b_J  \epsilon^{J}_{\xu MN})F^{\xu\xu MN}_{ab}  
 \, {^{(4)}}A^{MN}_a t^a D_b (-i\Tilde{E}^b_J   \epsilon^{J}_{\xu MN})  \Big) \,.
}
\subsection{Canonical Momenta}
One may begin to notice now that our underlying Lagrangian density of the action is taking the canonical ``$p\dot{q} - H$" form, which allows us to identify our configuration variable and canonically conjugate momenta. We define our configuration variable as the connection $A^{MN}_a$ and the previously ``highlighted" term to be the canonical momenta
\eq{
\tilde{\Pi}^a_{MN} := -i \tilde{E}^a_I \epsilon^I_{\xu MN} \,.
}
After defining the Lie derivative with respect to time-like vector field as the ``time derivative" with a dotted notation, the action becomes
\eq{
S &=
\int d^4x\, \Big( \tilde{\Pi}^a_{ MN}  \dot{A}_a^{MN}
+ \underaccent{\tilde}{N} \tilde{\Pi}^{a K}_{M} \tilde{\Pi}^b_{KN} F_{ab}^{\xu \xu MN}
- N^a \tilde{\Pi}^b_{MN} F^{\xu\xu MN}_{ab}  
+ \, {^{(4)}}A^{MN}_a t^a D_b \tilde{\Pi}^b_{MN}  \Big).
}
Since $A^{MN}_a$ is self-dual we can conclude from Eq. (\ref{-+sd}) that the canonical momenta $\tilde{\Pi}^a_{MN}$ must also be self-dual otherwise all the terms in the Hamiltonian would be zero. Utilizing the fact the momenta is self-dual we can expand it using Eq. (\ref{sd part}) as follows
\eq{
{^{+}}\tilde{\Pi}^a_{MN} &= -i \tilde{E}^a_{I} \epsilon^I_{\xu MN}
\nn \\ &=
\tfrac{1}{2} \big( 
-i \tilde{E}^a_{I} \epsilon^I_{\xu MN} -i*(-i \tilde{E}^a_{I} \epsilon^I_{\xu MN})
\big)
\nn \\ &=
\tfrac{1}{2} \big( 
-i \tilde{E}^a_{I} \epsilon^I_{\xu MN} -\tfrac{i}{2}\epsilon_{MN}^{\xu\xu\xu KL} (-i \tilde{E}^a_{I} \epsilon^I_{\xu KL})
\big)
\nn \\ &=
-\tfrac{i}{2} 
\tilde{E}^a_{I} \epsilon^I_{\xu MN} -\tfrac{1}{4} \epsilon_{MN}^{\xu\xu\xu KL} \epsilon^I_{\xu KL} \tilde{E}^a_{I}
}
where in the second term
\eq{
\epsilon_{MN}^{\xu\xu\xu KL} \epsilon^I_{\xu KL} &= \epsilon_{MN}^{\xu\xu\xu KL} \epsilon^{\xu\xu IJ}_{ KL} n_J
\nn \\&=
(-2!)(\delta^I_M \delta^J_N - \delta^I_N \delta^J_M) n_J
\nn \\&=
-2(\delta^I_M n_N - \delta^I_N n_M).
}
Using the previous result gives us
\eq{
{^{+}}\tilde{\Pi}^a_{MN} &= 
-\tfrac{i}{2} 
\tilde{E}^a_{I} \epsilon^I_{\xu MN} -\tfrac{1}{4} (-2)(\delta^I_M n_N - \delta^I_N n_M) \tilde{E}^a_{I}
\nn \\ &=
\tfrac{i}{2} 
\tilde{E}^a_{I} \epsilon^I_{\xu MN} +\tfrac{1}{2} (\tilde{E}^a_{M} n_N - \tilde{E}^a_{N} n_M)
\nn \\ &=
\tfrac{i}{2} 
\tilde{E}^a_{I} \epsilon^I_{\xu MN} + \tilde{E}^a_{[M} n_{N]} 
\nn \\ &=
-\tfrac{i}{2} 
\tilde{E}^a_{I} \epsilon^{\xu\xu\xu IJ}_{MN} n_J + \tilde{E}^a_{[M} n_{N]}
\nn \\ &=
-\tfrac{i}{2} 
 \epsilon^{\xu\xu\xu IJ}_{MN} \tilde{E}^a_{[I} n_{J]}  + \tilde{E}^a_{[M} n_{N]}
 \nn \\ &=
-i* \tilde{E}^a_{[M} n_{N]}  + \tilde{E}^a_{[M} n_{N]}
\nn \\ &=
\tilde{E}^a_{[M} n_{N]} -i* \tilde{E}^a_{[M} n_{N]}  
\nn \\ {^+}\tilde{\Pi}^a_{MN}&=
2 \tilde{E}^a_{[M} n_{N]}
}
where in the last step we again use Eq. (\ref{sd part}) and now identify $\tilde{\Pi}^a_{MN}$ as the self-dual part of $2\tilde{E}^a_{[M} n_{N]}$.

Now that we have identified the canonically conjugate pair of the self-dual theory, the self-dual connection $A^a_{MN}$ and its self-dual conjugate momenta $\tilde{\Pi}^a_{MN}$ which we just showed, in the second to last line, can be expressed as 
\eq{
\tilde{\Pi}^a_{MN} = \tilde{E}^a_{[M} n_{N]} -\tfrac{i}{2} 
 \epsilon^{\xu\xu\xu IJ}_{MN} \tilde{E}^a_{[I} n_{J]}, \label{pi expand}
}
we can define the basic Poisson bracket
\eq{
\{A^{IJ}_b(y),\tilde{\Pi}^a_{MN}(x) \} &= \int d^3z\, \left(
\frac{\delta A_b^{IJ}(y)}{\delta A_c^{KL}(z)} 
\frac{\delta \tilde{\Pi}^a_{MN}(x)}{\delta \tilde{\Pi}^c_{KL}(z)} - \frac{\delta A_b^{IJ}(y)}{\delta \tilde{\Pi}^c_{KL}(z)} \frac{\delta \tilde{\Pi}^a_{MN}(x)}{\delta A_c^{KL}(z)} \right) 
\nn \\ &=
\int d^3z\, \left(
\delta^c_b\delta^I_{[K}\delta^J_{L]} \delta^3(y-z) \,
\frac{\delta \Big( \tilde{E}^a_{[M} n_{N]} -\tfrac{i}{2} 
 \epsilon^{\xu\xu\xu IJ}_{MN} \tilde{E}^a_{[I} n_{J]}\Big)(x) }{\delta 2\tilde{E}^c_{[K} n_{L]}(z) }
 - 0 \right)
 \nn \\ &=
 \delta^c_b\delta^I_{[K}\delta^J_{L]}
 \frac{\delta \Big( \tilde{E}^a_{[M} n_{N]} -\tfrac{i}{2} 
 \epsilon^{\xu\xu\xu IJ}_{MN} \tilde{E}^a_{[I} n_{J]}\Big)(x) }{\delta 2\tilde{E}^c_{[K} n_{L]}(y) }
 \nn \\ &=
 \delta^c_b\delta^I_{[K}\delta^J_{L]} (\tfrac{1}{2}) \delta^a_c \Big( 
 \delta^{[K}_M \delta^{L]}_N - \tfrac{i}{2}\epsilon_{MN}^{\xu\xu \xu ST} \delta^{[K}_S \delta^{L]}_T
 \Big) \delta^3(x-y)
 \nn \\  \{A^{IJ}_b(y),\tilde{\Pi}^a_{MN}(x) \} &=
 \tfrac{1}{2}\delta^a_b  (\delta^I_{[M} \delta^J_{N]} - \tfrac{i}{2}\epsilon_{MN}^{\xu\xu \xu IJ} 
 ) \delta^3(x-y)
}
where in the second line we expand the canonical momenta in its equivalent forms and simply put $(x)$ outside of the parentheses since all functions in the parentheses are functions of $x$. The extra term is to make the Poisson bracket self-dual with respect to the internal indices.

\section{``Internal Pullback''}
Now that we have the external indices completely spatial i.e. pulled back to the spatial hypersurface, the next step is to do a similar procedure for the internal indices which we would like to call it the ``internal pullback''.
To simplify this process we will exploit the gauge freedom of the internal normal vector $n^I$. We choose the gauge so that
\eq{
n^I &:= \delta^I_0 \label{n delta} = \begin{bmatrix}
    1 \\ 0 \\ 0 \\ 0
\end{bmatrix}
}
and in this gauge since $\Tilde{E}^a_I n^I = 0$ (identity from the previous section in box \ref{triad proofs}) we can show that
\eq{
\Tilde{E}^a_I n^I = 0 &= \Tilde{E}^a_0 n^0 + \Tilde{E}^a_i n^i 
\nn \\ &=
\Tilde{E}^a_0 \delta^0_0 + \Tilde{E}^a_i \delta^i_0
\tilde{E}^a_0 
\nn \\ 
0 &= \Tilde{E}^a_0  \label{gauge E} \,
}
Where $\delta^i_0$ vanishes identically. This immediately allows us to identify the internal indices of the triad as purely spatial since in this gauge
\eq{
\tilde{E}^a_I &= \tilde{E}^a_i \label{id:e0}\,.
}

To begin the internal pullback of the connection we will explore the self-duality of the connection and identify some important relations. We begin by expanding the connection
\eq{
A_a^{IJ} &= -i*A_a^{IJ}
\nn \\ &=
-\tfrac{i}{2} \epsilon_{MN}^{\xu\xu IJ} A_a^{MN}
\nn \\ &=
-\tfrac{i}{2} (\epsilon_{0N}^{\xu\xu IJ} A_a^{0N} + \epsilon_{mN}^{\xu\xu IJ} A_a^{mN} )
\nn \\ &=
-\tfrac{i}{2} (\epsilon_{0n}^{\xu\xu ij} A_a^{0n} + \epsilon_{m0}^{\xu\xu ij} A_a^{m0} + \epsilon_{mn}^{\xu\xu IJ} A_a^{mn} )
\nn \\ &=
-\tfrac{i}{2} (2 \epsilon_{0nij} A_a^{0n} +  \epsilon_{mn}^{\xu\xu IJ} A_a^{mn} )
\nn \\ &=
-\tfrac{i}{2} (2 \epsilon_{nij} A_a^{n0} +  \epsilon_{mn}^{\xu\xu IJ} A_a^{mn} )
\nn \\ &=
-i \epsilon_{ijk} A_a^{k0} - \tfrac{i}{2}  \epsilon_{mn}^{\xu\xu IJ} A_a^{mn}
}
where in the third and fourth lines we separate the connection into its time and space parts with respect to the internal indices and take advantage of the fact that
\eq{
\epsilon_{0IJK} = \epsilon_{0ijk}\, := \epsilon_{ijk}\,,
}
since the epsilon tensor would vanish if $i, j, k$ were to be zero. It should also be noted that the spatial indices can be raised or lowered freely due to the sign convention of the metric we are using.

Now if we set $J=0$, in the previous expression for $A_a^{IJ}$, we can see that the \textit{electric part} or \textit{boost part} of the connection, with respect to the internal indices, is equal to the following
\eq{
A^{I0}_a = A_a^{i0}=
-i \epsilon_{i0k} A_a^{k0} - \tfrac{i}{2}  \epsilon_{mn}^{\xu\xu i0} A_a^{mn} = 
 - \tfrac{i}{2} \epsilon_{mn}^{\xu\xu i0} A_a^{mn} = \tfrac{i}{2} \epsilon_{mn i0} A_a^{mn} = \tfrac{i}{2} \epsilon_{ijk} A_a^{jk}
}
where $A^{I0}_a = A^{i0}_a $ because the antisymmetry of the internal indices ($A_a^{00}$ vanishes) and the epsilon in the first term vanishes because we are plugging zero into its spatial index $j$. Now taking a look at the \textit{magnetic part} or \textit{rotation part} of $A_a^{IJ}$ i.e. $I \rightarrow i,J \rightarrow j,$ we have
\eq{
A_a^{ij} = -i \epsilon_{ijk} A_a^{k0} - \tfrac{i}{2}  \epsilon_{mn}^{\xu\xu ij} A_a^{mn}
= -i \epsilon_{ijk} A_a^{k0}
}
where the second term vanishes because all of the indices in the 4-d epsilon tensor are spatial  (the four indices can only take on values of 1, 2, and 3 therefore there will always be indices that have the same value causing it to vanish). 

Putting these results on hold we will now split the connection into its boost and rotation parts again but in another way. Similar to how we found the canonical momenta $\tilde{\Pi}_{IJ}^a$ to be the self-dual part of $2\tilde{E}^a_{[I}n_{J]}$ we can say the connection $A^{IJ}_a$ is the self-dual part of $2A^a_{[I}n_{J]}$. This allows us to expand the connection using Eq. (\ref{sd part}) in a similar fashion to how we expanded the momenta in Eq. (\ref{pi expand}) as follows
\eq{
A_a^{IJ} &:= i( A_a^{[I}n^{J]} - \tfrac{i}{2} \epsilon_{KL}^{\xu\xu IJ} A_a^{[K}n^{L]} )
\nn \\&=
i( \tfrac{1}{2}(A_a^{I}n^{J} - A_a^{J}n^{I}) - \tfrac{i}{2} \epsilon_{KL}^{\xu\xu IJ} \tfrac{1}{2}(A_a^{K}n^{L} - A_a^{L}n^{K} ))
\nn \\&=
\tfrac{i}{2}(A_a^{I}\delta_0^{J} - A_a^{J}\delta_0^{I}) + \tfrac{1}{4} \epsilon_{KL}^{\xu\xu IJ} (A_a^{K}\delta_0^{L} - A_a^{L}\delta_0^{K} )
\nn \\&=
\tfrac{i}{2}(A_a^{I}\delta_0^{J} - A_a^{J}\delta_0^{I})
+ \tfrac{1}{4} 
( \epsilon_{K0}^{\xu\xu IJ} A_a^{K} - \epsilon_{0L}^{\xu\xu IJ} A_a^{L} )
\nn \\&=
 \tfrac{i}{2}(A_a^{I}\delta^{J}_0 - A_a^{J}\delta_0^{I}) + \tfrac{1}{2} \epsilon_{K0}^{\xu\xu IJ} A_a^K 
}
where in the third line we use our gauge choice  (Eq. \ref{n delta}) to get the Kronecker deltas. Also, an artificial factor of $i$ is inserted to simplify reality conditions which we will briefly touch upon at the very end.

Now, taking a look at the boost part of the connection we have
\eq{
A^{I0}_a = A^{i0}_a = \tfrac{i}{2}(A_a^{i}\delta^{0}_0 - A_a^{0}\delta_0^{i}) + \tfrac{1}{2} \epsilon_{K0}^{\xu\xu I0} A_a^K 
= \tfrac{i}{2} A^i_a \,.
} 
Looking at the rotation part again we have
\eq{
A^{ij}_a = \tfrac{i}{2}(A_a^{i}\delta^{j}_0 - A_a^{j}\delta_0^{i}) + \tfrac{1}{2} \epsilon_{K0}^{\xu\xu ij} A_a^K  = \tfrac{1}{2} \epsilon_{ijk} A_a^k
}
where the Kronecker deltas vanish identically. In summary, between the two times, we split the connection into its boost and rotation parts we have  
\eq{
A^{i0}&=\tfrac i2 \epsilon_{ijk}A^{jk} \qquad 
A^{ij}_a = -i \epsilon_{ijk} A^{k0}_a \nn \\
A^{i0}&=\tfrac i2 A^i_a \qquad \quad \xu
A^{ij}_a = \tfrac 12 \epsilon_{ijk} A^{k}_a
}
Setting the two results we got for the boost part equal to each other gives us
\eq{
A^{i0}_a = \tfrac{i}{2} \epsilon_{ijk} A_a^{jk} &= \tfrac{i}{2} A^i_a
\nn \\ 
 \epsilon_{ijk} A_a^{jk} &:= A^i_a
}
which we will consider a definition, or equivalently
\eq{
A_a^{ij} &:= \tfrac{1}{2} \epsilon_{ijk} A^k_a.
}
Then, setting the two results we got for the rotation part equal to each other gives us
\eq{
A^{ij}_a = -i \epsilon_{ijk}A_a^{k0} &= \tfrac{1}{2} \epsilon_{ijk} A^k_a,
\nn \\
A_a^{k0} &:= \tfrac{i}{2} A^k_a.
} 
To summarize the previous results we have
\begin{center}
 \framebox{ 
\parbox[t][3.7cm]{15.50cm}{
\addvspace{0.2cm} %\centering%
\begin{itemize}
    \item \textbf{Connection Definitions}
\end{itemize}
\vspace{-0.3cm}
\eq{
A_a^i &:= \epsilon_{ijk} A^{jk}_a ,\label{a3}
\\
A^{i0}_a &:= \tfrac{i}{2}A^i_a \label{a1},
\\
A_a^{ij} &:= \tfrac{1}{2} \epsilon_{ijk} A^k_a \label{a2}.
}
}}
\end{center}
These re-definitions will prove to be essential in rewriting the self-dual action in its most efficient form. In particular, the combination of Eq. (\ref{a1}) and Eq. (\ref{a2}) implies that the self-dual connection $A_a^{IJ}$ is characterized completely by an $SU(2)$ connection $A_a^{i}$, just like that of a Yang-Mills theory! This new connection will eventually take the place of the configuration variable in the self-dual formulation. 

Before we finish this subsection, it's worth pointing out a striking similarity here. The properties of the self-dual connection $A_a^{IJ}$ almost make it seem like an internal analog of the field strength tensor $F^{\mu\nu}$ from the electromagnetism. Recall the electric field and magnetic field vectors in relation to the field strength tensor, namely $E^i=F^{0i}, B_i = \frac 1 2 \epsilon_{ijk} F^{jk}$. This makes the earlier terminology with the ``electric part'' and ``magnetic part'' etc. quite appropriate.

\section{Self-Dual Action}
Utilizing our new definitions we can obtain the final form of the action known as the self-dual action. Starting from before we defined the canonical momenta $\Pi^a_{MN}$ we had
\eq{
S_{SD}&=
\int d^4x \, \Big( (- i \Tilde{E}^a_J   \epsilon^{J}_{\xu MN} )  \dot{A}_a^{MN}
+ \underaccent{\tilde}{N} \Tilde{ E}^a_I \Tilde{E}^b_J F^{\xu\xu IJ}_{ab}
-N^a (-i \Tilde{E}^b_J  \epsilon^{J}_{\xu MN})F^{\xu\xu MN}_{ab}  
\nn \\
&\quad + \, {^{(4)}}A^{MN}_a t^a D_b (-i\Tilde{E}^b_J   \epsilon^{J}_{\xu MN})  \Big).
}
Using our new definitions we can rewrite the first term as follows
\eq{
-i \tilde{E}^a_I \epsilon^I_{\xu MN} \dot{A}_a^{MN} &=
-i \tilde{E}^a_i \epsilon^i_{\xu MN} \dot{A}_a^{MN}
\nn \\ &= 
-i \tilde{E}^a_i \epsilon^i_{\xu MNJ} n^J \dot{A}_a^{MN}
\nn \\ &= 
-i \tilde{E}^a_i \epsilon^i_{\xu MNJ} \delta^J_0 \dot{A}_a^{MN}
\nn \\ &= 
-i \tilde{E}^a_i \epsilon^i_{\xu mn0} \dot{A}_a^{mn}
\nn \\ &= 
-i \tilde{E}^a_i \epsilon^i_{ jk} \dot{A}_a^{jk}
\nn \\ &= 
-i \tilde{E}^a_i \dot{A}_a^{i}
}
where in the first line we use Eq. (\ref{id:e0}), in the second line we use identity (\ref{id:epsi}), in the third we use identity (\ref{n delta}), and in the last line we use Eq.  (\ref{a3}). Next, the second term in the self-dual action can be rewritten as follows
\eq{
\underaccent{\tilde}{N} \Tilde{ E}^a_I \Tilde{E}^b_J F^{\xu\xu IJ}_{ab} &=
- \tfrac{i}{2} \underaccent{\tilde}{N} \Tilde{ E}^a_I \Tilde{E}^b_J \epsilon_{MN}^{\xu\xu\xu\, IJ} F^{\xu\xu MN}_{ab}
\nn \\ &=
- \tfrac{i}{2} \underaccent{\tilde}{N} \Tilde{ E}^a_i \Tilde{E}^b_j \epsilon_{MN}^{\xu\xu\xu\,ij} F^{\xu\xu MN}_{ab}
\nn \\ &=
- \tfrac{i}{2} \underaccent{\tilde}{N} \Tilde{ E}^a_i \Tilde{E}^b_j ( \epsilon_{0N}^{\xu\xu\xu ij}  F^{\xu\xu 0N}_{ab} + \epsilon_{mN}^{\xu\xu\xu ij}  F^{\xu\xu mN}_{ab} )
\nn \\ &=
- \tfrac{i}{2} \underaccent{\tilde}{N} \Tilde{ E}^a_i \Tilde{E}^b_j ( \epsilon_{0n}^{\xu\xu ij}  F^{\xu\xu 0n}_{ab} + \epsilon_{m0}^{\xu\xu\, ij}  F^{\xu\xu m0}_{ab} )
\nn \\ &=
- \tfrac{i}{2} \underaccent{\tilde}{N} \Tilde{ E}^a_i \Tilde{E}^b_j ( \epsilon_{nij0}  F^{\xu\xu n0}_{ab} + \epsilon_{mij0}  F^{\xu\xu m0}_{ab} )
\nn \\ &=
-i \underaccent{\tilde}{N} \Tilde{ E}^a_i \Tilde{E}^b_j \epsilon_{mij} F_{ab}^{\xu\xu mo}
}
where in the second line we use identity (\ref{id:e0}), and in the third and fourth lines we separate the time and spatial components with respect to the internal indices. Note the $F_{ab}^{\xu\xu 00}$ term does not appear since it vanishes due to the anti-symmetry of the internal indices and the $F_{ab}^{\xu\xu mn}$ terms do not appear since the 4-d epsilon tensor cannot have all spatial indices.
Now we define the ``boost-part curvature tensor" as follows 
\eq{
F_{ab}^{\xu\xu i0} &= \del_a A^{i0}_b - \del_b A^{i0}_a + A^{ik}_a A_{bk}^{\xu\xu 0} - A^{ik}_b A_{ak}^{\xu\xu 0}
\nn \\ &=
\del_a (\tfrac{i}{2}A^{i}_b) - \del_b (\tfrac{i}{2}A^{i}_a) + \tfrac{1}{2}\epsilon_{ikj}A^j_a ( \tfrac{i}{2} A^k_b) - \tfrac{1}{2}\epsilon_{ikj}A^j_b ( \tfrac{i}{2} A^k_a)
\nn \\ &=
\tfrac{i}{2} ( \del_a A^i_b - \del_b A^i_a + \tfrac{1}{2}\epsilon_{ikj}A^j_a A^k_b - \tfrac{1}{2}\epsilon_{ikj}A^j_b A^k_a )
\nn \\ &=
\tfrac{i}{2} ( \del_a A^i_b - \del_b A^i_a + \tfrac{1}{2}\epsilon_{ikj}A^j_a A^k_b + \tfrac{1}{2}\epsilon_{ikj}A^k_b A^j_a )
\nn \\ &=
\tfrac{i}{2} ( \del_a A^i_b - \del_b A^i_a + \epsilon_{ikj}A^j_a A^k_b )
\nn \\
&:= \tfrac{i}{2}F^i_{ab} 
}
where in the second line we use Eqs. (\ref{a1}) and (\ref{a2}), and in the last line we define $F^i_{ab} $. Using this result we can finish rewriting the second term in the action as
\eq{
\underaccent{\tilde}{N} \Tilde{ E}^a_I \Tilde{E}^b_J F^{\xu\xu IJ}_{ab} &= -i \underaccent{\tilde}{N} \Tilde{ E}^a_i \Tilde{E}^b_j \epsilon_{mij} F_{ab}^{\xu\xu mo}
\nn \\
&= \tfrac{1}{2} \underaccent{\tilde}{N} \Tilde{ E}^a_i \Tilde{E}^b_j \epsilon_{ijk}  F_{ab}^{\xu\xu k} 
}
Next, the third term in the action can be written as follows
\eq{
-N^a( -i \tilde{E}^b_I \epsilon^I_{\xu MN} ) F_{ab}^{\xu\xu MN} &=
i N^a \tilde{E}^b_I \epsilon^I_{\xu MNL} n^L F_{ab}^{\xu\xu MN}
\nn \\ &=
i N^a \tilde{E}^b_I \epsilon^I_{\xu MNL} \delta^L_0 F_{ab}^{\xu\xu MN}
\nn \\ &=
i N^a \tilde{E}^b_I \epsilon^I_{\xu MN0} F_{ab}^{\xu\xu MN}
\nn \\ &=
i N^a \tilde{E}^b_i \epsilon^i_{\xu mn} F_{ab}^{\xu\xu mn}
\nn \\ &=
iN^a \tilde{E}^b_i F_{ab}^i
}
where in the first line we use identity (\ref{id:epsi}) and in the second line we use the gauge choice of Eq. (\ref{n delta}) and Eq.  (\ref{id:e0}) to justify that the indices become spatial. Additionally in the last line  $\epsilon^i_{\xu mn} F_{ab}^{\xu\xu mn} = F_{ab}^i$ as proven below
\eq{
\epsilon^i_{\xu mn} F_{ab}^{\xu\xu mn} &= \epsilon^i_{\xu mn} ( \del_a A^{mn}_b - \del_b A^{mn}_a + A^{ml}_a A_{bl}^{\xu\xu n} - A^{ml}_b A_{al}^{\xu\xu n} )
\nn \\ &=
\del_a A^{i}_b - \del_b A^{i}_a 
+ \epsilon^i_{\xu mn} (A^{ml}_a A_{bl}^{\xu\xu n} 
-  A^{ml}_b A_{al}^{\xu\xu n})
\nn \\ &=
\del_a A^{i}_b - \del_b A^{i}_a 
+ \epsilon_{imn} (A^{ml}_a A_{b}^{l n} 
-  A^{ml}_b A_{a}^{ln})
\nn \\ &=
\del_a A^{i}_b - \del_b A^{i}_a 
+ \epsilon_{imn} \Big( (\tfrac12 \epsilon_{mlk} A^{k}_a)
(\tfrac12 \epsilon_{lnj} A^{j}_b)
-  (\tfrac12 \epsilon_{mlk} A^{k}_b)
(\tfrac12 \epsilon_{lnj} A^{j}_a)
\Big)
\nn \\ &=
\del_a A^{i}_b - \del_b A^{i}_a 
+ \tfrac14
\epsilon_{imn} (\epsilon_{mlk} \epsilon_{lnj}) ( 
A^{k}_a A^{j}_b -   A^{k}_b A^{j}_a)
\nn \\ &=
\del_a A^{i}_b - \del_b A^{i}_a 
+ \tfrac14
\epsilon_{imn} (-2(\delta_{mn}\delta_{kj}-\delta_{mj} \delta_{kn})) ( 
A^{k}_a A^{j}_b -   A^{k}_b A^{j}_a)
\nn \\ &=
\del_a A^{i}_b - \del_b A^{i}_a 
- \tfrac12
\epsilon_{imn} ( -\delta_{mj} \delta_{kn}) ( 
A^{k}_a A^{j}_b -   A^{k}_b A^{j}_a)
\nn \\ &=
\del_a A^{i}_b - \del_b A^{i}_a 
+ \tfrac12
\epsilon_{ijk} ( 
A^{k}_a A^{j}_b -   A^{k}_b A^{j}_a)
\nn \\ &=
\del_a A^{i}_b - \del_b A^{i}_a 
+ 
\epsilon_{ijk} 
A^{k}_a A^{j}_b
\nn \\ &= F_{ab}^i,
}
where in the second line we used Eq. (\ref{a3}) on the partial derivative terms, in the third line we freely raised and lowered spatial indices, in the fourth line we used Eq. (\ref{a2}), and in the seventh line the first pair of Kronecker deltas vanish because they create repeated indices on the epsilon tensor.

Finally, the fourth term in the action can be rewritten as follows
\eq{
{^{(4)}}A^{MN}_a t^a D_b (-i\Tilde{E}^b_I   \epsilon^{I}_{\xu MN}) &=
{^{(4)}}A^{MN}_a t^a D_b (-i\Tilde{E}^b_I   \epsilon^{I}_{\xu MNL}n^L)
\nn \\ &=
{^{(4)}}A^{MN}_a t^a D_b (-i\Tilde{E}^b_I   \epsilon^{I}_{\xu MNL}\delta^L_0)
\nn \\ &=
{^{(4)}}A^{MN}_a t^a D_b (-i\Tilde{E}^b_I   \epsilon^{I}_{\xu MN0})
\nn \\ &=
-i({^{(4)}}A^{mn}_a t^a) D_b (\Tilde{E}^b_i   \epsilon^{i}_{\xu mn})
\nn \\ &=
-i(\epsilon^{i}_{\xu mn} {^{(4)}}A^{mn}_a t^a) D_b \Tilde{E}^b_i   
\nn \\ &=
-i ({^{(4)}}A^{i}_a t^a) D_b \Tilde{E}^b_i 
\nn \\ &=
-i \lambda^{i} D_b \Tilde{E}^b_i 
}
where we use a similar procedure as the third term and use Eq. (\ref{a3}) in the third to last line. Additionally in the final line, we define $A^i_at^a := \lambda^i$ and drop the raised four since it is a Lagrange multiplier and thus a free parameter. With the previous re-definitions, we can write the final form of the self-dual action as follows
\eq{
S_{SD} &=\int d^4x \left( 
-i \tilde{E}^a_i \dot{A}_a^{i} 
+ \tfrac{1}{2} \underaccent{\tilde}{N} \Tilde{ E}^a_i \Tilde{E}^b_j \epsilon_{ijk}  F_{ab}^{k} 
+ iN^a \tilde{E}^b_i F_{ab}^i
- i \lambda^{i} D_b \Tilde{E}^b_i 
\right)
\nn \\ &=
\tfrac{1}{i} \int d^4x \left( 
 \tilde{E}^a_i \dot{A}_a^{i} 
+  \tfrac{i}{2} \underaccent{\tilde}{N} \epsilon_{ijk}\Tilde{ E}^a_i \Tilde{E}^b_j   F_{ab}^{k} 
- N^a \tilde{E}^b_i F_{ab}^i
+ \lambda^{i} D_a \Tilde{E}^a_i 
\right).\label{SD action}
}
This action turned out to be a special case of the more general action
\eq{
S=\tfrac{1}{\beta} \int d^4x \left( 
 \tilde{E}^a_i \dot{A}_a^{i} 
+  \tfrac{i}{2} \underaccent{\tilde}{N} \epsilon_{ijk}\Tilde{ E}^a_i \Tilde{E}^b_j   F_{ab}^{\xu\xu k} 
- N^a \tilde{E}^b_i F_{ab}^i
+ \lambda^{i} D_a \Tilde{E}^a_i 
\right),
}
where $\beta$ is known as the Barbero-Immirzi parameter. Comparing the two actions, we see that the self-dual action of Ashtekar corresponds to a specific choice of $\beta =i$. This choice would lead to a much simpler expression for the total Hamiltonian but at the cost of defining a complex theory of GR instead of the real GR, as we will discuss more in the end. In general, the Barbero-Immirzi parameter does not necessarily need to take on this value, as shown in later developments of the framework in the 1990s \cite{Barbero, Immirzi}. We hope to discuss more about that part of the development in a follow-up work in the future.

\section{Final Piece of The Puzzle}
The last step is to relate the self-dual formulation to the Palatini formulation via the tetrad and spin connection. From the original setup in Eq. (\ref{w+}), the self-dual connection is now the self-dual part of the 3-dimensional spin connection from the standard Palatini theory, i.e.
\eq{
A_{a}^{jk} = {^+}\omega_a^{ij}.
}
This lets us express the self-dual connection as follows
\eq{
A^i_a &= \epsilon_{ijk} A^{ij}_a
\nn \\ &=
\epsilon_{ijk} {^+}\omega_a^{ij}
\nn \\ &=
\epsilon_{ijk} \tfrac12 ( \omega_a^{ij} - \tfrac{i}{2} \epsilon_{MN}^{\xu\xu jk} \omega_a^{MN})
\nn \\ &=
\tfrac12\epsilon_{ijk} \omega_a^{ij} - \tfrac{i}{4} \epsilon_{ijk} ( \epsilon_{0N}^{\xu\xu jk} \omega_a^{0N} + \epsilon_{mN}^{\xu\xu jk} \omega_a^{mN} )
\nn \\ &=
\tfrac12\epsilon_{ijk} \omega_a^{ij} - \tfrac{i}{4} \epsilon_{ijk} ( \epsilon_{0n}^{\xu\xu jk} \omega_a^{0n} + \epsilon_{m0}^{\xu\xu jk} \omega_a^{m0} )
\nn \\ &=
\tfrac12\epsilon_{ijk} \omega_a^{ij} - \tfrac{i}{4} \epsilon_{ijk} ( 2\epsilon_{0n jk} \omega_a^{0n})
\nn \\ &=
\tfrac12\epsilon_{ijk} \omega_a^{ij} - \tfrac{i}{4} \epsilon_{ijk} ( -2\epsilon_{njk0} \omega_a^{0n})
\nn \\ &=
\tfrac12\epsilon_{ijk} \omega_a^{ij} + \tfrac{i}{2} \epsilon_{ijk} \epsilon_{njk} \omega_a^{0n}
\nn \\ &=
\tfrac12\epsilon_{ijk} \omega_a^{ij} + \tfrac{i}{2} (2\delta_{in}) \omega_a^{0n}
\nn \\ &=
\Gamma_a^i + i \omega_a^{0i}
\label{gamma w}
}
where in the third line we use Eq. (\ref{sd part}), and in the fourth and fifth lines we separate the spatial and boost parts of the spin connection with respect to the internal indices (where the $\omega_a^{mn}$ term vanishes due to totally antisymmetric spatial indices on the epsilon tensor). Additionally, in the last line, we define
\eq{
\Gamma_a^i := \tfrac12\epsilon_{ijk} \omega_a^{ij} .
}

Recall the definition of the tetrad-compatible spin connection (Eq. (\ref{compact1})) from Section 5 
\eq{
\omega^I_{a J} &= -e^{b}_J( 
\del_a e_{b}^I + \Gamma^c_{a b} e_{c}^I),
}
which lets us write
\eq{
\omega_{a}^{0i} = \omega_{ai}^{0} = -e^{b}_i( 
\del_a e_{b}^0 + \Gamma^c_{a b} e_{c}^0),
}
where we freely lower the spatial index $i$. To further simplify the previous expression we can show that
\eq{
e^0_a &= \eta^{I0}e_{Ia} 
\nn \\&= 
\eta^{00}e_{0a} 
\nn \\&= 
- e_{0a} 
\nn \\&= 
-e_0^b g_{ab}
\nn \\&= 
-e_0^b( q_{ab} - n_a n_b )
\nn \\&= 
-E_0^a + n_a n_0
\nn \\&= 
0 - n_a n^0
\nn \\&= 
-n_a \delta^0_0
\nn \\ 
e^0_a &= -n_a 
}
where in the fifth line we use identity (\ref{id: qab}), and in the seventh line the triad vanishes due to Eq. (\ref{gauge E}). Subbing this result into the spin connection gives us
\eq{
\omega_{a}^{0i} = \omega_{ai}^{0} &= -e^{b}_i( 
\del_a (-n_b) + \Gamma^c_{a b} (-n_c))
\nn \\ &=
e^{b}_i( 
\del_a n_b + \Gamma^c_{a b} n_c))
\nn \\ &=
e^{b}_i \Del_a n_b.
}
This expression looks rather close to the extrinsic curvature (refer to Eq. (\ref{extrinsic2})) of a hypersurface. It suggests us to define a quantity $K^i_a$ in the following
\eq{
K_a^i &= E^b_i K_{ab}
\nn \\ &=
( q^b_e e^e_i )( q^c_a q^d_b \Del_c n_d )
\nn \\ &=
e^e_i q^c_a \delta^d_e \Del_c n_d
\nn \\ &=
e^e_i \Del_a n_e,
}
where $K_{ab}$ is the extrinsic curvature of a spatial hypersurface with spatial metric $q_{ab}$ and a normal vector $n^a$.
The last expression shows that $K_a^i$ is exactly the boost part of the spin connection
\eq{
K_a^i &= \omega_a^{0i}.
}
Using this new quantity in Eq. (\ref{gamma w}) we can now express the self-dual connection as
\eq{
A^i_a = \Gamma^i_a + i K^i_a.
}
This form of the self-dual connection, together with the densitized triad $\tilde{E}_i^a$ constitute Ashtekar's new variables for General Relativity, where the connection plays the role of the configuration variable, similar to the gauge potential in electromagnetism, and the densitized triad plays the role of the canonically conjugate momentum, similar to the role of the ``electric field". 

As mentioned before, later developments extended Ashtekar's framework with a general Barbero-Immiriz parameter $\beta$ that is not necessarily complex such that 
\eq{
A^i_a = \Gamma^i_a + \beta K^i_a,
}
which reduces back to Ashtekar's self-dual connection upon setting $\beta = i$ once again.

With this last piece of the puzzle unraveled, we have finally reached the point of obtaining the Hamiltonian theory of GR using Ashtekar's new variables. From the self-dual action Eq. (\ref{SD action}), we immediately read off the total Hamiltonian 
\eq{
H_T = \tfrac{1}{i} \int d^3x \left( 
 -  \tfrac{i}{2} \underaccent{\tilde}{N} \epsilon_{ijk} \Tilde{ E}^a_i \Tilde{E}^b_j  F_{ab}^{\xu\xu k} 
+ N^a \tilde{E}^b_i F_{ab}^i
- \lambda^{i} D_a \Tilde{E}^a_i 
\right),
}
and the fundamental Poisson brackets in terms of the canonically conjugate variables $A_a^i$ and $E_i^a$
\eq{
\{A^i_a(x) , A^b_j(y)\} &= 0
\\
\{\tilde{E}^i_a(x) , \tilde{E}^b_j(y)\} &= 0
\\
\{A^i_a(x) , \tilde{E}^b_j(y)\} &= i \delta^i_j \delta^b_a \delta^3(x-y) .
}
Unmistakably, the total Hamiltonian is a linear combination of constraints where $\underaccent{\tilde}{N}, N^a, \lambda^i$ are all Lagrange multipliers. In terms of the new variables, the constraints are 
\eq{
S &= \epsilon_{ijk} \tilde{E}^a_i \tilde{E}^b_j F_{ab}^k \approx 0,
\\
V_a &= \tilde{E}^b_i F_{ab}^i \approx 0,
\\
G_i &= D_b \tilde{E}^b_i \approx 0.
}

The forms of the total Hamiltonian and the set of constraints are very similar to those in the Palatini case. But do not be deceived! The implications of these two formulations are drastically different.
First of all, the self-dual formulation has the clear advantage that the total Hamiltonian, or equivalently, the set of all the constraints, is in simple polynomial form in terms of the new phase space variables. Besides that, the canonical momenta $\tilde{E}^a_i$ in the self-dual formulation are completely arbitrary and thus do not lead to any additional constraints that require further Dirac analysis. 

The next step is to evaluate the constraint algebra. Here, one needs to perform the usual trick of introducing smeared constraints below \cite{Romano}
\eq{
C(\underaccent{\tilde}{N}) &:= \frac {1}{2} \int d^3x\, \underaccent{\tilde}{N} \epsilon_{ijk} \tilde{E}^a_i \tilde{E}^b_j F_{ab}^k, \\
C'(\vec N) &:= -i\int d^3x\, N^a \tilde{E}^b_i F_{ab}^i, \\
G(v) &:=-i\int d^3x\, v^i (D_a \tilde{E}^a_i).
}
Let's talk about the geometric interpretations of these constraints first. Starting from the most obvious one, the Gauss constraint $G(v)$ generates internal $SU(2)$ rotations as ordinary gauge transformations similar to that in Yang-Mills theory. Then, the scalar constraint $C(\underaccent{\tilde}{N})$ generates ``time evolutions'' in the direction normal to the spatial hypersurface $\Sigma$, for which reason it is also commonly referred to as the Hamiltonian constraint. The vector constraint $C'(\vec N)$, however, does not have a direct geometric interpretation on its own. In fact, it would take the linear combination of the vector and the Gauss constraints to define another constraint that carries a direct geometric interpretation:
\eq{
C(\vec N):=C'(\vec N) -G(N^i),
}
where $N^i:=N^a A^i_a$. This constraint is called the diffeomorphism constraint. It can be shown that it generates ``time evolutions'' that are tangent to the spatial hypersurface $\Sigma$, namely a diffeomorphism generated by the shift vector $N^a$. 

Next, by evaluating the Poisson brackets between $G(v), C(\underaccent{\tilde}{N})$, and $C(\vec N)$, one can show that they indeed form a closed Poisson algebra \cite{Ashtekar}
\eq{
\{G(v), G(w)\} &= G([v,w]),\\
\{G(v), C(\underaccent{\tilde}{N})\} &= 0,\\
\{C(\vec N), G(v)\} &= C(\mathcal{L}_{\vec N} v)\\
\{C(\vec N), C(\vec M)\} &= C([\vec N, \vec M])\\
\{C(\vec N), C(\underaccent{\tilde}{M})\} &= C(\mathcal{L}_{\vec N}\underaccent{\tilde}{M})\\
\{C(\underaccent{\tilde}{N}), C(\underaccent{\tilde}{M})\} &= C'(\vec K) = C(\vec K) + G(\vec K),
}
where $M$ serves the same role as the lapse or shift and $\vec K = \tilde{E}^a_i \tilde{E}^{bi}(\underaccent{\tilde}{N}\del^a \underaccent{\tilde}{M}-\underaccent{\tilde}{M}\del^a \underaccent{\tilde}{N})$. Therefore the constraints are all of first class, as expected.

At this point, it may seem that we have finally arrived at a Hamiltonian theory of general relativity that resembles a Yang-Mills gauge theory. However, we are not quite done yet. As mentioned at the beginning of this section, formulating GR in terms of Ashtekar's new, self-dual variables, which are complex variables by construction, means that what we have at hand is a complex theory of GR. The underlying metric tensor constructed by the triads would be generally complex as well. This of course does not describe the real 4-dimensional space-time. In order to connect back to the real GR, additional steps have to be taken. This was done by introducing additional conditions on the new variables known as reality conditions, originally by Ashtekar \cite{Ashtekar}. Specifically, we need 
\eq{
(\tilde{E}^a_i \tilde{E}^{bi})^* &= \tilde{E}^a_i \tilde{E}^{bi},\\ [\epsilon_{ijk}\tilde{E}^a_i D_c(\tilde{E}^c_j \tilde{E}^{bk})]^* &= -\epsilon_{ijk}\tilde{E}^a_i D_c(\tilde{E}^c_j \tilde{E}^{bk}),
}
where the first condition simply required the 3-dimensional spatial metric $q^{ab}$ to be real, and the second condition essentially required the extrinsic curvature of the spatial metric to be real as well \cite{Ashtekar}. 

Most of the important results in this section regarding the constraint algebra and reality conditions are worth exploring in greater detail. However, as pointed out in the beginning, the focus of this paper is to obtain the Hamiltonian theory of GR in terms of Ashtekar's new variables, up to the point of the total Hamiltonian, we are going to skip the detailed discussions on these interesting aspects for now and we hope to return to them in the future.

\section{Looking Ahead}

In summary, we have developed a step-by-step user guide to understand the self-dual formulation of GR in terms of the densitized triad and Ashtekar connection. We finally have a well-behaved Hamiltonian theory of GR that is ready for canonical quantization. Now, in terms of the progress made in the area of quantum gravity, the early canonical gravity framework is only a small step in history, but it was a step in the right direction crucial to the later developments in the field that eventually led to LQG. It is our belief that ANYONE taking on the journey of pursuing quantum gravity research should understand this foundational work to its great detail. In a similar sense, what we have achieved here is also a small step, but a necessary first step to bridge a long-awaited gap in the literature. 

This work is far from being complete. As a matter of fact, it is only the beginning of potentially a series of works following the same spirit of this one. Firstly, we are hoping to look into the aspect of constraint algebra of the self-dual formulation to give a more thorough account of the constraint analysis in a similar level of detail to the Maxwell and Yang-Mills sections. Secondly, we look to showcase another foundational part in the early LQG development in detail, namely the Loop Representation. One interesting application of the loop representation is the construction of quantum operators of area and volume \cite{Rovelli:1994, Rovelli:1995ac}. These lead to a discrete structure of space-time geometry, whose application in the context of black hole thermodynamics reproduced the famous Bekenstein-Hawking entropy relation, giving the black hole entropy a clear geometric interpretation and microscopic explanation \cite{Rovelli:1996dv}. 
Lastly, we would like to discuss the exciting works on quantum Hamiltonian constraint in the loop representation, which upon acting on the quantum wavefunction of the universe gives reformulations of the Wheeler-DeWitt equations, whose solutions should hypothetically describe the quantum nature of the universe itself! We look forward to exploring all these exciting developments of LQG in the near future.

\subsection*{ACKNOWLEDGMENTS}
We are grateful to Robert Echols, Sean Echols, and Benjamin Shlaer, for many helpful discussions during the preparation of this work. This work is supported in part by the 2023 Frost Summer Undergraduate Research Program of California Polytechnic State University.

\end{document}